# Um Estudo sobre Atividades Participativas para Soluções IoT para o *Home care* de Pessoas Idosas

## - Relatório Técnico -

Este documento corresponde ao Relatório Técnico sobre atividades participativas desenvolvidas no âmbito do Projeto de Pesquisa da aluna Renata de Podestá Gaspar no curso de Mestrado em Ciência da Computação da Faculdade Campo Limpo Paulista.

Campo Limpo Paulista, 20 de Outubro de 2018.

Renata de Podestá Gaspar

Rodrigo Bonacin (Orientador)
Vinícius Gonçalves (Co-Orientador)

***Resumo**. O envelhecimento populacional no Brasil e no mundo é uma realidade e ocorre ao mesmo tempo e em velocidade proporcional aos avanços e evoluções na tecnologia. Desta forma, surgem oportunidades de novas soluções para o público idoso, tais como inovações em Home Care e selfcare. A Internet das Coisas (do inglês, Internet of Things – IoT) é uma revolução tecnológica que tem por objetivo conectar dispositivos eletrônicos utilizados no dia-a-dia (como aparelhos eletrodomésticos, eletro portáteis, máquinas industriais, meios de transporte etc) à Internet. Com ela, é possível promover maior autonomia, segurança e qualidade de vida. Entretanto, o design de soluções de IoT para Home Care de pessoas idosas traz novos desafios a área Interação Humano-Computador. Idosos são um grupo de usuários heterogêneos, exigentes e em geral resistentes quanto ao uso de tecnologia e soluções interativas. Portanto, acredita-se que desenvolver soluções para esse público exige o planejamento e aplicação de métodos, atividades, materiais específicos, que promovam interesse e engajamento para o desenvolvimento de soluções que sejam efetivamente úteis e usáveis. Diante disto, este relatório técnico tem o objetivo de detalhar atividades desenvolvida como estudo de caso para avaliação do Método IoT-PMHCS, desenvolvido no contexto do projeto intitulado "Um Método Baseado em Design Participativo, Personas e Semiótica para Soluções IoT para o Home Care de Pessoas Idosas" no programa de Mestrado em Ciências da Computação da Faculdade de Campo Limpo Paulista. O relatório inclui o planejamento e resultados de entrevistas, workshops participativos, pesquisas de validação, simulação de soluções, dentre outras atividades. De maneira geral, este documento relata a experiência prática da aplicação do Método IoT-PMHCS, sua avaliação e evolução.*

## Glossário

AAL - Ambiente Assistido (do inglês *Ambient Assisted Living*)

AmI - Inteligência Ambiental (do inglês, *Ambient Intelligence*)

DCU - Design Centrado no Usuário

DP - Design Participativo

DT – Design Thinking

IBGE - Instituto Brasileiro de Geografia e Estatística

IHC - Interação Humano-Computador

IoT - Internet das Coisas (do inglês *Internet of Things*)

IoT-PMHCS - *IoT Participatory Method for Home Care Solutions*

OMS - Organização Mundial da Saúde

RFID - Identificação por radiofrequência (do inglês Radio-Frequency IDentification)

TI – Tecnologia da Informação

TIC – Tecnologia da Informação e Comunicação

# Sumário

## Lista de Tabelas

# Lista de Figuras

# 1 Introdução

Este relatório é referente ao detalhamento de atividades na pesquisa do Mestrado em Ciências da Computação da Faculdade de Campo Limpo Paulista no projeto intitulado “Um Método Baseado em Design Participativo, Personas e Semiótica para Soluções IoT para o *Home care* de Pessoas Idosas”. Este documento tem por objetivo relatar todas as atividades realizadas durante o estudo de caso para experimentar a propostas do Método IoT-PMHCS.

Os resultados obtidos são apresentados nesse documento, conforme a seguir: a seção 2 sintetiza a participantes da avaliação e as principais atividades realizadas no estudo de caso do método IoT-PMHCS; a seção 3 apresenta o registro das atividades realizadas na Etapa 0 do método IoT-PMHCS; a seção 4 apresenta o registro das atividades realizadas na Etapa 1 do método IoT-PMHCS; as seções 5 e 6 apresentam o registro das atividades realizadas na Etapa 2 do método IoT-PMHCS; e por fim, a seção 7 apresenta o registro das atividades realizadas na Etapa 3 do método IoT-PMHCS.

## 2 Estudo de Caso

Para a validação da proposta do método IoT-PMHCS, foi realizado estudo de caso contendo vários processos participativos e diferentes *stakeholders*.

Os participantes do estudo de caso pertencem aos estados de Minas Gerais ( cidade de Muzambinho), São Paulo (cidade de Campinas) e Pará (cidades de Belém e Tucuruí). Sempre que possível foi mantido o mesmo grupo durante todo o processo.

Seguem os principais grupos focais participantes dessa pesquisa:

- Um grupo de 9 idosos na cidade de Muzambinho, MG; (ver Tabela 1);
- Um time técnico com 10 profissionais de tecnologia (engenheiros, analistas, cientistas, estudantes) dos estados de São Paulo e Pará; (ver
- Tabela 2);
- 1 profissional de saúde na cidade de Muzambinho, MG; (ver Tabela 3);
- 15 familiares dos idosos (pesquisa pela internet).

A seguir serão apresentados os participantes de cada grupo focal: a Tabela 1, detalha o perfil do grupo de idosos, a Tabela 2, detalha o perfil dos profissionais de tecnologia e a Tabela 3, detalha o perfil do profissional de saúde.

**Tabela 1. Grupo de Idosos, de Muzambinho-MG**

| Identificação | Idade | Familiaridade com TI | Com quem vive | Escolaridade | Profissão |
|---|---|---|---|---|---|
| Maria Gabriela (Bela) | 81 | Não | Filhas | 2o grau | Do Lar |
| Teresa | 75 | Sim | Cônjuge | 2o grau | Do Lar |
| Elizabete (Beta) | 68 | Sim | Cônjuge | Graduação | Do Lar |
| Luiz | 67 | Sim | Cônjuge | Graduação | Bancário Aposentado e Empresário |
| Noel | 70 | Sim | Cônjuge | Graduação | Bancário Aposentado |
| Edson | 75 | Sim | Cônjuge | Graduação | Bancário Aposentado e Advogado |
| Maria José (Zezé) | 68 | Não | Filho | Graduação | Professora Aposentada |
| Elga | 64 | Sim | Cônjuge | Graduação | Dentista |
| Lúcia | 72 | Sim | Cônjuge | Magistério | Professora Aposentada |

**Tabela 2. Grupo de Profissionais de Tecnologia, dos estados de São Paulo e Pará**

| Nome | Idade | Faculdade | Trabalho | Conhece IoT? | Convive com idosos? | Etapa do Método |
|---|---|---|---|---|---|---|
| Julia | 22 | Ciências da Computação - USP | Samsung-SIDI | Apenas ouviu falar | Não | 2 e 3 |
| Erick | 33 | Ciências da Computação - Anhanguera | Samsung-SIDI | Já trabalhou com Arduino e *Rasberry* | Não | 2 e 3 |
| Matheus | 19 | Engenharia de Produção | Estudante | Apenas ouviu falar | Sim, avós | 2 e 3 |
| Zanatta | 34 | Ciências da Computação - Unip | Samsung-SIDI | Apenas ouviu falar | Não | 2 e 3 |
| Sócrates | 29 | Engenharia da Computação - Unicamp | Samsung-SIDI | Já fez alguns projetos iniciais na faculdade | Sim, os pais | 2 e 3 |
| Dennis | 49 | Análise de Sistemas Unaerp | Samsung-SIDI | Apenas ouviu falar | Sim, pais e sogros | 2 e 3 |
| Filipe | 25 | Mestrado em Computação na UFPA | Estudante | Já fez alguns projetos iniciais na faculdade | Não | 3 |
| Hugo | 20 | Computação na UFPA | Estudante | Já fez alguns projetos iniciais na faculdade | Não | 3 |
| Eduardo | 20 | Computação na UFPA | Estudante | Já fez projetos na faculdade | Não | 3 |
| Lídia | 19 | UFPA - Engenharia da Computação | Estudante | Já fez projetos na faculdade | Não | 3 |

**Tabela 3. Profissional de Saúde, de Muzambinho-SP**

| Nome | Formação | Idade | Familiaridade com TI? | Trabalha com idosos? |
|---|---|---|---|---|
| Márcia | Fisioterapeuta | 48 | Experiente | Sim |

Foram realizadas várias atividades participativas e individuais com os grupos focais. O agendamento e organização das atividades foi realizado segundo proposta do método IoT-PMHCS. A Tabela 4 apresenta as principais atividades participativas realizadas para aplicação do método.

**Tabela 4. Atividades participativas do estudo de caso do método IoT-PMHCS**

| Etapas | Objetivo | *Stakeholders* | Local | Duração | Método |
|---|---|---|---|---|---|
| Etapa 0 | **Entrevistas e Questionários** | Profissionais de Sáude, idosos e familiares | Muzambinho, MG e Campinas, SP | 16h | - Questionário semiestruturado<br>- Formulário na internet |
| Etapa 1 | ***Workshop* Valores** | 9 idosos de 60 e 80 anos | Muzambinho, MG | 3 h | - Sessão Generativa - *Brainstorm* |
| Etapa 2 | ***Workshop* Ideação** | 5 Profissionais de TI | Campinas, SP | 3 h | - *Brainstorm*<br>- Co-Criação |
| | ***Workshop* Prototipação** | 9 idosos entre 60 e 80 anos | Muzambinho, MG | 5h | - *Brainstorm*<br>- *Storytelling*<br>- Análise de Normas<br>- Jornada do Usuário |
| Etapa 3 | ***Workshop* Técnico** | 10 Profissionais de TI | Parte 1: Campinas, SP<br>Parte2: Online | 4h | - *Brainstorm*<br>- Jornada do Usuário<br>- Co-Criação |
| | ***Workshop* de Validação** | 6 idosos entre 60 e 80 anos | Muzambinho, MG | 6h | - Cenários<br>- Encenação<br>- *Think Aloud* |

Após esse estudo de caso, todas as adaptações pertinentes foram realizadas na proposta do Método IoT-PMHCS. A versão final encontra-se no documento de Dissertação de Mestrado (Podestá, Bonacin e Gonçalves, 2018).

# 3 Etapa 0 – Estudo Exploratório e Planejamento

Durante a Etapa 0 do método IoT-PMHCS foram realizados entrevistas e questionários com idosos com profissionais de tecnologia, da saúde e familiares. Além disso, foram realizadas atividades de pesquisa exploratória e pesquisa *desk.* Com isso, foram produzidos artefatos para organização e condução das próximas atividades.

As principais atividades dessa etapa estão brevemente apresentadas no Apêndice I – Etapa 0: Estudo Exploratório e Planejamento. As subseções seguintes detalham os resultados.

## 3.1 Entrevista com Idosos

Foi realizada entrevista individual com 8 idosos com idades entre 60 e 80 anos na cidade de Muzambinho-MG com o objetivo de explorar temas como desafios da idade, familiaridade com tecnologia e soluções para apoiá-los em casa.

Para organização da entrevista foi preparado um questionário semiestruturado contendo 19 perguntas. Além disso, os idosos também eram aconselhados a explorar o tema independentemente do questionário.

As entrevistas individuais tiveram duração de aproximadamente 1 hora por pessoa. Todas as entrevistas foram registradas, para facilitar a coleta de resultados e análises, conforme a seguir.

### 3.1.1 Questionário

O questionário aplicado durante a entrevista com os idosos era composto pelas seguintes questões:

1. Nome
2. Sexo
3. Idade
4. Qual seu grau de escolaridade?
5. Você mora sozinho?
6. Fica muito tempo sozinho em casa?
7. Qual sua maior dificuldade/riscos dentro de casa?
8. Tem dificuldades de visão?

9. Tem dificuldades de audição?
10. Você tem algum problema crônico de saúde?
11. Exige cuidados especiais?
12. Em situações de perigo, qual recurso você usa ou usaria para chamar ajuda?
13. Tem Plano de Saúde?
14. Você usa o celular com qual frequência?
15. Você usa o celular para que?
16. Qual sua maior dificuldade com a tecnologia?
17. Como você interage com seu celular atualmente?
18. Como você imagina que a tecnologia pode te ajudar em situações de perigo dentro de casa?
19. Você concorda em participar desse projeto, fornecendo e validando informações em atividades participativas?

### 3.1.2 Resultados e Análises

A maioria dos entrevistados (5 pessoas) eram do sexo feminino, como apresentado na Figura 1. Isso também reflete a maior população idosa do sexo feminino.

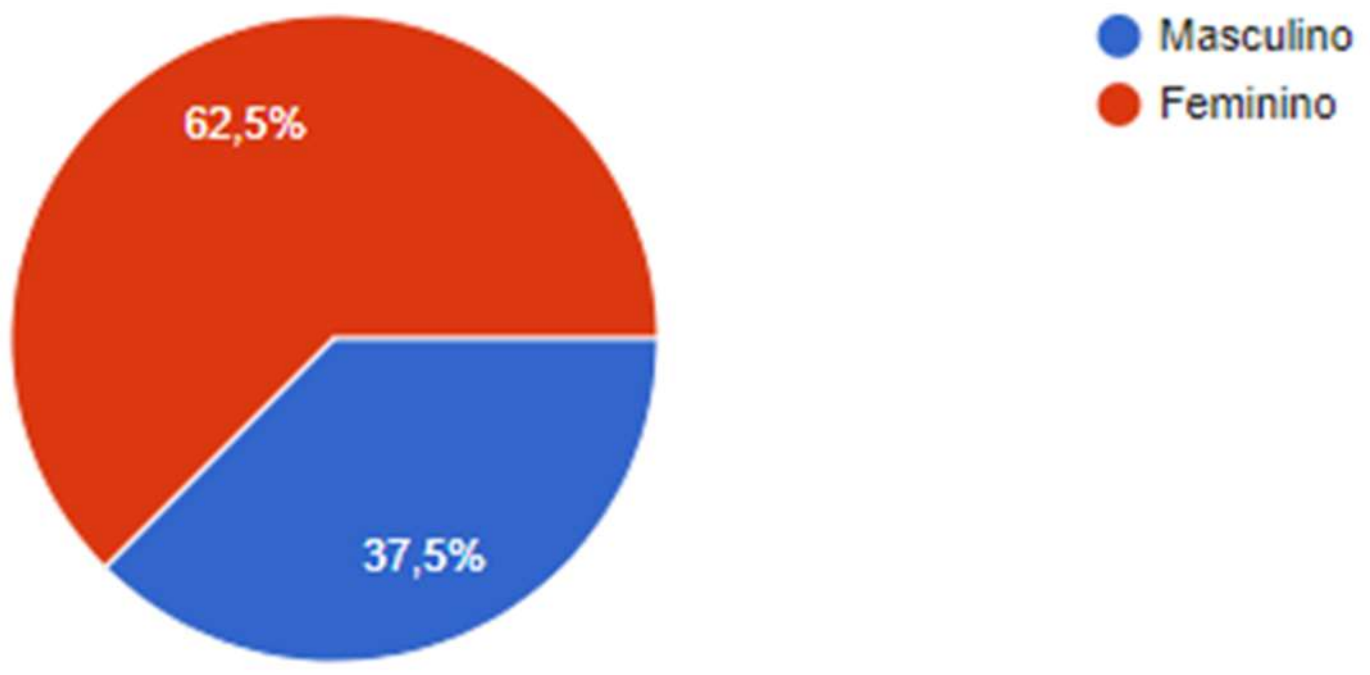

**Figura 1. Entrevista idosos: percentual de distribuição dos participantes por sexo**

A maior parte dos entrevistados (4 pessoas) possui entre 70 e 80 anos de idade, conforme apresentado na Figura 2.

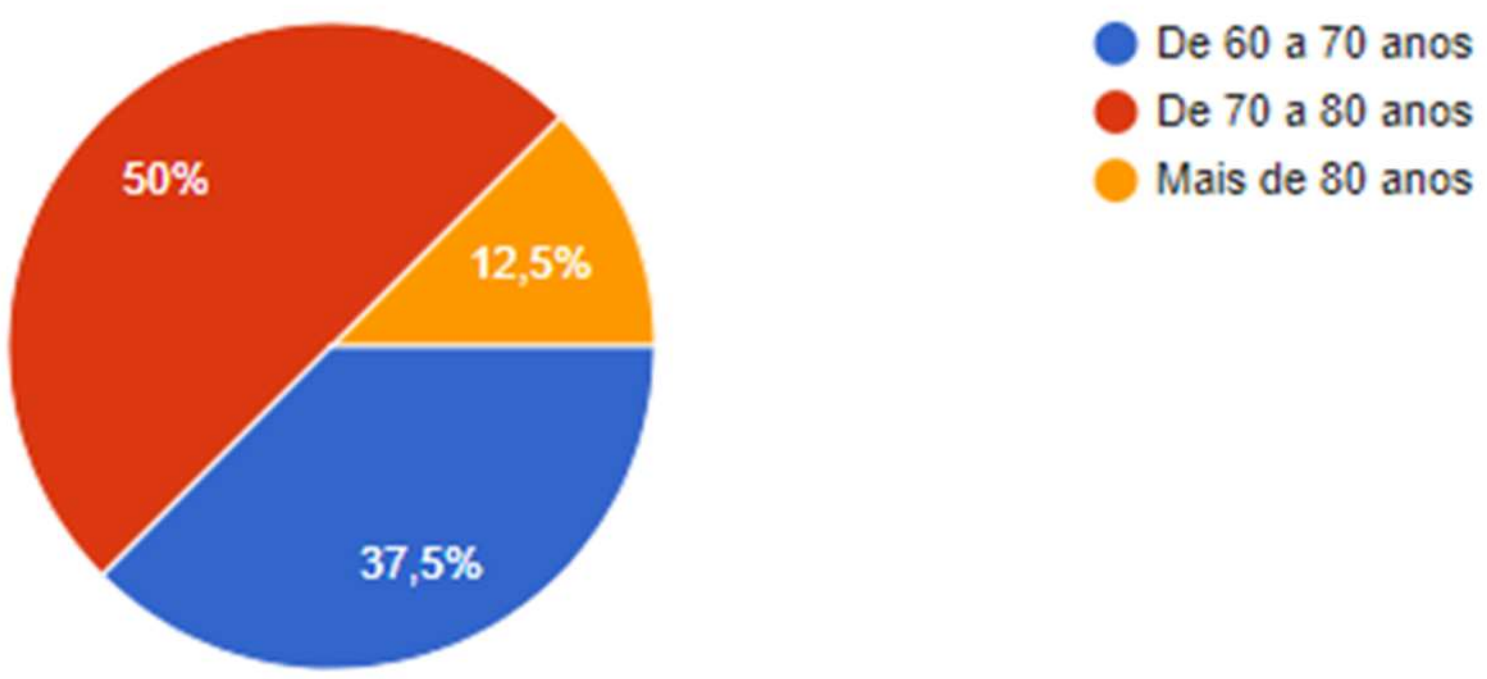

**Figura 2. Entrevista idosos: percentual de distribuição dos participantes por idade**

Cerca de 62,5% dos entrevistados (5 pessoas) possuem Ensino Superior, como apresentado na Figura 3.

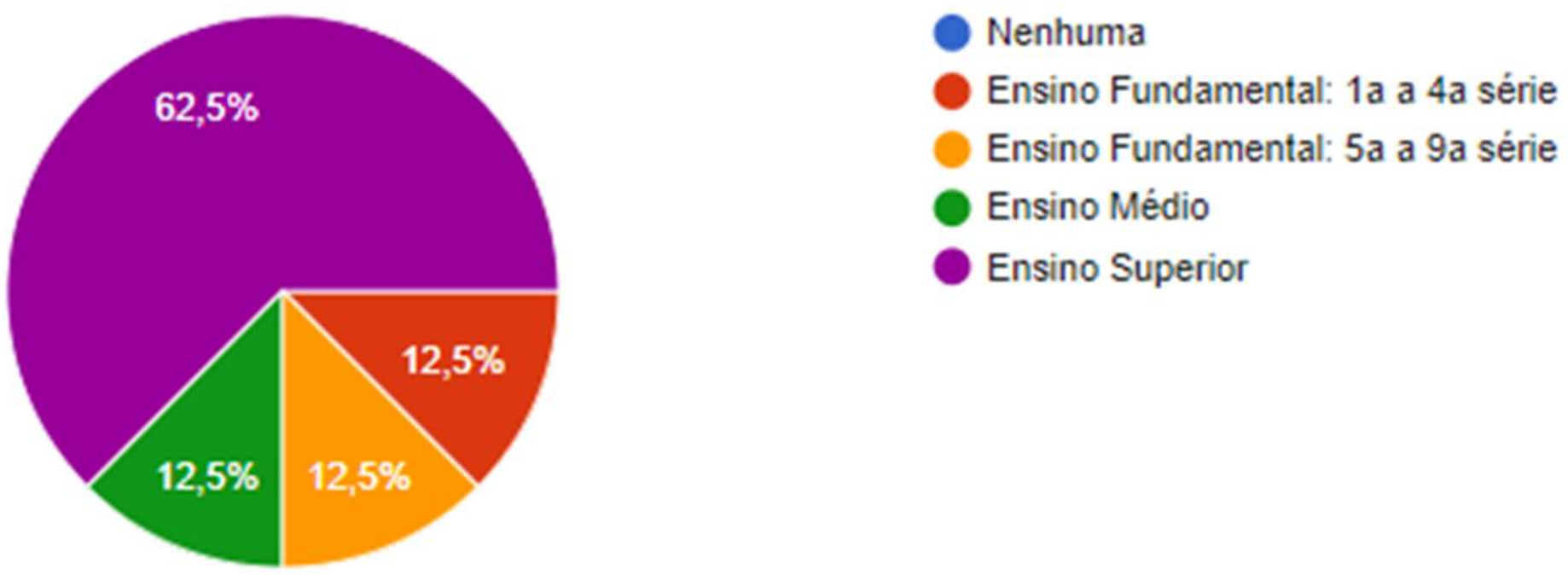

**Figura 3. Entrevista idosos: percentual de distribuição dos participantes por escolaridade**

Apenas 1 participante mora sozinho. Os demais moram com o cônjuge ou familiares, e ficam no máximo 4 horas sozinho em casa, como apresentado na Figura 4.

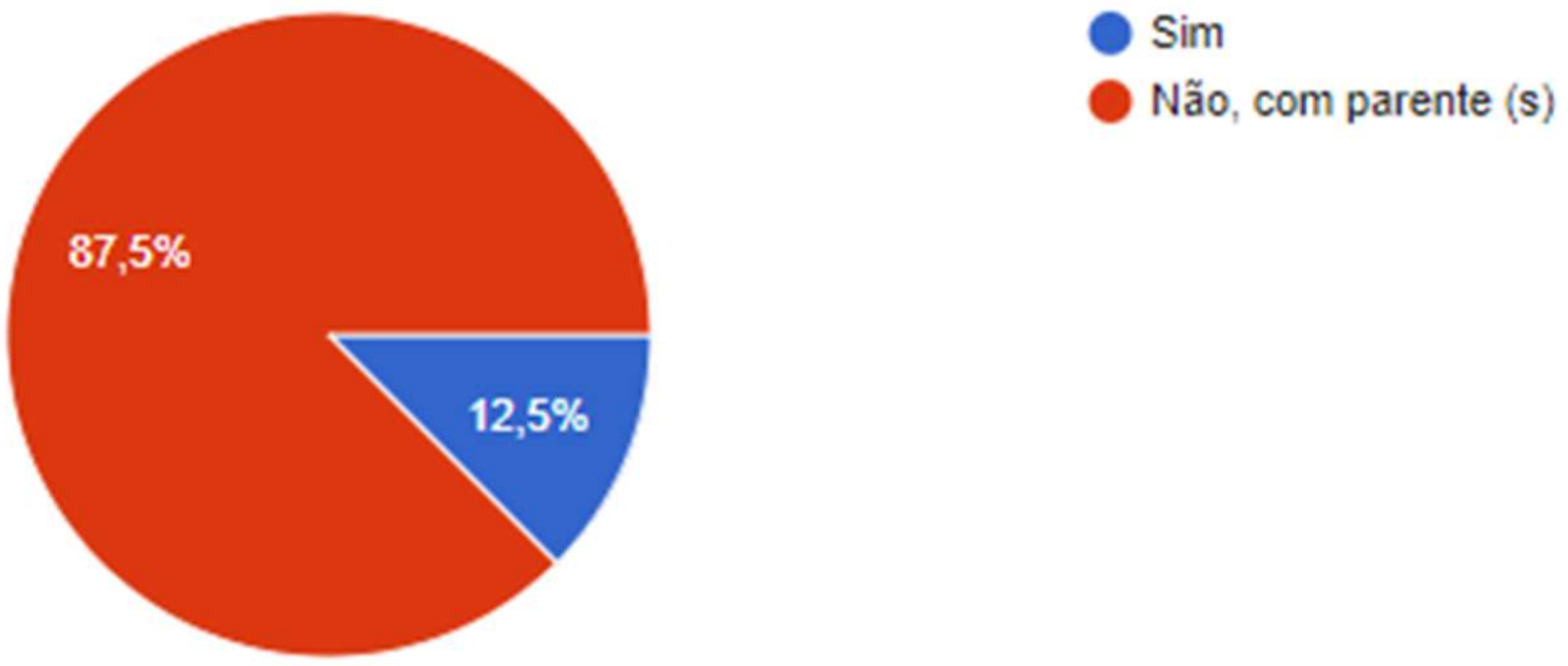

**Figura 4. Entrevista idosos: percentual de distribuição dos participantes por composição do lar**

Foi questionado aos entrevistados, quais suas maiores dificuldades ou riscos encontrados dentro de suas casas, dado a idade avançada. Os maiores desafios apontados pelos entrevistados foram com relação a quedas e esquecimentos, como apresentado na Figura 5.

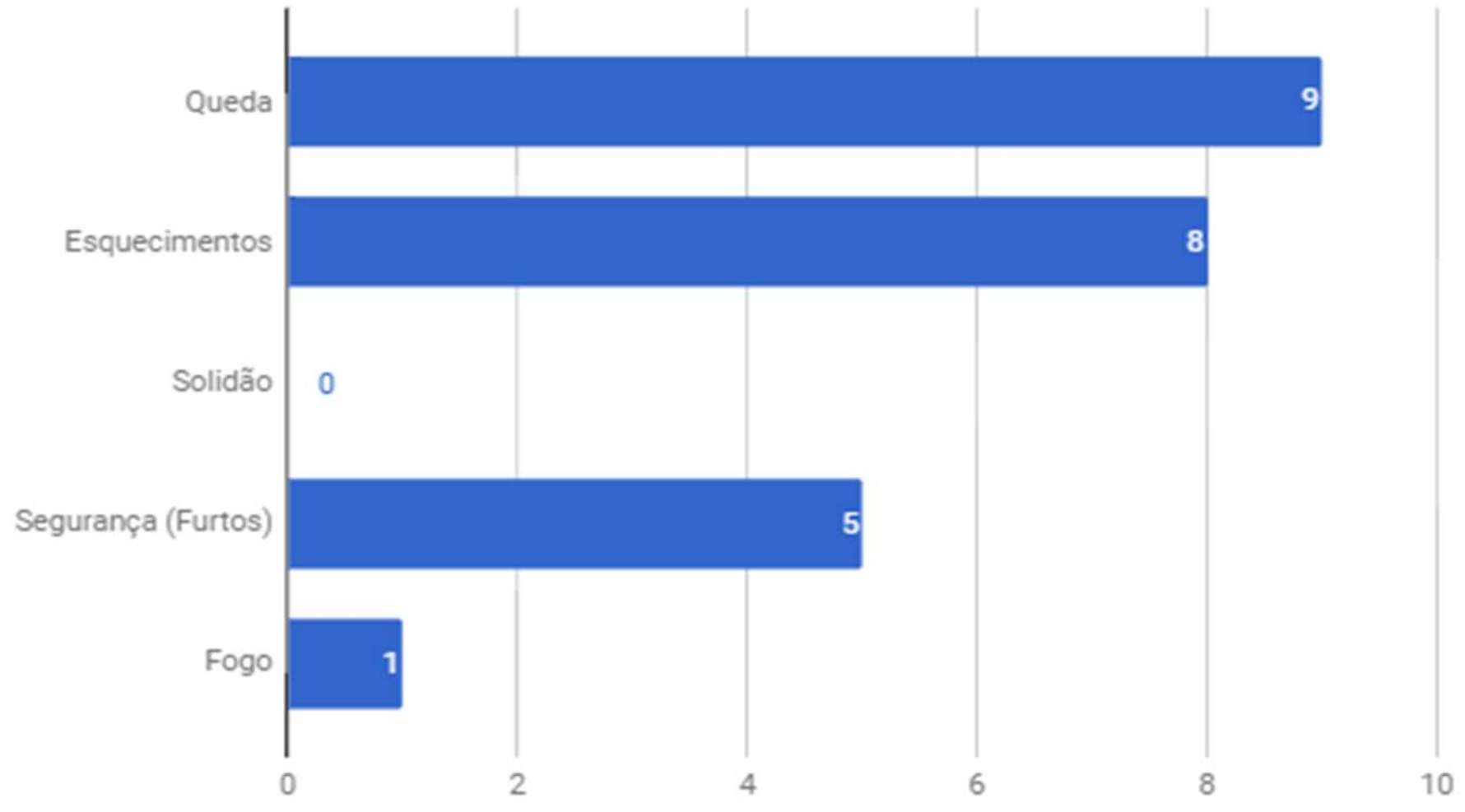

**Figura 5. Entrevista idosos: maiores Riscos e Desafios para os idosos dentro de casa**

Dentre os problemas de saúde apresentados, o mais comum entre os idosos foram os relacionados a pressão (62,5% dos idosos) e audição (50% dos idosos), como apresentado na Figura 6. Apenas um entrevistado não relatou qualquer problema de saúde. Nenhum dos entrevistados demanda cuidados especiais ou a presença de cuidadores profissionais.

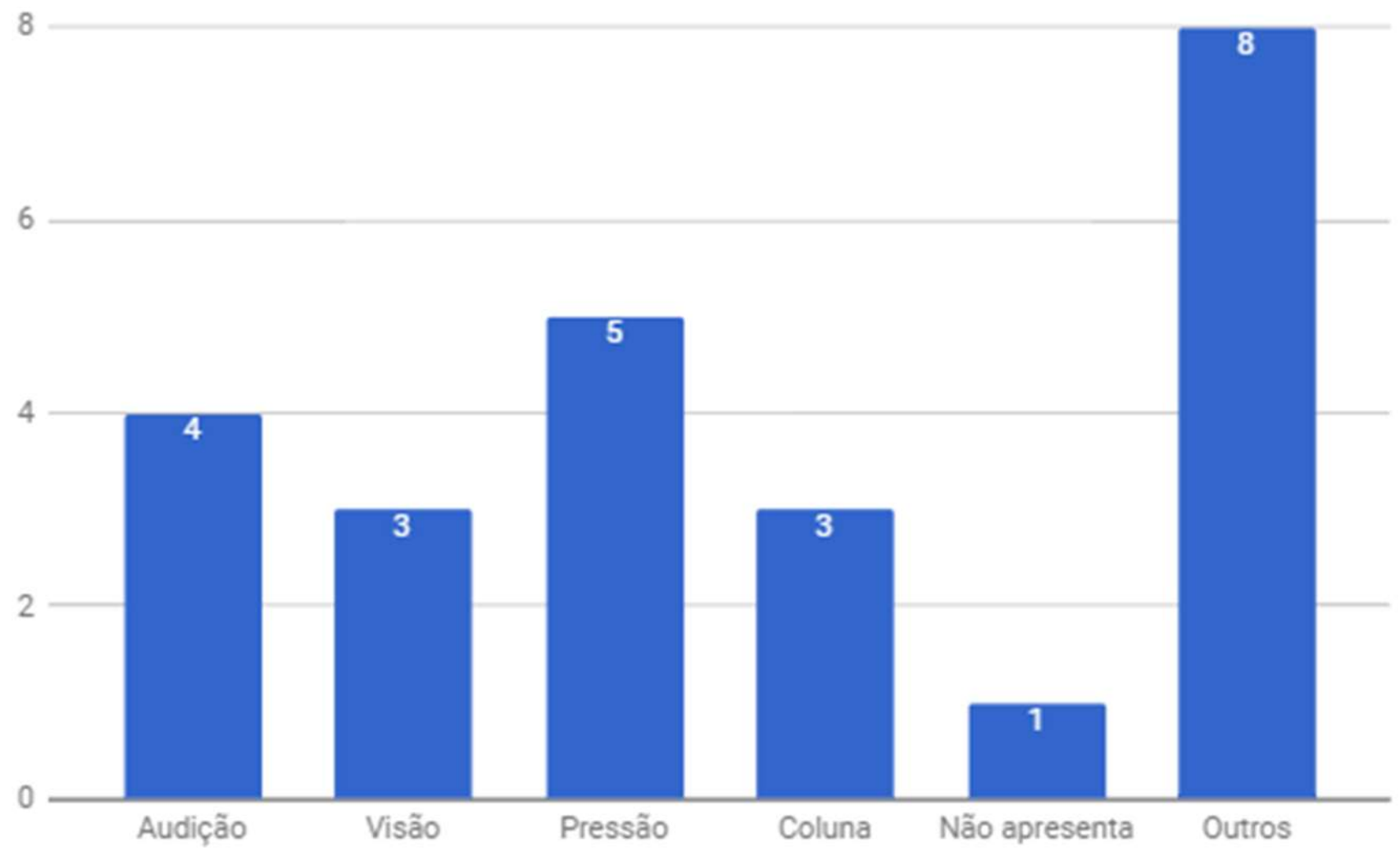

**Figura 6. Entrevista idosos: problemas de saúde apresentados pelos idosos**

Quando questionados sobre situações de perigo, a grande maioria reportou que usaria o telefone celular para chamar socorro, como apresentado na Figura 7. Cerca de 57,1% dos idosos entrevistados reportou que usa o aparelho com frequência.

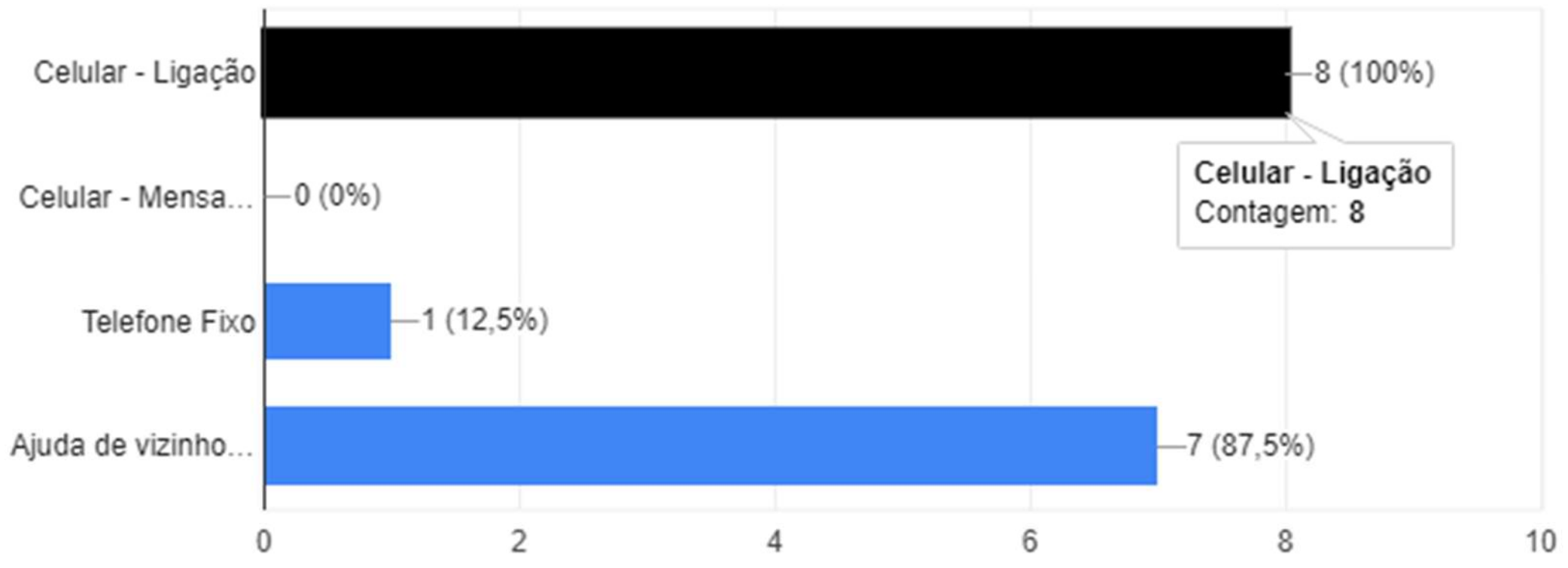

**Figura 7. Entrevista idosos: Recursos utilizados para chamar socorro pelos idosos**

Quando questionados sobre a maior dificuldade no uso da tecnologia, dois resultados se destacaram: 2 idosos (25%) disseram que não tem nenhuma dificuldade com tecnologia e gostam muito. Por outro lado, 2 idosos (25%) responderam que não gostam de novas tecnologias, mas quando se acostumam passam a utilizar com frequência, conforme a Figura 8 apresenta.

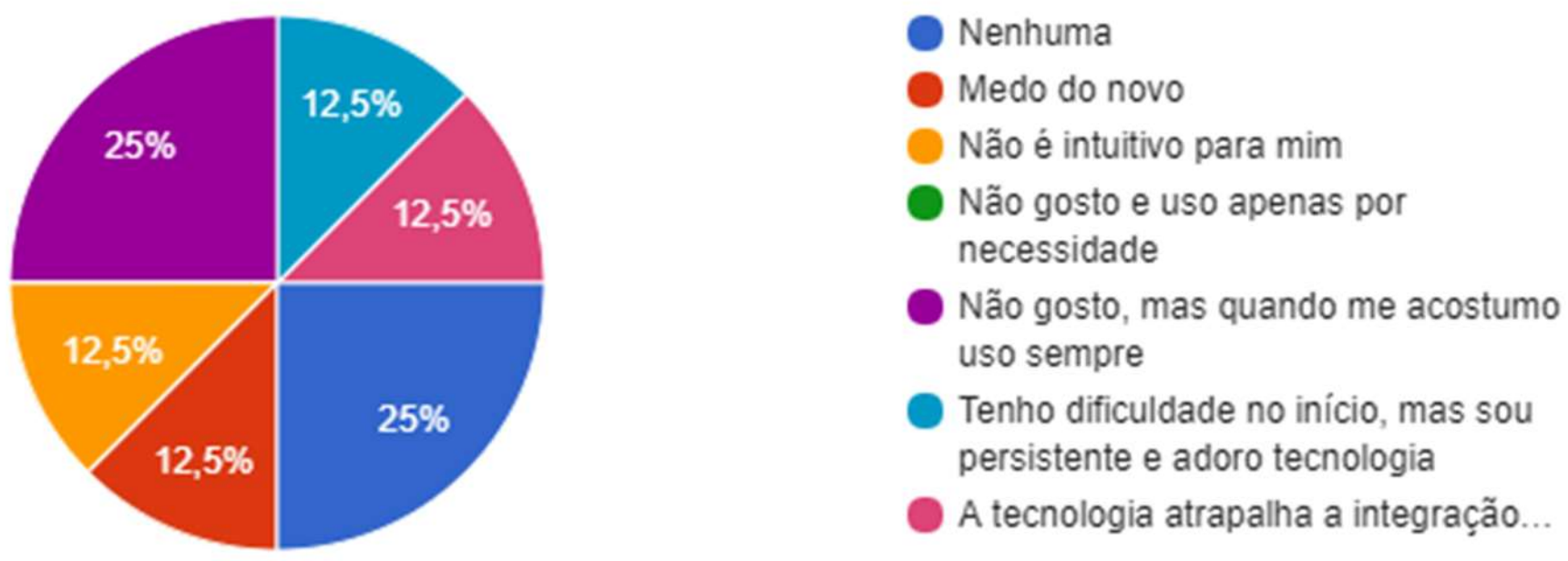

**Figura 8. Entrevista idosos: Familiaridade dos idosos com novas tecnologias**

A questão 18 “Como você imagina que a tecnologia pode te ajudar em situações de perigo dentro de casa?”, foi aberta para estimular a criatividade dos idosos. Nessa questão, alguns dos idosos apresentaram alguma dificuldade, então foram apresentados estudos de caso para facilitar o processo de ideação. Seguem os resultados:

- Sensor que pode identificar um estranho dentro de casa e acionar a polícia;
- Reconhecimento de face ao entrar em casa, e acesso a base da polícia para identificar se é um ladrão já cadastrado;
- Câmera para monitorar cuidadores, para evitar maus tratos;
- Acionamento de alarmes e câmeras por comando de voz;
- Sensores em determinados itens (chaves, celular, óculos, carteira, bolsa) com acionamento por voz ou acionamento por celular. Emite alerta no objeto esquecido para facilitar localização. Exemplo: "Localizar Celular", "Localizar Óculos" etc;
- Acionamento de energia e eletroeletrônicos por comando de voz. Em caso de falta de energia, sensor aciona energia emergencial em toda a casa. O acionamento também poderia ser por voz;
- Sensores nas escadas: luminoso ou por comando de voz;
- Alarme monitorado com câmera em casa e comando de voz para acionar socorro em caso de queda;
- O sistema de alarme em casa poderia, em caso de acidentes, enviar uma mensagem de alerta por celular de familiares;
- Pulseira para análise de sintomas de perigo (pulsação acelerada, pressão arterial alta ou alta): emissão de alerta via mensagem para o responsável ou grupo de familiares;
    - Acionamento do médico ou plano de saúde em caso de perigo. Nesse caso, o plano de saúde entra em contato com o cliente para verificar se existe algum problema de saúde;
    - Histórico de saúde na nuvem, para integrar com a pulseira. Ajuda no registro de histórico e no resgate de informações para análise de riscos e problemas de saúde.

### 3.1.3 Resultado Consolidado

A partir da entrevista com os idosos foi possível identificar os principais desafios dentro de casa. Os resultados foram agrupados de acordo com as categorias de AAL de Li, Lu e Maier (2015). Foi incluída a categoria Segurança, dado o cenário brasileiro e as respostas fornecidas pelos entrevistados (Tabela 5).

**Tabela 5. Diagrama de Afinidades de principais desafios no *Home care*, a partir da entrevista com idosos**

| Categoria | Subcategoria | Desafios |
|---|---|---|
| **Segurança** | Roubos | -Furtos e roubos<br>-Fogo |
| **Facilitação diária de tarefas** | Esquecimento | Esquecimentos |
| **Mobilidade** | Quedas | Quedas |
| **Cuidados de saúde e Reabilitação** | Problemas de Saúde | O mais comum entre todos os idosos foram problemas relacionados a pressão (62,5% dos idosos) e audição (50% dos idosos) |
| **Inclusão social e comunicação** | Nenhum | Solidão |

Os maiores desafios dentro de casa apontados pelos entrevistados foram com relação a quedas (9 citações em 23, ou seja 40%) e esquecimentos (8 citações em 23, ou seja 35%). Esse resultado compõe uma das saídas esperadas da Etapa 0: Diagrama de Afinidades de Riscos e Desafios para idade.

A Tabela 6 apresenta o resultado consolidado das principais sugestões fornecidas pelos idosos para uso da tecnologia no *Home Care.* Esse resultado compõe uma das saídas esperadas da Etapa 0: Diagrama de Afinidades de Principais sugestões AAL.

**Tabela 6. Principais sugestões AAL por categoria, criado a partir da entrevista com idosos**

| Categoria | Subcategoria | Sugestão de Tecnologia |
|---|---|---|
| **Segurança** | Roubos | - Sensor que pode identificar um estranho dentro de casa e acionar a polícia;<br>- Reconhecimento de face ao entrar em casa, e acesso a base da polícia para identificar se é um ladrão já cadastrado;<br>- Câmera para monitorar cuidadores, para evitar maus tratos;<br>- Acionamento de alarmes e câmeras por comando de voz. |
| **Facilitação diária de tarefas** | Esquecimento | - Sensores em determinados itens (chaves, celular, óculos, carteira, bolsa) com acionamento por voz ou celular para localização. Emite alerta no objeto esquecido para facilitar localização. Ex: "Localizar Celular", "Localizar Óculos" etc;<br>- Acionamento de energia e eletroeletrônicos por comando de voz. |
| **Assistência de mobilidade** | Quedas | - Sensores nas escadas: luminoso ou por voz;<br>- Alarme monitorado com câmera em casa e comando de voz para acionar socorro em caso de queda;<br>- O sistema de alarme em casa poderia, em caso de acidentes, enviar uma mensagem de alerta por celular de familiares;<br>- Em caso de falta de energia, sensor aciona energia emergencial |
| | | em toda a casa. O acionamento também poderia ser por voz. |
| **Cuidados de saúde e Reabilitação** | Problemas de Saúde | - Pulseira para análise de sintomas de perigo (pulsação acelerada, pressão arterial alta ou alta): emissão de alerta via mensagem para o responsável ou grupo de familiares;<br>- Acionamento do médico ou plano de saúde em caso de perigo. Nesse caso, o plano de saúde entra em contato com o cliente para verificar se existe algum problema de saúde;<br>- Histórico de saúde na nuvem, para integrar com a pulseira. Ajuda no registro de histórico e no resgate de informações para análise de riscos e problemas de saúde. |
| **Inclusão social e comunicação** | Nenhum | Não identificado. |

## 3.2 Entrevista com Profissional de Saúde

Foi realizada uma entrevista individual com uma profissional de saúde que tem contato direto com idosos com o objetivo de explorar temas como desafios da idade, familiaridade com tecnologia e soluções para apoiá-los em casa.

A profissional entrevistada é uma Fisioterapeuta e reside na cidade de Muzambinho-MG. Ela respondeu a um questionário semiestruturado contendo 5 perguntas durante uma entrevista de aproximadamente 1 hora.

### 3.2.1 Questionário

A seguir o questionário realizado com a profissional de saúde

1- Quais os principais problemas enfrentados pelos idosos em casa?
2- Quais itens de segurança são recomendados dentro de casa?
3- Do ponto de vista comportamental, quais os principais desafios dos idosos?
4- Os idosos de um modo geral usam a tecnologia?
5- Como você imagina que a tecnologia pode apoiar os idosos nos cuidados dentro de casa?

### 3.2.2 Análises

A seguir os resultados do questionário aplicado com a profissional de saúde:

**1- Quais os principais problemas enfrentados pelos idosos em casa?**

Os principais desafios enfrentados pelos idosos dentro de casa são:

- Quedas: muitas ocorrências de fisioterapia devido a fratura de fêmur.
    - Homens: quedas noturnas ao irem ao banheiro;
    - Mulheres: quedas causadas por conta da Osteoporose e Vertigem;
    - As escadas não são a maior causa das quedas;
    - Calçadas em frente as casas e subir em bancos e cadeiras são grandes desafios.
- Saúde: muitos casos de constipação intestinal
    - Vasos baixos dificultam.
- Cuidadores não são preparados:
    - Muitos não têm cursos, apesar de que na cidade existe curso gratuito no Instituto Federal: apenas um curso técnico para cuidadores já ajudaria;
    - Fazem a tarefa da casa e junto cuidam dos idosos;
    - Baixa renda dificulta contratação de enfermeiros, ficando os idosos muitas vezes aos cuidados de cuidadores informais (familiares) ou não capacitados;
    - Existem muitos casos de maus tratos de cuidadores;
    - Médicos e Fisioterapeutas devem conscientizar mais a família.
- Alzheimer: existem muitos casos atualmente
    - Família tira muitas atividades do idosos pelo medo do perigo causando sedentarismo;
    - Precisam ser mais estimulados com atividades mesmo que dentro de casa.

**2- Quais itens de segurança são recomendados dentro de casa?**

Os itens mínimos que são recomendados dentro de casa são:

- Tapete no banheiro;
- Iluminação noturna;
- Apoio de cama;
- Barras no banheiro.

**3- Do ponto de vista comportamental, quais os principais desafios dos idosos?**

Os principais itens observados pela profissional são:

- Corpo não acompanha a cabeça: geralmente eles têm uma capacidade intelectual

e psicológica muito boa, porém problemas de saúde muitas vezes impactam as atividades diárias e causam acidentes;

- Ausência de memória é o mais observado.

**4- Os idosos de um modo geral usam a tecnologia?**

Sim, a maioria dos idosos utiliza celular, na maioria dos casos para voz (chamadas). O uso de redes sociais e WhatsApp está aumentando a cada dia.

**5-Como você imagina que a tecnologia pode apoiar os idosos nos cuidados dentro de casa?**

Seguem as principais sugestões de TI discutidas com profissional:

- Quedas Noturnas
  - Iluminação automática com sensores.
- Uso de sensores dentro de casa:
  - Precisam ser camuflados (*ubíquos)*, pois idosos são muito teimosos e não aceitam bem a tecnologia;
  - Pulseiras (*wearable*) não seriam bem aceitas num primeiro momento;
  - Uma vez sendo bem aceitos, poderia ter seu acionamento por voz ou celular para pedidos de socorro;
  - Principais locais recomendados: banheiro, quarto e próximo a escadas.
- Câmeras de segurança
  - Poderiam ser utilizadas com monitoramento pelos familiares pelo celular;
  - Sensores para monitorar sinais de perigo com envio de alertas por celular;
- Pulseiras (*wearable*): uma vez aceitos, poderiam ser utilizados para medir temperatura, frequência cardíaca, pressão arterial, frequência respiratória.
  - Emissão de alertas de perigo, em função de combinação das medições.

Além das alternativas de TI, a profissional também sugeriu a criação de uma clínica para cuidados de idosos diários, de tal forma que eles pudessem passar o dia e dormir em casa. Nela poderiam ser realizadas atividades para estimular o corpo e mente, promovendo maior relacionamento social, cuidados com fisioterapeutas e uma dieta balanceada por nutricionistas.

### 3.2.3 Resultado Consolidado

A partir da entrevista foi possível criar o Diagrama de Afinidades dos desafios em casa segundo a Tabela 7. Esse resultado compõe uma das saídas esperadas da Etapa 0: Diagrama de Afinidades de Riscos e Desafios para idade.

**Tabela 7. Diagrama de Afinidades de principais desafios no *Home care*, criado a partir da entrevista com profissional de saúde**

| Categoria | Subcategoria | Desafios |
|---|---|---|
| **Segurança** | Maus tratos | -Cuidadores muitas vezes não são preparados para apoiar os idosos, não tem cursos;<br>-Existem muitos casos de maus tratos. |
| **Facilitação diária de tarefas** | Esquecimento | -Falhas de memória;<br>-Muitas ocorrências de Alzheimer, em estágio inicial e avançado, causando sedentarismo pois a família tira muitas atividades do idoso. |
| **Mobilidade** | Quedas | -Quedas noturnas;<br>-Subir em bancos e cadeiras;<br>-Calçadas em frente as casas;<br>-Escadas não são um problema muito crítico e observado. |
| **Cuidados de saúde e Reabilitação** | Problemas de Saúde | -Constipação intestinal; |
| **Inclusão social e comunicação** | Nenhum | Não identificado. |

A partir da entrevista foi possível criar o Diagrama de Afinidades das principais sugestões de AAL segundo a Tabela 8. Esse resultado compõe uma das saídas esperadas da Etapa 0: Diagrama de Afinidades de Principais sugestões AAL.

**Tabela 8. Principais sugestões AAL por categoria, criado a partir da entrevista com profissional de saúde**

| Categoria | Subcategoria | Desafios |
|---|---|---|
| **Segurança** | Maus tratos | -Câmeras com monitoramento pelo celular por familiares<br>-Sensores para monitorar sinais de perigo com envio de alertas por celular |
| **Facilitação diária de tarefas** | Nenhum | Não identificado. |
| **Mobilidade** | Quedas | -Iluminação automática com sensores<br>-Sensores com acionamento por voz no banheiro, quarto e próximo a escadas |
| **Cuidados de saúde e Reabilitação** | Problemas de Saúde | -Pulseira, quando aceitas, para monitorar temperatura, frequência cardíaca, pressão arterial, frequência respiratória, emissão de alertas de perigo;<br>-Em função de combinação das medições emite alerta de socorro para médicos e familiares. Acionamento por comando de voz; |
| **Inclusão social e comunicação** | Nenhum | Nenhum |

## 3.3 Entrevista com Familiares

Durante a Etapa 0 do método IoT-PMHCS foi aplicado também um questionário semiestruturado com familiares. O questionário foi encaminhado por e-mail para 22 pessoas e respondido por 14 participantes. Seu objetivo foi compreender o universo dos idosos, seus valores e necessidades sob a ótica dos filhos, que convivem diretamente ou não com pais idosos.

### 3.3.1 Questionário

O questionário aplicado com os familiares na pesquisa semiestruturada era composto pelas seguintes questões:

1. Quem é você?
    - Nome Completo, Idade, Sexo, Profissão
2. Vida e Saúde
    - Qual a idade do seu pai?
    - Qual a idade da sua mãe?
    - Com quem seus pais vivem?
    - Quanto tempo seus pais ficam sozinhos em casa?
    - Seus pais possuem algum problema de saúde?
    - Seus pais demandam algum cuidado especial em casa?
    - Qual a distância que você vive dos seus pais?
3. Valores e Necessidades
    - Qual dos valores abaixo você acredita ser o mais importante para seus pais?
    - Qual a sua maior preocupação com seus pais atualmente?
    - Qual sua maior preocupação com seus pais dentro de casa?

- Seus pais possuem alguém próximo que pode auxiliar em situações de perigo?

4. A Tecnologia e o *Home care*
    - Qual a relação dos seus pais com a tecnologia?
    - Quais tecnologias seus pais estão mais familiarizados?
    - Como a tecnologia auxilia seus pais dentro de casa atualmente?
    - Numa situação de perigo ou emergência, você gostaria de ser notificado? Qual a forma mais rápida de te localizar?
    - Você gostaria de monitorar seus pais? Quais das tecnologias inovadoras abaixo você concorda em utilizar.
    - No futuro, em situações de perigo, como a tecnologia poderia auxiliar seus pais e te manter informado ou te acionar?

### 3.3.2 Análises

Esta seção apresenta a análise dos resultados do questionário descrito na última seção:

*(1) Quem é você?*

Dos 22 convidados apenas 14 responderam ao questionário, sendo 7 homens e 7 mulheres. A grande maioria possui entre 30 e 40 anos, como mostra a Figura 9

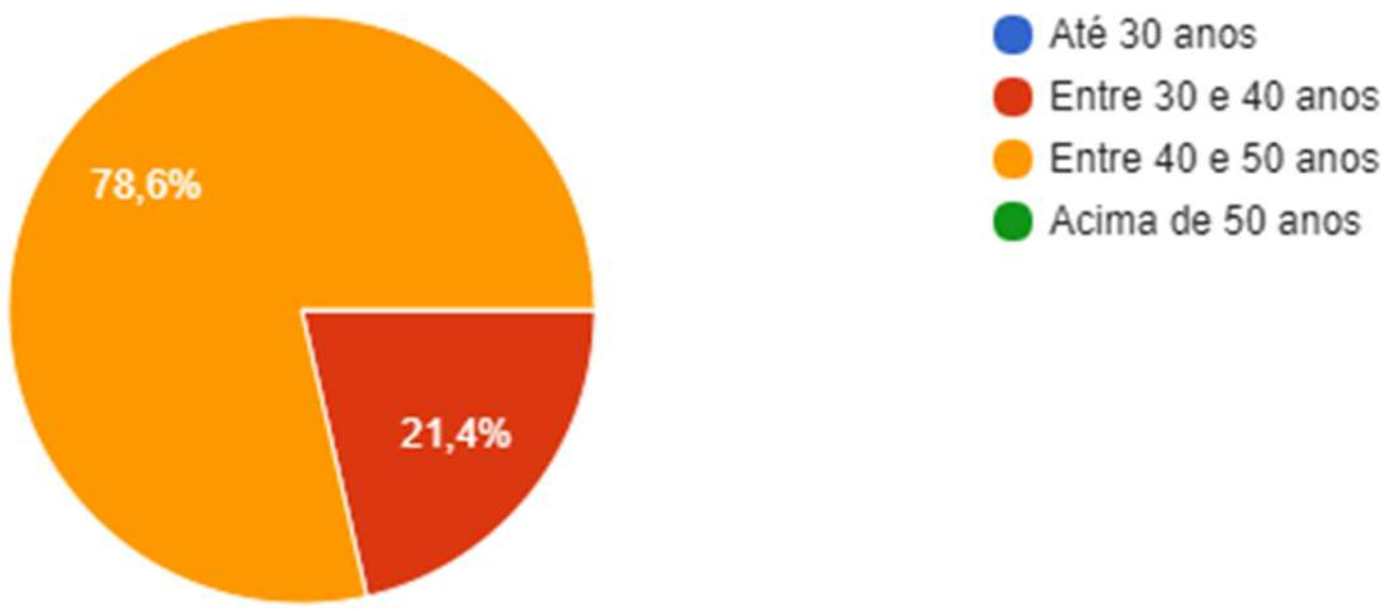

**Figura 9. Questionário familiares: idade dos participantes**

Dentre as profissões, destacam-se os professores (7, ou seja 50% dos participantes). Participaram do processo também 2 Analistas de Sistemas e 1 Fisioterapeuta, o que complementa o olhar do participante: além de familiar, profissionais de TI e de saúde que são *stakeholders* selecionados para esse projeto de pesquisa.

*(2) Vida e Saúde*

A maior parte dos entrevistados tem pais com idades acima de 70 anos (8, ou seja, cerca de 57%).

Quando questionados sobre estilo de vida, cerca de 50% apontou que os pais não vivem sozinhos e sim com um familiar, conforme apresentado na Figura 10. Além disso 46,2% apontou que os pais ficam mais 4 horas sozinhos em casa, o que representa um risco para acidentes.

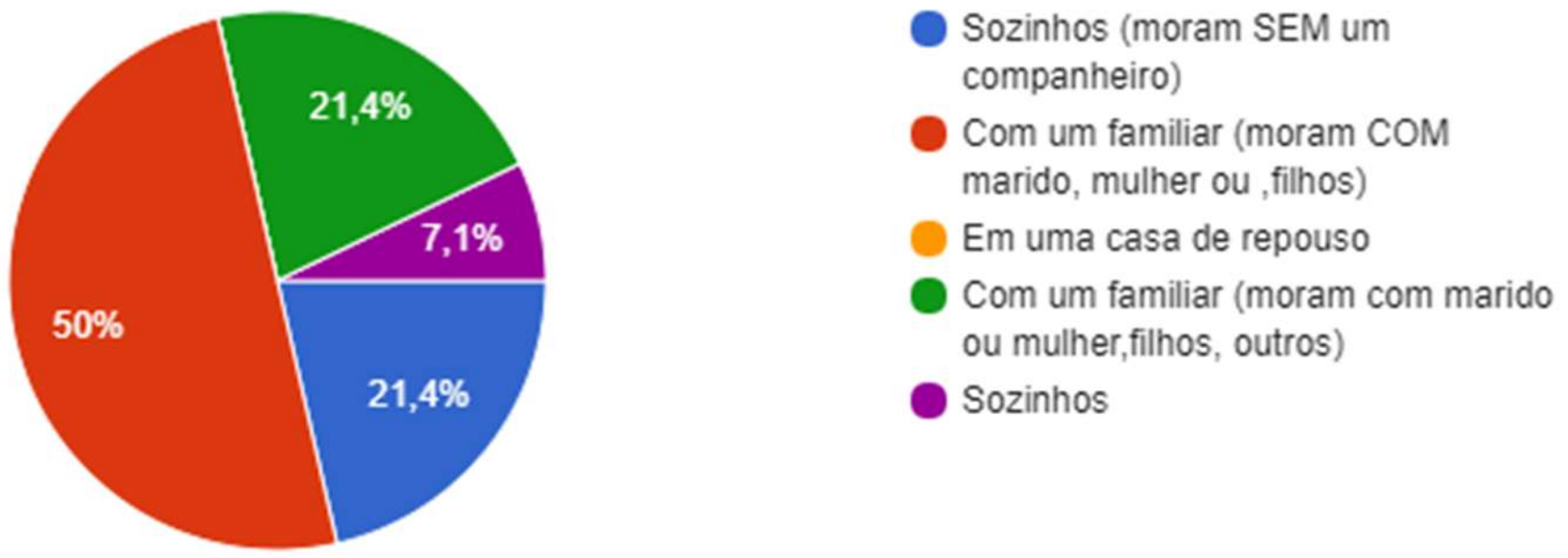

**Figura 10. Questionário familiares: com quem os pais moram**

Quando questionados sobre a maior preocupação com pais, cerca de 64% dos entrevistados apontam os problemas relacionados de saúde. Os mais expressivos reportados foram problemas nas articulações tais como artrites e artroses (64,3%), seguidos de problemas cardíacos. Nenhum dos entrevistados reportou que os pais demandam cuidados especiais, como por exemplo a presença de cuidadores formais ou informais. Com base nisto, observa-se que tais problemas aumentam a probabilidade de acionamento emergencial dos familiares devido a quedas e problemas de coração.

Porém, quando questionados sobre a distância que vivem dos pais (Figura 11), a maior parte dos entrevistados (92%) aponta que não vivem junto com os pais, sendo que 50% vivem até mesmo em cidades diferentes, o que dificultaria a atenção e socorro desses familiares em casos de perigo.

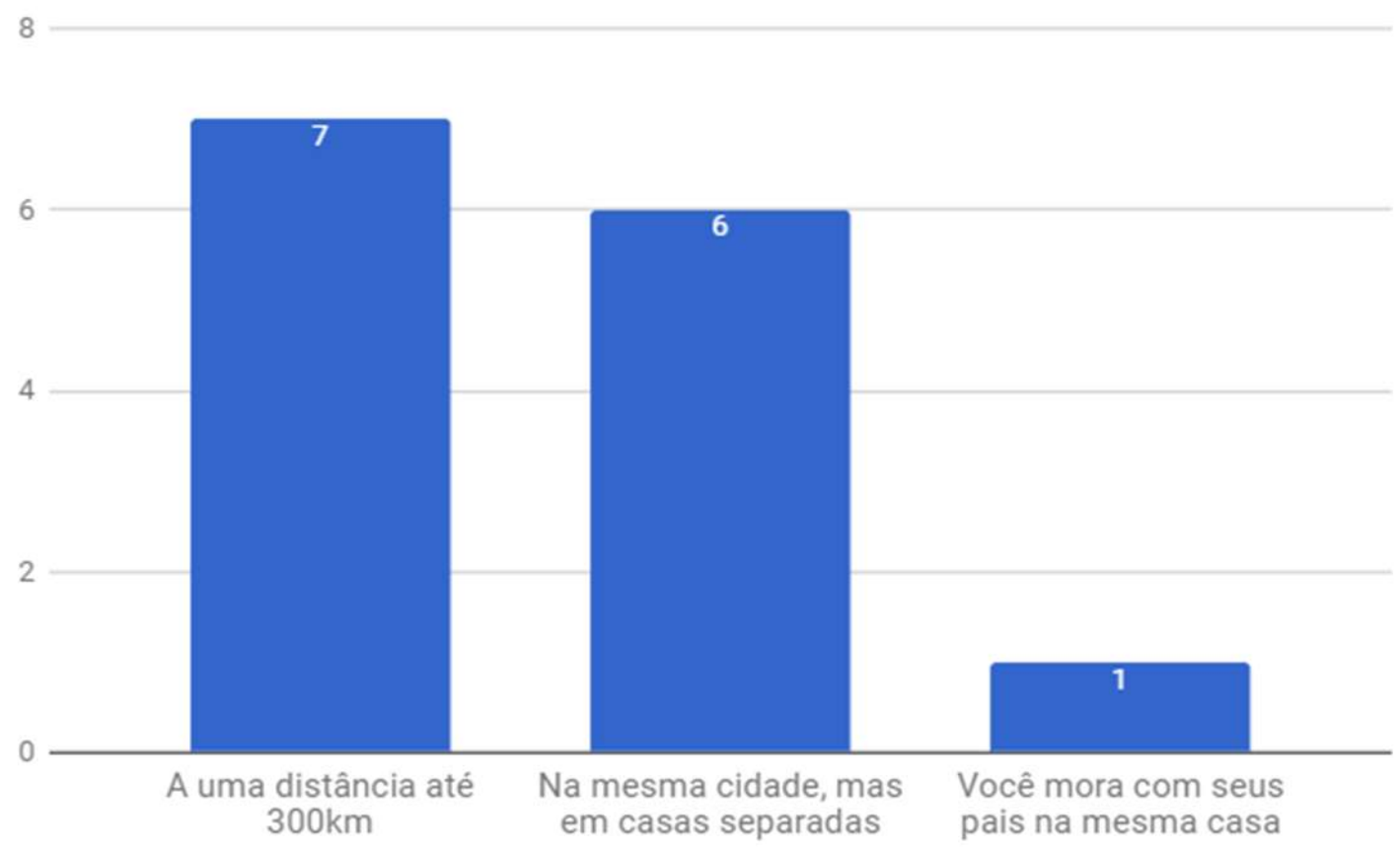

**Figura 11. Questionário familiares: distância que familiares vivem dos pais**

*(3) Valores e Necessidades*

Quando questionados sobre valores, usando como referência os resultados da pesquisa de Leong e Robertson (2016), a maioria dos entrevistados (9, ou seja 64%) acreditam que seus pais prezam pela Independência como principal valor, como mostra a Figura 12.

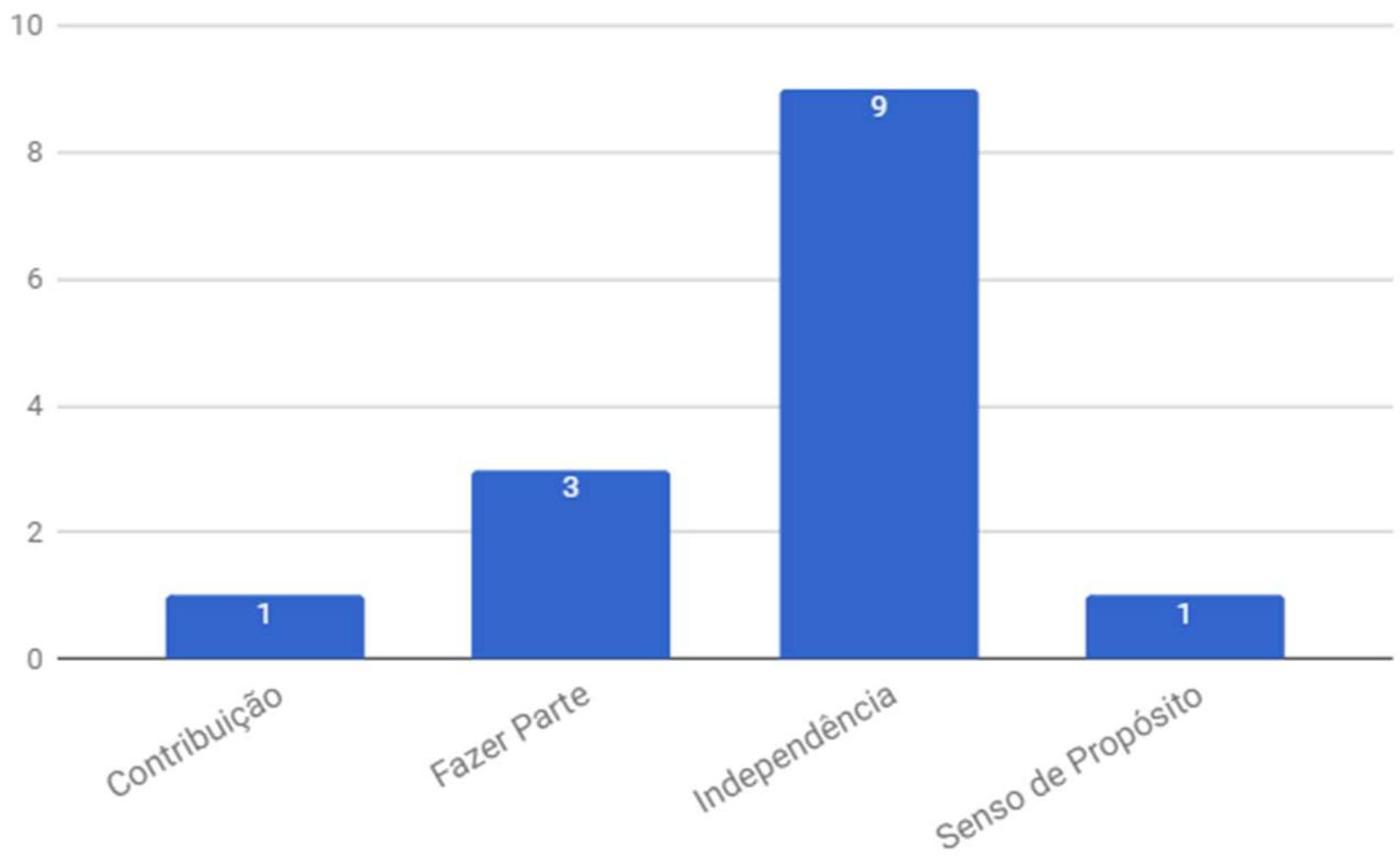

**Figura 12. Questionário familiares: Valores dos pais na visão dos filhos**

Ou seja, por um lado os idosos precisam de ajuda com a idade avançada e os riscos eminentes devido a problemas de saúde, porém prezam pela independência na visão dos filhos. Por outro lado, os filhos muitas vezes não estão próximos, nem mesmo na mesma cidade, o que dificulta o suporte. E mesmo os que estão na mesma cidade enfrentam dificuldades de apoiar na medida em que encontram barreiras nos próprios pais que não reconhecem a necessidade de ajuda, pois querem ser independentes. Aqui um forte indício de que tecnologias que apoiem os familiares de forma ubíqua e pervasiva seriam de grande diferencial para as duas partes.

Quando questionados sobre preocupações e desafios com os dentro de casa, o mais citado foi a questão de segurança (roubos e assaltos), mencionado por 50% dos participantes. Os principais problemas citados são apresentados na Figura 13.

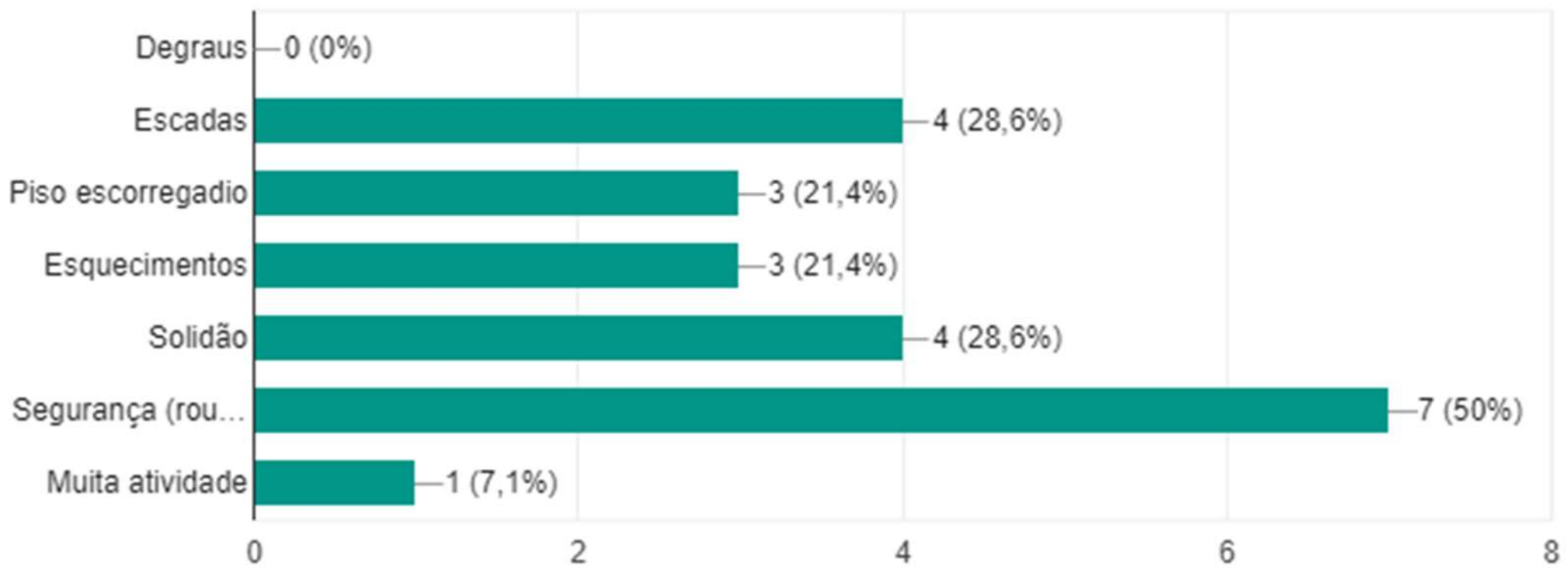

**Figura 13. Questionário familiares: Principais desafios e necessidades dentro de casa**

Nas situações de perigo, cerca de 84% dos entrevistados acreditam que seus pais iriam recorrer a ajuda de vizinho ou familiar que vive na mesma cidade.

*(4) A Tecnologia e o Home Care*

Com relação a tecnologia, a maioria dos entrevistados aponta que os pais usam pouco, não gostam ou não usam, sendo que as mulheres aparecem como um grupo menos resistente do que os homens.

A tecnologia mais usada é o celular, para comunicação, inclusão social (93% dos entrevistados) e uso de redes sociais do tipo *Facebook Instagram* e *WhatsApp*. Em emergências, 93% dos entrevistados apontam que gostariam de ser notificados através de ligação pelo celular.

Foram apresentadas aos entrevistados 4 opções de tecnologias AAL que poderiam ser utilizadas no suporte ao monitoramento dentro de casa. São elas:

1. Sensores de ambiente para monitorar movimento e sinais de queda, acionados por comando de voz pelos seus pais;

2. Sensores de presença para reconhecer estranhos não autorizados;
3. Pulseiras de monitoramento de saúde, com medição de sinais vitais e alertas em caso de anomalias;
4. Sensores luminosos capazes de iluminar a casa, escadas e degraus, acionados por comando de voz.

A Figura 14 apresenta os principais resultados:

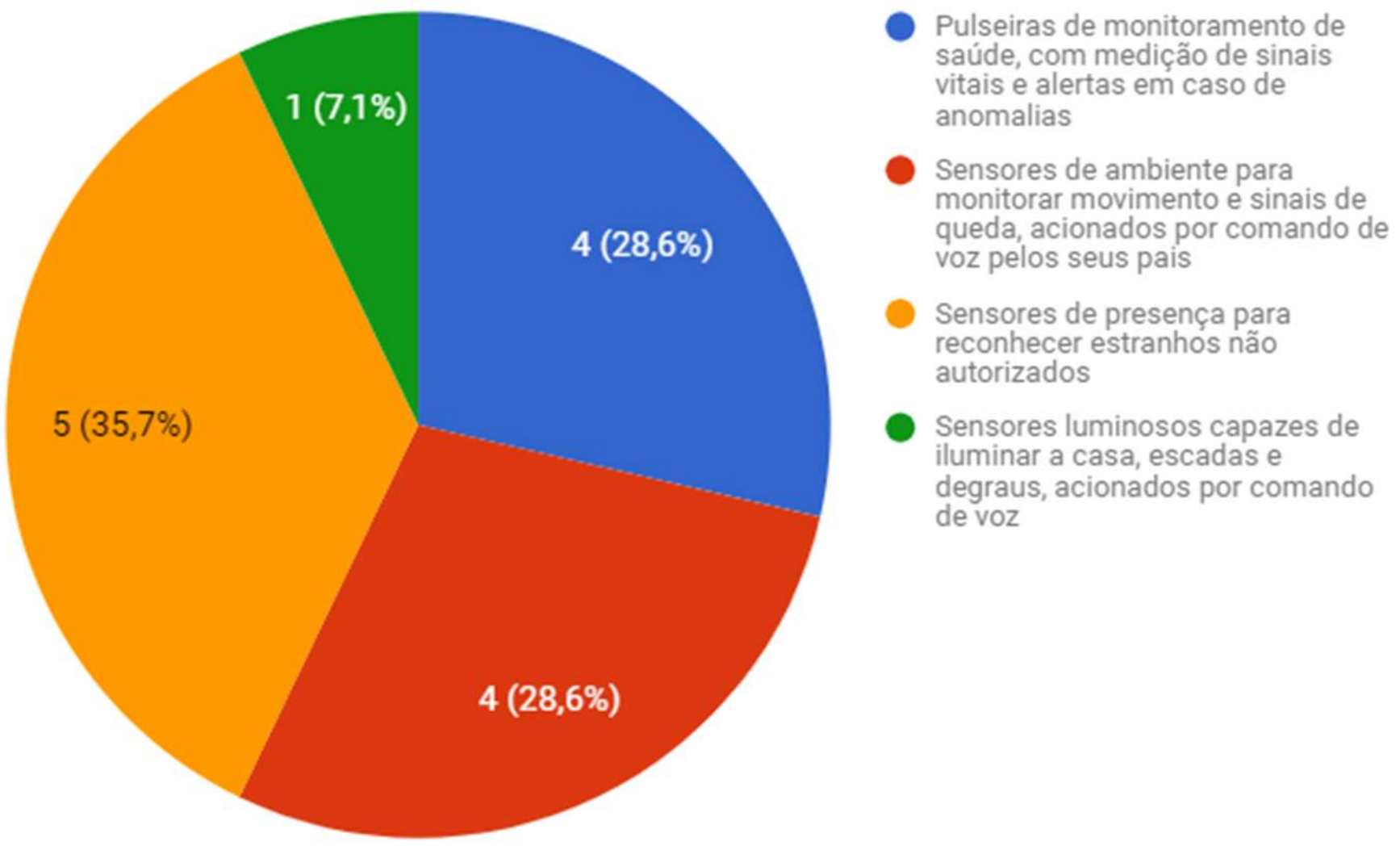

**Figura 14. Questionário familiares: Tecnologias AAL que podem ajudar nos cuidados com os idosos dentro de casa**

Como a maior preocupação dos filhos é com relação à Segurança, cerca de 36% dos entrevistados apontaram soluções que atendam a essa necessidade.

A pergunta “No futuro, em situações de perigo, como a tecnologia poderia auxiliar seus pais e te manter informado ou te acionar?” trouxe algumas sugestões e ideias para tecnologias AAL, categorizados conforme a seguir:

<u>Pulseira *wearable*</u>

- Uma pulseira que ao detectar algum tipo de perigo ela automaticamente filmaria o ambiente e a pessoa que está usando-a e ao mesmo tempo acionaria os familiares e ou médico responsável;
- Minha ideia é uma Pulseira de Comunicação: minha mãe já caiu e por conta do problema de coluna que tem não conseguiu se levantar para pedir socorro nem se comunicar com ninguém. Graças a Deus não demorou muito até que chegasse alguém e tudo ficou bem. O celular e outros meios de tecnologia nem sempre estão conosco em todos os momentos. Teria que ser algo que estivesse fixo ao nosso corpo e que fosse fácil acionar em uma eventualidade, com um simples aperto ou clique por exemplo;

- Talvez a pulseira que monitora anomalias ter conectividade e inteligência para avisar os familiares caso algo aconteça. Exemplo sensor de queda, botão emergência capaz de enviar mensagens;
- Já pesquisei soluções e entre elas, vi câmeras de monitoramento que sozinhas não resolvem, uma vez que falta o aviso. Então pesquisei e vi um serviço que fornece um colar ou pulseira que são botões de emergência. Ao apertar, o número desejado (até 5 números) é chamado. Dependendo do caso, dá para ver pela câmera. Mas isso não resolve problemas como desmaio etc. Não é o caso no momento, mas nunca se sabe;
- As pessoas que moram com idosos devem saber primeiros socorros, até que chegue ajuda de ambulância de um futuro drone ambulância. Pulseira que mostre todos os sinais vitais do paciente direto ao hospital com endereço.

<u>Celular</u>

- Como estou longe dos meus pais e eles moram em uma cidade pequena no interior, demoraria pelo menos 4 horas para que eu pudesse chegar numa emergência. Seria muito confortável se houvesse algum cadastramento dos idosos da cidade, principalmente dos que moram sozinhos (que são maioria) e que de alguma maneira eles pudessem acionar algum órgão em emergência. Nesse cadastramento já teriam doenças, medicamentos e particularidades cadastradas. Não imagino nada melhor que o celular para ajudar. Penso que um aplicativo que pudesse avisar quando ocorresse queda, pressão alterada, aumento da diabetes, por exemplo, os idosos portando celular, o aplicativo detecta alguma alteração e envia imediatamente um alerta para o meu celular ou de alguém próximo ou até de alguma entidade na cidade que pudesse acompanhar as emergências;
- Gostaria de ser acionado com um movimento brusco de queda do aparelho celular, mesmo que tenha caído das mãos deles, com esse aviso eu tentaria entrar em contato com eles caso não consiga acionar a emergência;

<u>Sensores</u>

- Sensores que identificam perigo e que façam ligação caso ocorra alguma situação

de perigo;

- Acredito que criar um ambiente de monitoramento, que possa identificar que algo inesperado aconteceu e disparar algum tipo de alerta, seria de grande ajuda pois muitas vezes o acidente pode ocorrer em um momento em que o socorro poderia demorar demais para chegar e os danos causados pelo acidente serem irreversíveis;
- As casas poderiam ter um sistema interligado com o sistema de emergência, caso ocorra algum problema, um alerta seja acionado na central;
- Caso ocorra uma queda na escada, essa ocorrência gera um som diferente. Esse som poderia disparar um SMS ou mesmo uma mensagem no *WhatsApp* para me avisar;
- Pelo menos uma pessoa da casa que more com idoso deveria saber primeiros socorros para cada caso. Assim daria tempo de acionar uma ambulância ou um drone. Ter câmeras e sensores de voz que acionem algum hospital, porque quando estamos em situações de um ataque cardíaco ficamos assustados e não sabemos o que fazer.

<u>Drones</u>

- Esse drone do vídeo poderia ser acionado pelo próprio usuário caso se sentisse mal ou por alteração brusca na frequência cardíaca, através de uma pulseira com botão de pane, conectado à uma Central remota. Essa informação poderia ser replicada para parentes que moram próximo ou mesmo algum vizinho pré-cadastrado.

### 3.3.3 Resultado Consolidado

A partir do questionário foi possível criar o Diagrama de Afinidades dos desafios em casa segundo a Tabela 9. Esse resultado compõe uma das saídas esperadas da Etapa 0: Diagrama de Afinidades de Riscos e Desafios para idade.

**Tabela 9. Diagrama de Afinidades de principais desafios no *Home care*, criado a partir do questionário com familiares**

| Categoria | Subcategoria | Desafios |
|---|---|---|
| **Segurança** | Roubos | -Roubos e assaltos |
| **Facilitação diária de tarefas** | Esquecimentos | -Hiperatividade: o corpo não acompanha a cabeça<br>-Esquecimentos |
| **Mobilidade** | Quedas | -Escadas<br>-Piso escorregadio |
| **Cuidados de saúde e Reabilitação** | Saúde | -Problemas nas articulações<br>-Doenças cardíacas<br>-Diabetes<br>-Problemas na coluna |
| **Inclusão social e comunicação** | Solidão | -Solidão |

A partir da entrevista foi possível criar o Diagrama de Afinidades das principais sugestões de AAL segundo a Tabela 10, categorizados de acordo com a classificação de AAL de Li, Lu e Maier (2015).

**Tabela 10. Principais sugestões AAL por categoria, criado a partir da entrevista com familiares**

| Categoria | Subcategoria | Desafios |
|---|---|---|
| **Segurança** | Roubos | -Sistema interligado nas casas com sistema de emergência, caso ocorra algum problema, um alerta seja acionado na central;<br>-Sensor que identifica queda por som e dispara um *SMS* ou *WhatsApp* de alerta; |
| **Facilitação diária de tarefas** | Não identificado | Não identificado |
| **Mobilidade** | Quedas | -Dispositivo que mediante movimento brusco de queda do aparelho celular notifica familiares; |
| **Cuidados de saúde e Reabilitação** | Saúde | -Pulseira que ao detectar perigo automaticamente filmaria o ambiente e a pessoa que está usando-a e aciona familiares e ou médico responsável;<br>-Pulseira de comunicação, que pode ser acionada para chamar socorro com um clique ou aperto;<br>-Pulseira que mostre todos os sinais vitais do paciente direto ao hospital com endereço;<br>-Cadastramento dos idosos da cidade, com doenças, medicamentos e particularidades cadastradas. Aplicativo no celular para detectar sobre perigos (queda, pressão, diabetes etc) e avisa órgão, vizinho ou familiar sobre emergência;<br>- Ambiente de monitoramento que possa identificar perigo e disparar alerta;<br>- Câmeras e sensores de voz que acionem algum hospital;<br>-Drone acionado pelo próprio usuário em situações de perigo, através de uma pulseira com botão de pane, conectado à uma Central e alerta pessoas cadastradas; |
| **Inclusão social e comunicação** | Não identificado | Não identificado |

### 3.4 Diagrama de Afinidades de principais desafios no *Home care*

A partir do questionário e entrevistas foi possível criar o Diagrama de Afinidades dos desafios em casa segundo a Tabela 11, que será usado como referência nas próximas etapas do método IoT-PMHCS.

**Tabela 11. Diagrama de Afinidades de principais desafios no *Home care***

| Categoria | Subcategoria | Desafios |
|---|---|---|
| **Segurança** | Roubos | -Roubos e assaltos<br>- Maus tratos |
| **Facilitação diária de tarefas** | Esquecimentos | -Hiperatividade: o corpo não acompanha a cabeça<br>-Esquecimentos |
| **Mobilidade** | Quedas | -Quedas em Escadas<br>-Quedas por piso escorregadio<br>-Quedas noturnas<br>-Queda na calçada em frente de casa |
| **Cuidados de saúde e Reabilitação** | Saúde | -Problemas nas articulações<br>-Doenças cardíacas<br>- Pressão alta<br>-Diabetes<br>-Problemas na coluna |
| **Inclusão social e comunicação** | Solidão | -Solidão |

### 3.5 Diagrama de Afinidades de sugestões AAL

A partir do questionário e entrevistas foi possível criar o Diagrama de Afinidades das principais sugestões AAL categorizadas. Esse diagrama é composto da consolidação dos resultados das Tabela 6, Tabela 8 e Tabela 10, e foi usado como referência nas próximas etapas do método IoT-PMHCS.

O resultado consolidado por ser encontrado no Apêndice V – Diagrama de Afinidades de Soluções AAL Propostas.

### 3.6 Plano de Design

O Plano de Design é uma das saídas dessa etapa e foi descrito conforme Tabela 4.

## 4 Etapa 1 - Mapeamento de Personas, Valores e Necessidades

Durante a Etapa 1 do método IoT-PMHCS foi realizado um processo participativo com os idosos a produção de artefatos.

As principais atividades dessa etapa estão brevemente apresentadas no Apêndice II – Etapa 1: Mapeamento de Personas, Valores e Necessidades.

Segue o registro das principais atividades realizadas, análises e resultados gerados de acordo com a proposta do método IoT-PMHCS.

## 4.1 Relatório *Workshop* Valores e Necessidades

Para a realização desse primeiro *Workshop* foi selecionado um grupo de 9 pessoas entre 60 e 80 anos da cidade de Muzambinho, MG, conforme descrito na seção 0.

### 4.1.1 Organização do *Workshop*

A Tabela 12 descreve a organização do estudo de caso para o *Workshop* de Valores e Necessidades de acordo com a proposta do Método IoT-PMHCS, contendo uma análise de eficácia do método para cada item.

**Tabela 12. *Workshop* de Valores e Necessidades: organização das atividades**

| Proposta de Organização | | Análise |
|---|---|---|
| **Material** | Para a execução desse *Workshop* foram utilizados os seguintes materiais: *post-its* coloridos, cartolina para a organização dos resultados, vídeo para contextualizado do conceito IoT, música ambiente para receber os convidados;<br>O convite aos participantes foi feito pessoalmente pela mediadora e como agradecimento foi oferecido um lanche (comes-e-bebes) ao final do evento. | Todo material foi considerado adequado e suficiente para a execução dessa atividade. |
| **Material Adicional** | Foi solicitado que cada participante trouxesse um objeto pessoal que represente um valor para si.<br>Todos trouxeram e souberam explicar muito bem dentro do contexto, o que enriqueceu o trabalho, gerou maior sinergia e empatia entre o grupo. | Foi necessário explicar com detalhes e exemplos essa atividade durante o convite. |
| **Participantes** | Foram convidados 12 participantes, sendo que apenas 9 compareceram. Composição do grupo:<br>- 60 a 69 anos: 3 mulheres e 1 homem<br>-70 a 79 anos: 2 mulheres e 2 homens<br>-Acima de 80 anos: 1 mulher<br>Os voluntários vivem com autonomia em suas casas, acompanhados do cônjuge ou familiar.<br>Eles foram selecionados pela proximidade com a mediadora e por suas características semelhantes às 3 Personas selecionadas para análise.<br>Todos já se conheciam previamente, o que facilitou à sinergia do grupo. Foi convidado também um assistente | O convite foi feito pessoalmente e houve resistência na aceitação de alguns idosos, por não se sentirem à vontade na execução das atividades em grupo ou ao tema da pesquisa. |
| | para apoio às atividades, Analista de Sistemas. | |
| **Local** | A atividade foi realizada em Muzambinho, MG na casa de um dos voluntários, em ambiente aconchegante e familiar, o que facilitou a aceitação do convite. | O local trouxe desafios por não ser apropriado para o acompanhamento próximo de cada grupo durante as atividades, que ficaram trabalhando em cômodos diferentes da casa e de acesso difícil. |
| **Registro** | O registro das atividades foi feito por de forma textual e por fotos.<br>Todas as atividades executadas foram registradas em *post-its* e cartolinas. Todos os resultados foram armazenados para análise. | Os participantes não se sentiram à vontade em serem filmados.<br>Mas os registros realizados foram considerados satisfatórios para essa atividade. |
| **Duração** | A duração recomendada foi de 2 a 3 horas<br>A duração real da atividade foi de 2:15 e o café final de 1h. | Esse tempo foi adequado para à maioria dos participantes, porém alguns já se apresentavam impacientes e cansados nas atividades finais. |
| **Resultados Esperados** | Como resultado desse *Workshop* serão criados 2 Diagramas de Afinidades, um para valores e outro para desafios da idade (ou categorias de *Home Care*). | Foi possível produzir os resultados esperados com a condução desse *Workshop* e criar uma versão preliminar do Mapa de Persona. |

### 4.1.2 Atividades Propostas no *Workshop*

A Tabela 13 apresenta o planejamento de atividades para esse *Workshop*.

**Tabela 13. *Workshop* de Valores e Necessidades: planejamento de atividades**

| Objetivo | Atividades Sugeridas | Duração |
|---|---|---|
| **1-Contextualizar** | -Contextualizar sobre o projeto (pode ser usado um pôster ou slides) e desafio e design;<br>-Apresentar vídeo para facilitar a contextualização;<br>-Apresentar os valores identificados em pesquisa de Leong e Robertson (2016);<br>-Apresentar as categorias de *Home Care* de acordo de Li, Lu e Maier (2015); | 15 min |
| **2-Apresentação dos participantes** | -Pedir para que cada participante se apresente, respondendo às seguintes perguntas (adaptado de Leong e Robertson (2016 - p.32):<br>1-Quem eu sou (nome e idade); | 30 min |
| | 2-Estilo de Vida (onde vivem, família, atividades);<br>3- Familiaridade com tecnologia (gosta ou não);<br>3-Apresentar o objeto que trouxe e explicar porque ele representa um valor em sua vida. | |
| **3-Organização dos grupos de trabalho por Personas** | Selecionar as Personas descritas no trabalho de Gonçalves e Bonacin (2017) que mais se assemelham com o grupo de trabalho;<br>Organizar os grupos de trabalho de acordo com a semelhança com as Personas selecionadas;<br>A escolha pode ser feita pelo mediador ou por afinidade pelo próprio grupo. | NA |
| ***4-Brainstorm* para priorização de Valores do Grupo** | - Separar em grupos de 3-4 pessoas de acordo com a Persona selecionada;<br>- Avaliar os valores da pesquisa de Leong e Robertson (2016);<br>- Existem novos valores? Se sim, identificar em *Post-Its* de cores diferentes;<br>-Os valores devem ser priorizados (do mais para o menos importante);<br>-Anotar os resultados consolidados em *Post-Its* e colocar de forma visual e agrupada numa lousa ou parede;<br>- Definir por votação o valor mais importante do grupo. | 20 min |
| ***5-Brainstorm* para priorização de Desafios no *Home care* da Persona** | - Manter os grupos anteriores;<br>- Apresentar a Persona ao grupo, caso à escolha tenha sido feita pelo mediador;<br>- Solicitar que o grupo escolha o Valor que mais se identifique com a Persona;<br>- Discutir em grupo os maiores desafios no dia a dia, dentro de casa, para a Persona selecionada. Identificar cada desafio em um *Post-It*;<br>- Discutir em grupo as maiores necessidades no dia a dia dentro de casa para o grupo; identificar cada desafio em um *Post-It*, com cores diferentes;<br>- Agrupar os desafios de acordo com as categorias propostas de Li, Lu e Maier (2015);<br>- As categorias devem ser priorizadas (do mais até o menos relevante) para a Persona seleciona;<br>- Destacar o desafio mais importantes da Persona. | 30 min |
| **6-Apresentação dos resultados** | Um representante de cada grupo apresenta para todos o resultado do *Brainstorm*;<br>1- Apresentar os valores priorizados pelo grupo e justificar;<br>2- Apresentar os desafios priorizados para a Persona e justificar;<br>Podem ser realizados um debate em grupo sobre os resultados apresentados. | 30 min |
| **7-Encerramento** | -Agradecimentos;<br>-Coletar *feedback* dos participantes;<br>-Definir próximos passos. | 10 min |

### 4.1.3 Análise das Atividades Realizadas

A Tabela 14 apresenta uma breve análise das atividades realizadas para se atingir o objetivo desse *Workshop.*

**Tabela 14.** ***Workshop* de Valores e Necessidades: Análise do *Workshop* de Valores**

| Atividade Planejada | Análise |
|---|---|
| **1-Contextualização** | A contextualização foi realizada com sucesso, com uma apresentação em *Power Point* projetada numa TV da sala e todos confortavelmente acomodados. Durante a apresentação foram informados aos participantes:<br>- os objetivos da pesquisa<br>- as etapas do método<br>- os principais conceitos<br>- resultados esperados.<br>Apresentou-se também um vídeo didático sobre o conceito de IoT.<br>Foi enfatizada a importância da participação de todos, no sentido de ajudar o futuro da humanidade através desse projeto de pesquisa: isso ajudou no engajamento e melhor participação, sem a preocupação de serem julgados durante as atividades. |
| **2-Apresentação dos participantes** | Esta atividade foi muito bem-sucedida, os participantes se mostraram à vontade para falar sobre seus objetos e suas reflexões. |
| **3-Organização do Grupo de trabalho por Personas** | A organização dos grupos por Personas feita pela mediadora foi bem-sucedida e construiu grupos aderentes para à execução dos trabalhos.<br>**Grupo 1 - Antônio**: os 3 voluntários homens aposentados e que gostam de tecnologia. Único aspecto que não correspondeu foi à agressividade, uma vez que todos possuem bom relacionamento familiar.<br>**Grupo 2 - Maria de Lourdes**: 3 voluntárias mulheres independentes, com questões de saúde não impeditivas e que gostam de tecnologia. Porém, as 3 são casadas e não divorciadas, como à Maria de Lourdes.<br>**Grupo 3 - Dona Nair:** foi o grupo mais heterogêneo para esses exercícios, uma vez que eram mulheres com condições de vida muito diferentes, sendo apenas uma próxima da descrição da Persona. Porém, todas possuem restrições de saúde ou tecnológicas semelhantes, o que será positivo para os próximos *Workshops*. |
| **4-*Brainstorm* para priorização de Valores do Grupo** | Esta atividade exigiu um acompanhamento próximo dos grupos, mesmo após explicação, mas o trabalho foi bem-sucedido e os grupos rapidamente chegaram à um resultado.<br>**Importante**: o exercício de valores foi feito pensando no grupo de pessoas e suas prioridades e não nas Personas. |
| **5-*Brainstorm* para priorização de Desafios no *Home Care* da Persona** | Esta atividade foi a mais trabalhosa e exigiu uma presença maior da mediadora na sua execução. Todos os grupos tiveram dúvidas na execução. Além disso, os grupos estavam em cômodos diferentes, o que dificultou o suporte e convergência. O assistente não conseguiu dar apoio adequado nessa atividade.<br>Apesar disso, o resultado foi satisfatório e todos os grupos conseguiram chegar ao resultado esperado, mesmo o exercício tendo demorado mais que o previsto.<br>**Lição Aprendida:** observou-se que um tempo maior de explicação do exercício, blocos menores de entregas e impressão da sequência de atividades teriam otimizado a execução do exercício.<br>**Importante**: o exercício foi feito pensando na Persona, seus valores e necessidades. Todos os aspectos do grupo foram incluídos em *Post-Its* de cores diferentes para facilitar a identificação. |
| **6-Apresentação dos resultados** | Cada grupo apresentou com sucesso os seus resultados, através de um representante. O grupo já estava bem mais desinibido e integrado ao final das |
| | atividades. Porém, a apresentação foi mais rápida do que o planejado pois os participantes já estavam cansados, não havendo tempo hábil para discussão dos resultados. |
| **7-Encerramento** | O encerramento foi muito positivo, e foram coletados feedbacks dos participantes. |

### 4.1.4 Valores dos Participantes

A partir da apresentação dos objetos de valor de cada voluntário foi feita uma análise de acordo com os critérios identificados por Leong e Robertson (2016), conforme apresentado na Tabela 15.

**Tabela 15. *Workshop* de Valores e Necessidades: Diagrama de Afinidade de Valores apresentados pelo grupo**

| Nome | Idade | Objeto Apresentado | Significado | Categoria de Valor |
|---|---|---|---|---|
| Maria Gabriela | 80 | Foto do marido falecido | A importância do marido na sua vida e da família que eles construíram juntos | Fazer Parte |
| Teresa | 74 | Foto dos filhos e uma caixa de Remédio | A importância dos filhos e da sua saúde | Fazer Parte e Senso de Propósito |
| Elizabete | 67 | Foto da família | Manter a família completa e unida por muitos anos | Fazer Parte |
| Luiz | 66 | Remédio | Saúde | Senso de Propósito |
| Noel | 70 | Não trouxe, mas falou | Não deixar de sonhar, independentemente da idade | Senso de Propósito |
| Edson | 75 | Relógio | O carinho dos pais e sua realização profissional | Fazer Parte e Senso de Propósito |
| Maria José | 68 | Bíblia | A importância da religiosidade em sua vida | Senso de Propósito |
| Elga | 63 | Foto do Netos e remédio | A importância dos netos e da sua saúde | Fazer Parte e Senso de Propósito |
| Lúcia | 72 | Imagem de Nossa Senhora | A importância da família, dos amigos e da sua saúde, representada por Nossa Senhora da Saúde | Fazer Parte e Senso de Propósito |

Os idosos não apontaram valores adicionais que não pudessem ser adequados as categorias já identificadas por Leong e Robertson (2016).

A Figura 15 representa o resultado consolidado da análise, sendo que o item mais mencionado por todos é o Senso de Propósito, representado pela Saúde ou pela Religiosidade, o que corresponde aos resultados apresentados por Leong e Robertson (2016).

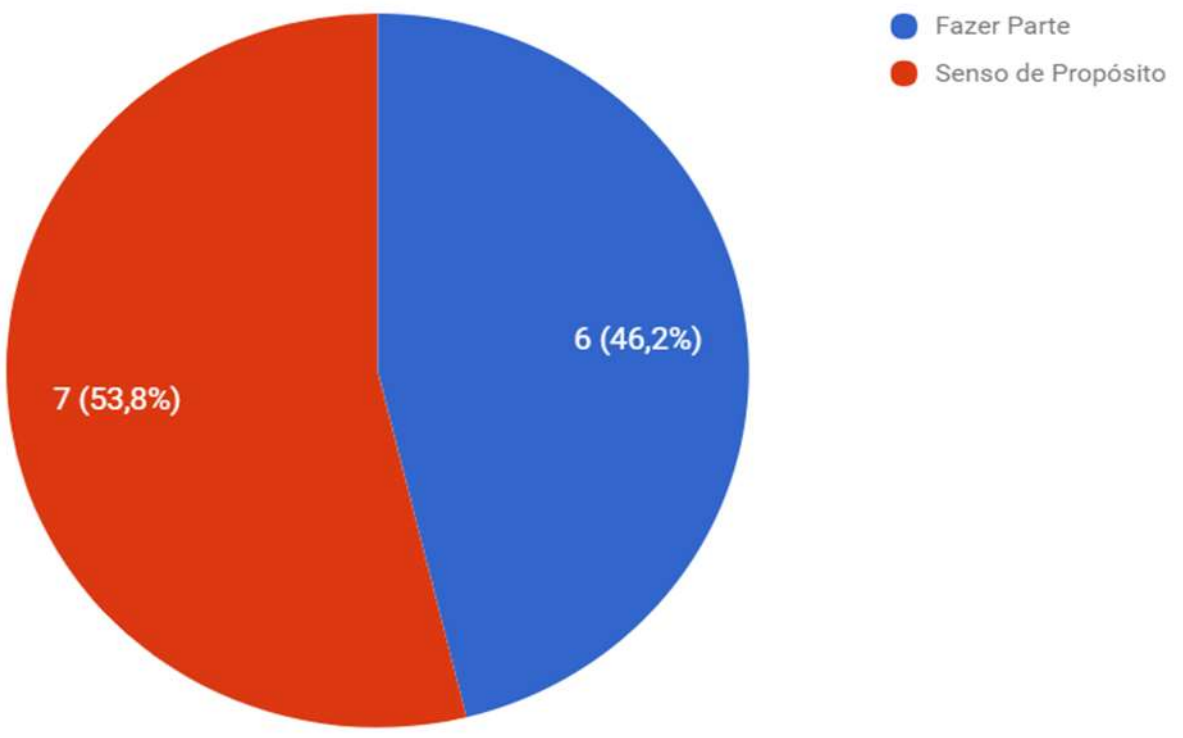

**Figura 15. *Workshop* de Valores e Necessidades: Valores Apresentados pelos Voluntários Individualmente**

### 4.1.5 *Brainstorm* para priorização de Valores do Grupo

Para a atividade 2 foi apresentado ao grupo o resultado do trabalho realizado por Leong e Robertson (2016) e as diferentes categorias de Valor. Foi solicitado ao grupo discutir sobre seus valores mais importantes e priorizar, sem deixar de considerar o valor representado pelos objetos trazidos.

A organização dos grupos foi realizada pela mediadora, usando critério de afinidade de cada voluntário com as Personas sugeridas por Gonçalves e Bonacin (2017), de acordo com a Tabela 16. Os voluntários até o momento não conheciam as Personas.

**Tabela 16. *Workshop* de Valores e Necessidades: organização dos grupos de voluntários por afinidade com as Personas**

| Grupo - Persona | Voluntários |
|---|---|
| Grupo 1 – Antônio | Luiz, Edson, Noel |
| Grupo 2 - Maria de Lourdes | Elizabete, Lúcia, Teresa |
| Grupo 3 – Nair | Maria José, Maria Gabriela, Elga |

A Tabela 17 apresenta o resultado consolidado por grupo, após o *Brainstorm de Valores*, considerando os valores do grupo e sem considerar as Personas.

**Tabela 17. *Workshop* de Valores e Necessidades: Valor Priorizados pelo Grupos**

| Grupo – Persona | Valores por Ordem | Valor Principal |
|---|---|---|
| Grupo 1 – Antônio | Senso de Propósito, Fazer Parte, Ser Valorizado, Contribuir, Independência | Senso de Propósito |
| Grupo 2 - Maria de Lourdes | Independência + Religiosidade, Fazer Parte, Senso de Propósito, Contribuir, ser valorizado | Independência Religiosidade |
| Grupo 3 – Nair | Senso de Propósito (Fé), Fazer Parte, Contribuir, Independência, Ser Valorizado, | Senso de Propósito (Fé) |

O Grupo 1 quando questionado sobre à Independência, afirmou que é algo já conquistado e que deve apenas ser mantido. Eles se referiam principalmente à independência financeira.

O Grupo 2 optou por associar à categoria Independência a Religiosidade, sendo essa responsável por completar também as categorias Senso de Propósito e Contribuição, indicando que a religiosidade é fundamental na vida das pessoas após os 60 anos para se conquistar a Independência, respalda seus Propósitos de Vida e é fundamental para se contribuir com o próximo.

O Grupo 3 optou por renomear a categoria Senso de Propósito para “Fé”, indicando o propósito principal de suas vidas é também a religiosidade.

Importante observar que 2 grupos formados por mulheres incluíram à religiosidade de forma direta ou indireta aos seus valores, sendo esse considerado um valor fundamental em suas vidas.

Observa-se que os grupos 1 e 3 mantiveram a análise inicial de Leong e Robertson (2016), de acordo com o resultado da Figura 15, sendo o Senso de Propósito o valor mais importante e representado pela Saúde e Religiosidade.

O grupo 2 optou pela Independência como principal valor, o que corresponde ao resultado da pesquisa realizada com familiares e as características da Persona Maria de Lourdes que esse grupo representa.

Se analisarmos as categorias definidas em grupo em comparação com os valores apresentados individualmente, existe uma correspondência em 55% dos casos (5 das 9 pessoas). Ou seja, a maioria dos voluntários manteve suas convicções mesmo após apresentados às categorias de valores e o debate em grupo.

Conclui-se, portanto, que o “Senso de Propósito” é ainda o valor mais relevante para as pessoas acima de 60 anos, sendo esse principalmente representado pelos cuidados com a saúde ou pela religiosidade. Em segundo lugar, encontra-se o valor “Fazer Parte” de uma família ou grupo, sendo um consenso para todo os grupos como segundo critério mais relevante

Importante reforçar que a “Independência” deve ser também considerada como um valor de peso para os idosos, uma vez que foi apontada pelos familiares em pesquisa e priorizada pelo Grupo 2.

### 4.1.6 *Brainstorm* para priorização de Desafios no *Home care*

Para esta atividade foram apresentadas aos grupos as Personas sugeridas por Gonçalves e Bonacin (2017). A divisão das pessoas por grupo foi feita pela mediadora, usando critério de afinidade de cada voluntário com a Persona selecionada como descrito acima.

Foi solicitado aos grupos apresentar o principal valor para cada Persona e os principais desafios no *Home Care*, agrupados e priorizados pelas categorias sugeridas por Li, Lu e Maier (2015). As impressões pessoais também deveriam ser mencionadas e explicitadas em cores diferentes.

Importante mencionar que a categoria Segurança foi incluída pela mediadora devido aos resultados da pesquisa com familiares, que trouxe essa preocupação enfatizada dado o cenário atual no Brasil. A Tabela 18 apresenta o resultado consolidado por grupo.

**Tabela 18. *Workshop* de Valores e Necessidades: Desafios no *Home care* Priorizados para as Personas**

| Grupo - Persona | Valor da Persona | Categorias Priorizadas por Persona | Desafios no *Home Care* da Persona | Desafios adicionais do Grupo |
|---|---|---|---|---|
| **Grupo 1 Antônio** | Independência | 1- Saúde<br>2- Mobilidade | 1-Controle da Medicação | 2-Mobilidade<br>3-Cultivar amigos |
| | | 3- Inclusão Social<br>4- Segurança<br>5- Facilitar dia-dia | | 4- Roubos e assaltos<br>5-Esquecimento |
| **Grupo 2 Maria de Lourdes** | Independência | 1- Saúde<br>2- Inclusão Social<br>3- Segurança<br>4- Facilitar dia-dia<br>5- Mobilidade | 1-Esquecer de tomar remédios, cuidar da alimentação<br>2-Melhorar autoconfiança<br>3-Preocupação com panelas ligadas<br>4-Esquecimentos<br>5-Dificuldade com mobilidade | 2-Solidão<br>3-Assalto e roubos |
| **Grupo 3 Nair** | Fazer Parte | 1- Inclusão Social<br>2- Saúde<br>3- Facilitar dia-dia<br>4- Mobilidade<br>5- Segurança | 1- Necessidade de companhia ou cuidador<br>2- Controle de medicamentos, atividades diárias, alimentação | 2-Depressão, dificuldade de enxergar<br>3-Esquecimento<br>4-Dificuldade com mobilidade |

Observa-se, na Tabela 18, que ao ser questionado por Valores, 2 Personas apresentam a “Independência” como principal valor.

A Persona “Maria de Lourdes” foi à única que manteve o valor mencionado pelo grupo na Atividade 1. Os outros 2 grupos classificaram valores diferentes entre o grupo e a Persona, o que demonstra grande afinidade do Grupo 2 com a Persona selecionada.

Com relação ao “Grupo 1- Antônio”, houve mudança do valor Senso de Propósito para Independência, uma vez que essa análise foi feita em função da descrição da Persona, e não mais somente dos valores do grupo. Porém, observa-se que a categoria assistiva priorizada por esse grupo foi a saúde, o que vai ao encontro do significado de Senso de Propósito para o grupo na atividade 3 (senso de propósito ligado à saúde). Isto nos permite fazer uma analogia nesse contexto entre independência e senso de propósitos, uma vez que ambos estão ligados a questões de cuidado e manutenção da saúde.

Observa-se que **saúde** é o maior desafio ou um dos maiores na vida das pessoas acima de 60 anos, independentemente de seu perfil.

O Figura 16 apresenta o resultado consolidado dos maiores desafios dentro de casa mencionados pelos grupos, independente da Persona ou Grupo, sendo o **esquecimento** o mais citado pelos grupos.

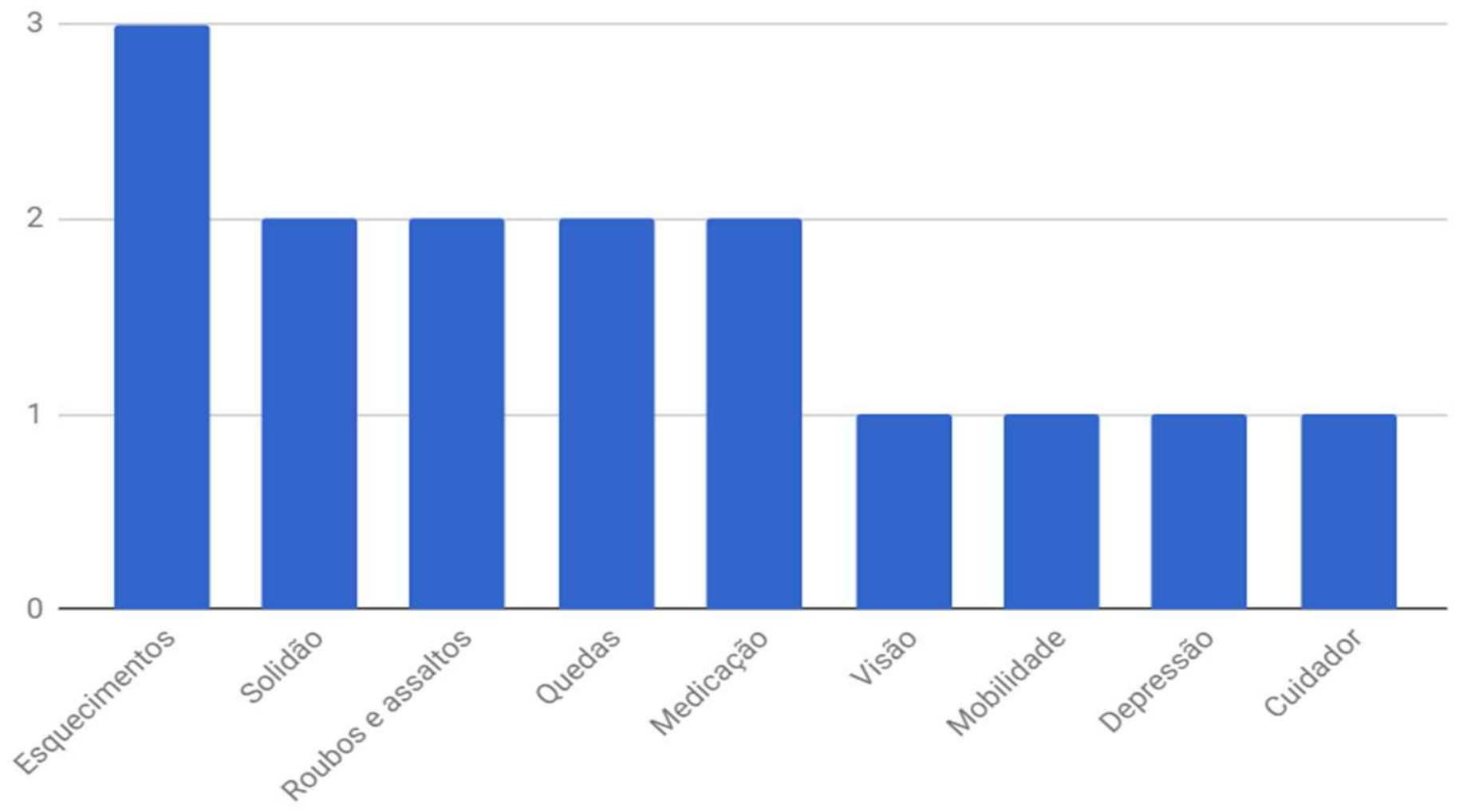

**Figura 16.** ***Workshop*** **de Valores e Necessidades: Maiores Desafios no** ***Home care***

Esse resultado confirma o mesmo resultado obtido durante as entrevistas com os idosos descrita na seção 3.1.

## 4.2 Personas

Foram escolhidas as seguintes Personas para a realização das atividades deste projeto, a partir dos resultados da pesquisa de Gonçalves e Bonacin (2017). Tais Personas são aquelas que mais possuem afinidade com o grupo de voluntários selecionado.

Estão destacadas em cada descrição abaixo as características da Personas que não se assemelham ao grupo de voluntários, e que foram consideradas como possíveis adaptações as características dessa Persona para maior afinidade com o grupo.

### 4.2.1 Antônio

A Figura 17 apresenta a descrição da Persona, com destaques em vermelho para as características da Persona que não se assemelham ao grupo de voluntários selecionado.

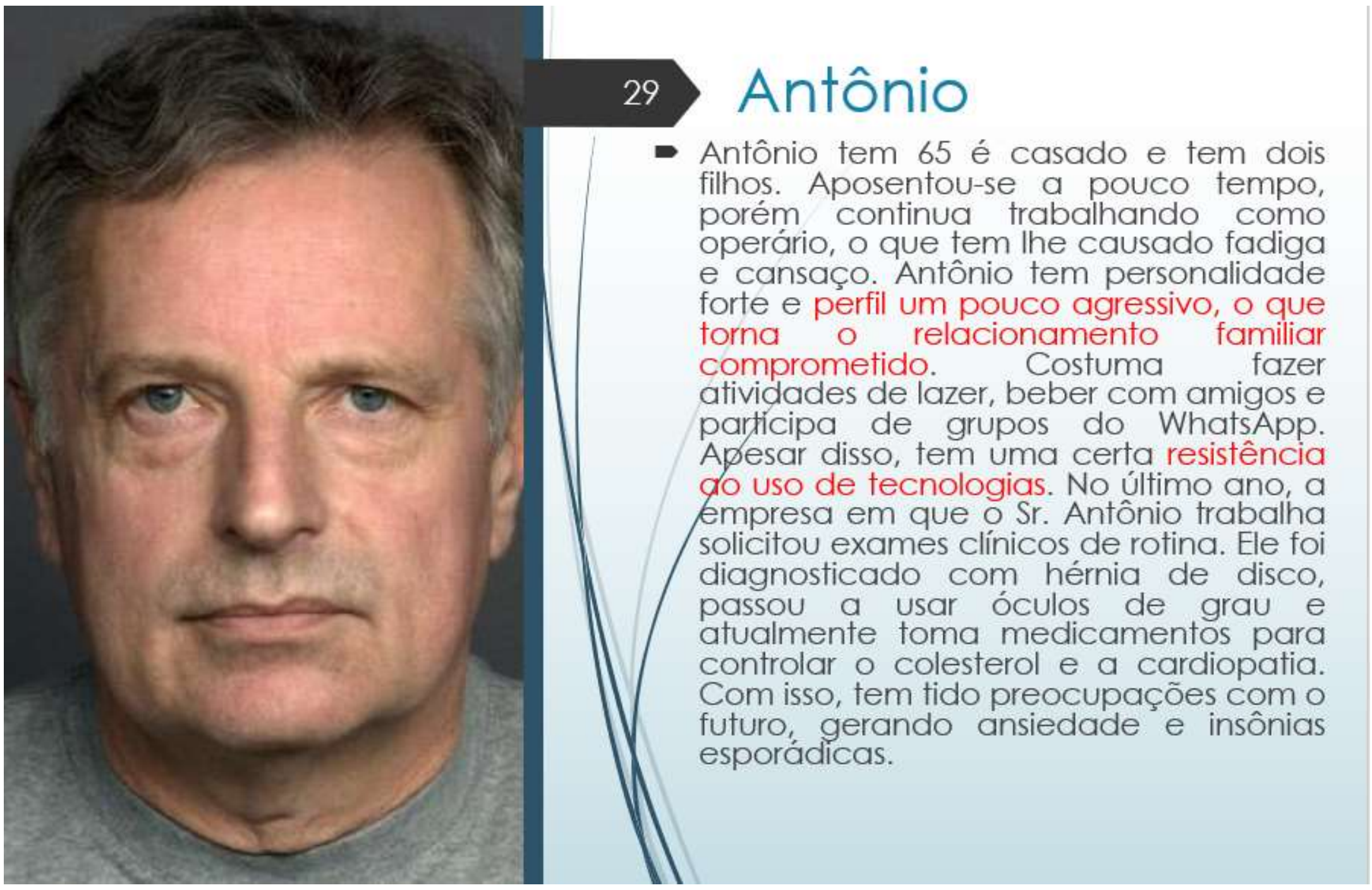

**Figura 17. Persona Antônio**

### 4.2.2 Maria de Lourdes

A Figura 18 apresenta a descrição da Persona, com destaques em vermelho para as características da Persona que não se assemelham ao grupo de voluntários selecionado.

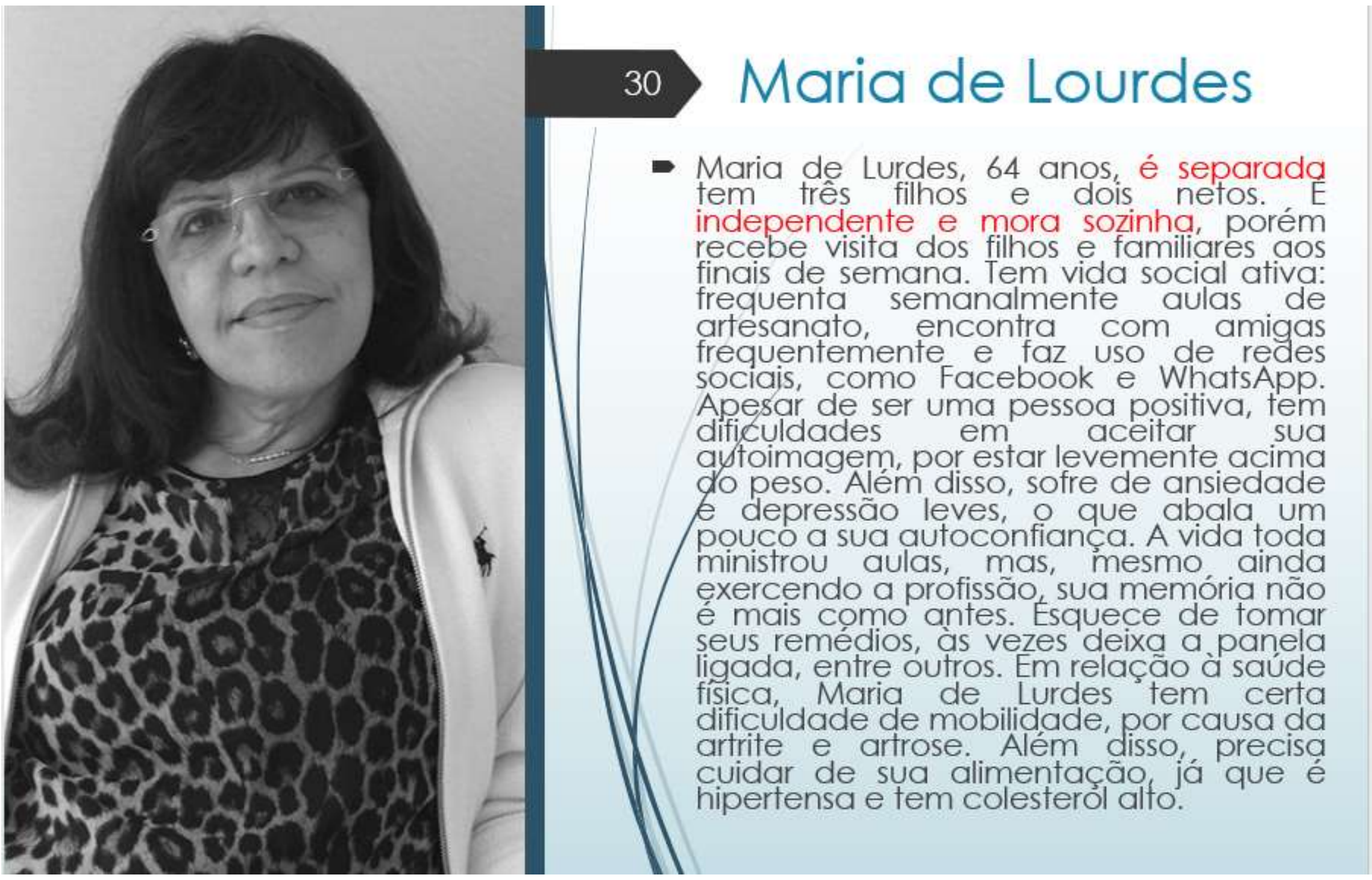

**Figura 18. Persona Maria de Lourdes**

### 4.2.3 Nair

A Figura 19 apresenta a descrição da Persona, com destaques em vermelho para as características da Persona que não se assemelham ao grupo de voluntários selecionado.

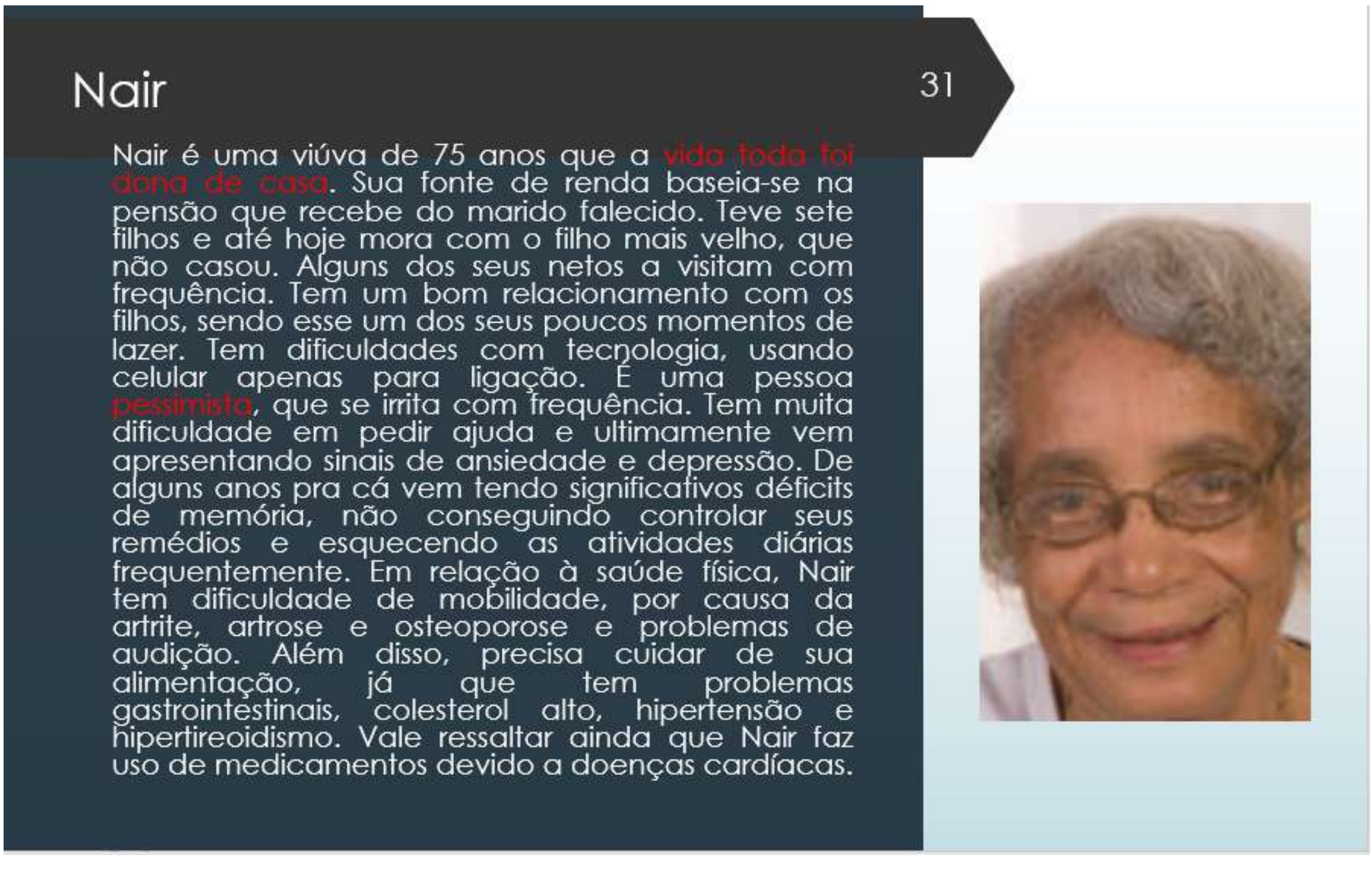

**Figura 19. Persona Nair**

## 4.3 Mapa de Persona

A partir das atividades participativas desse método foram gerados resultados para construir uma versão preliminar do Mapa de Persona para cada Persona selecionada para esse estudo de caso, conforme a seguir:

### 4.3.1 Antônio

A Figura 20 apresenta o mapeamento da Persona Antônio a partir dos resultados gerados pelo Grupo 1.

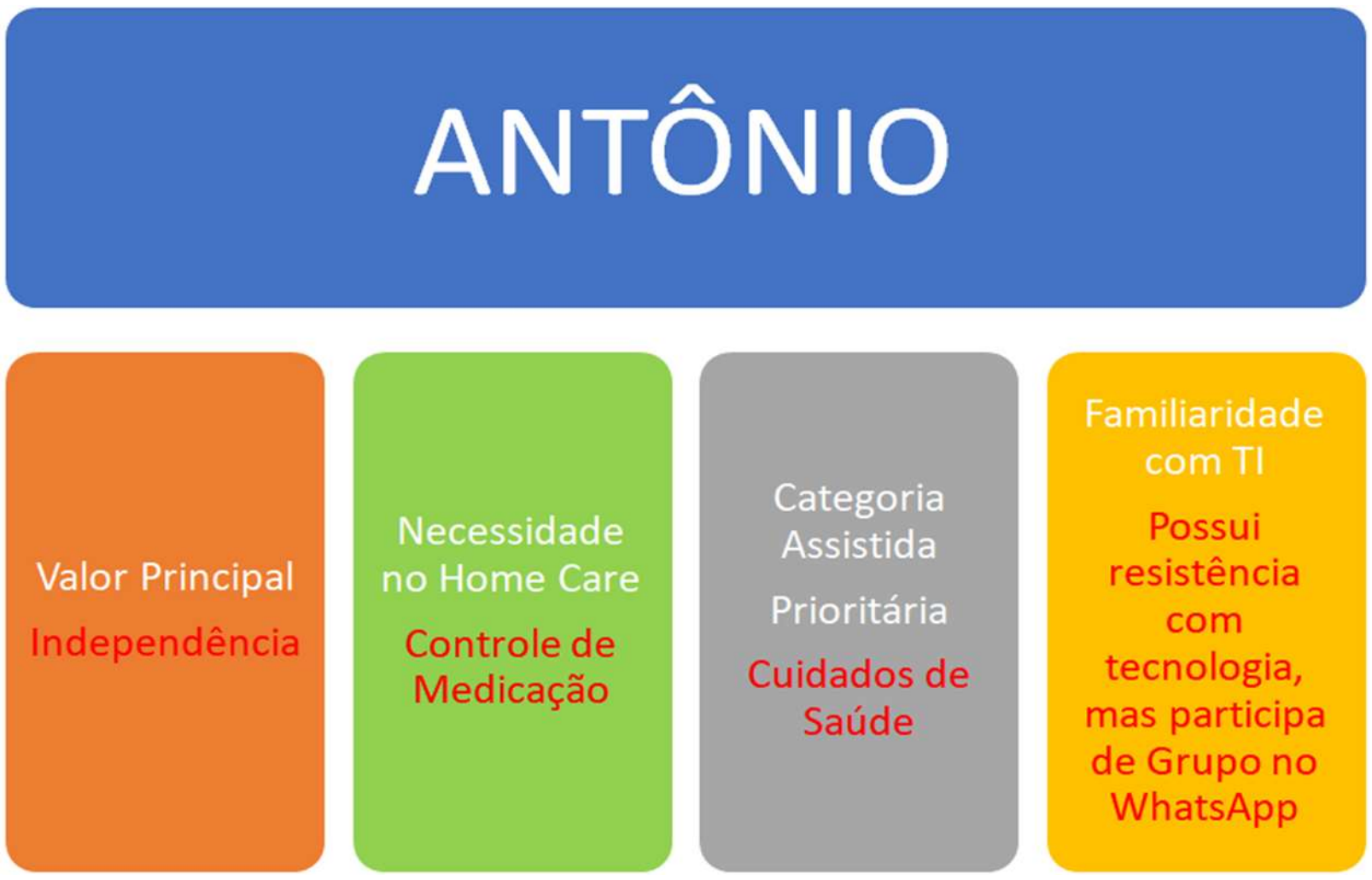

**Figura 20. Etapa 1: Mapa da Persona Antônio**

### 4.3.2 Maria de Lourdes

A Figura 21 apresenta o mapeamento da Persona Maria de Lourdes a partir dos resultados gerados pelo Grupo 2.

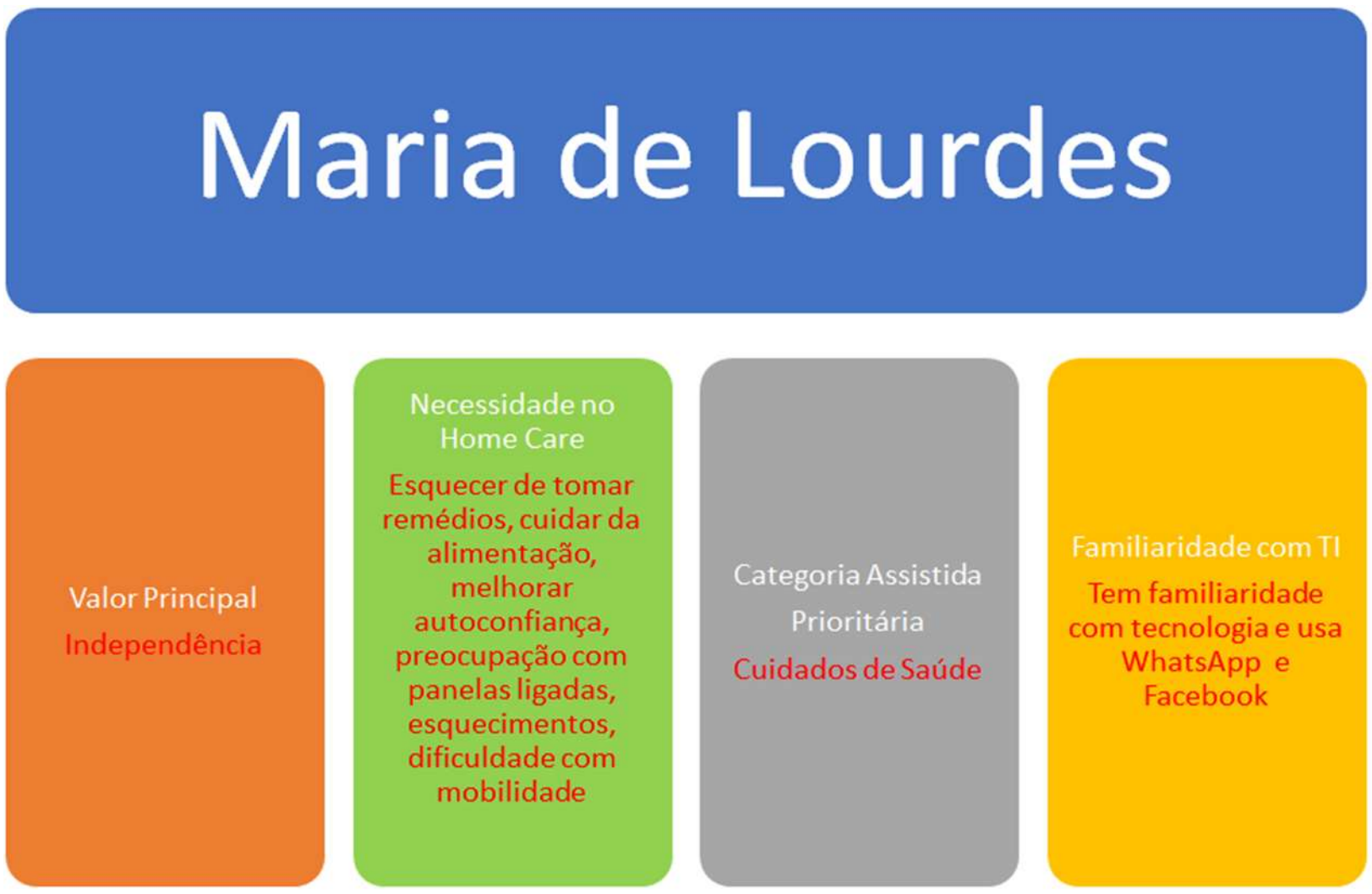

**Figura 21. Etapa 1: Mapa da Persona Maria de Lourdes**

### 4.3.3 Nair

A Figura 22 apresenta o mapeamento da Persona Nair a partir dos resultados gerados pelo Grupo 3.

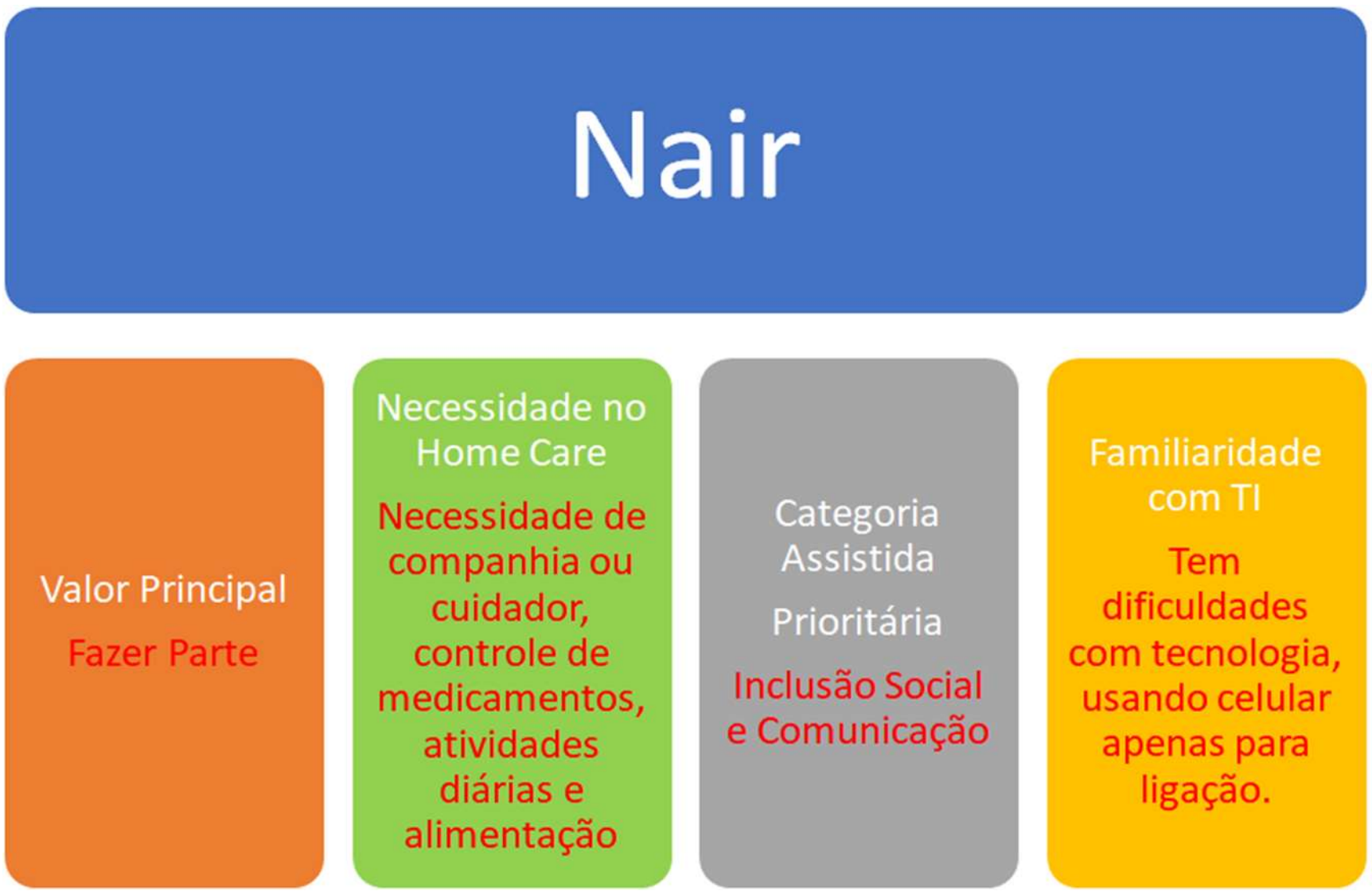

**Figura 22. Etapa 1: Mapa da Persona Nair**

## 4.4 Critérios Norteadores

Após a execução desta etapa foi possível verificar os seguintes Critérios Norteadores iniciais a serem utilizados nas próximas etapas para identificação de soluções inovadoras para auxílio no *Home Care* (Tabela 19).

**Tabela 19. Etapa 1: Critérios Norteadores**

| Critério | Objetivo | Valor |
|---|---|---|
| Aderência ao Valor de cada Persona | Esse é o critério de maior peso e deve ser considerado para que exista aceitação da tecnologia pelas idosos. Portanto deve-se observar se a tecnologia não irá ferir o valor identificado para a Persona. Por exemplo, se a Persona tem como principal valor Independência, colocar câmeras por toda a casa irá ferir esse valor e, portanto, a solução não terá boa aceitação. | Valor de 0 a 4<br>Quanto maior o valor maior à aderência ao valor da Persona. |
| Categoria de *Home Care* prioritária para a Persona | Esse critério indica o quanto a solução trará benefícios para a Persona dentro das categorias priorizadas no *Home Care* durante o *Workshop* de Valores. | Valor de 0 a 5.<br>Quanto maior o valor maior à aderência à Categoria de *Home Care* da Persona. |
| Facilidade de uso e assimilação por idosos | Esse critério indica a facilidade de uso pelos idosos e a rápida assimilação para a Persona analisada, em função de sua familiaridade e gosto pela tecnologia. | Valor de 0 a 2.<br>Quanto maior o valor maior à facilidade de uso. |
| Viabilidade Técnica de Implementação | Esse critério indica se à solução apresenta alguma restrição técnica de ser implementada, inviabilizando ou dificultando sua prototipação. | Valor de 0 ou 1.<br>Sendo que 1 indica que é viável de ser implementada e 0 muito difícil ou inviável. |

A Figura 23 apresenta os Critérios Norteadores identificados a partir das pesquisas realizadas até o momento.

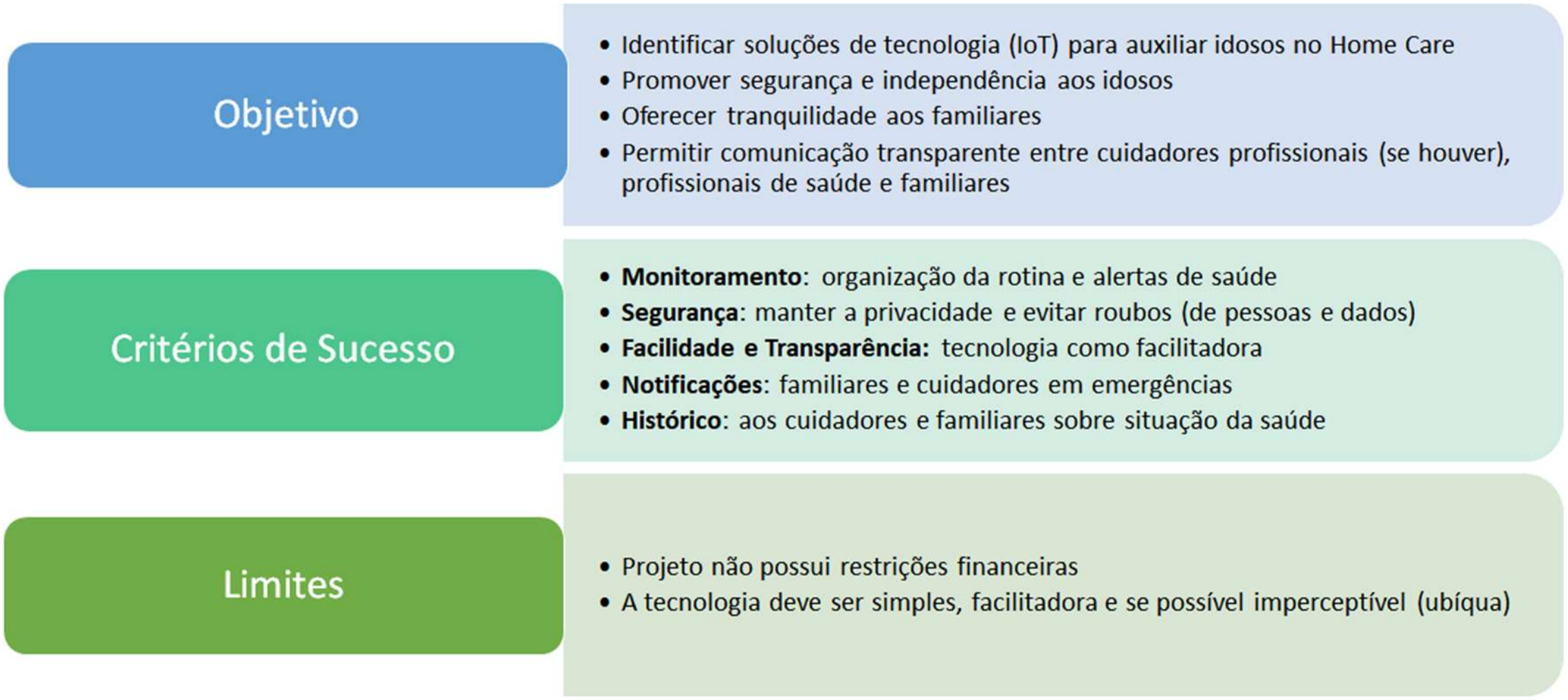

**Figura 23. Critérios Norteadores identificados durante a pesquisa**

## 4.5 Pesquisa de Participação

Após o encerramento das atividades, foi realizada uma coleta de *feedbacks* do grupo. Seguem os principais resultados obtidos:

- Realizar apresentação intermediárias de cada resultado;
- O tempo foi suficiente para a maioria dos participantes;
- A organização dos grupos foi satisfatória;
- A orientação da Atividade 2 não foi satisfatória, sendo necessário maior tempo de explicação e material impresso para apoiar no desenvolvimento;
- O assistente orientou a execução de algumas atividades de forma incorreta o que gerou retrabalho;
- Uma pequena pausa de 10 min após 1 hora de *Workshop*: essa pausa foi sugerida pela mediadora, mas suprimida a pedido dos voluntários, que optaram por uma pausa apenas no final;
- O resultado foi muito satisfatório e todos disseram que gostaram de participar e se disponibilizaram a voltar para o próximo *Workshop*. Seguem algumas frases e comentários que confirma essa impressão:
    - "Foi uma oportunidade de voltarmos no tempo de escola e trabalharmos em grupo: foi divertido";
    - "Ao sermos questionados por nossos valores, tivemos que buscar internamente o que era realmente importante em nossas vidas e isto nos trouxe grandes reflexões";
    - "Foi uma oportunidade de refletirmos sobre à nossa realidade hoje e sobre o nosso futuro";
    - "Foi oportunidade de estar entre amigos de uma forma diferente".
- Apesar de terem gostado da atividade, não ficou claro para os idosos ao final desse *Workshop* a ligação dos resultados obtidos com os próximos passos da pesquisa, mesmo a mediadora tendo explicado no início da apresentação.
    - Reforçar ao final de cada Etapa os resultados obtidos e os próximos passos.

# 5 Etapa 2 - Mapeamento de Soluções IoT - Ideação

Durante a Etapa 2 do método IoT-PMHCS foi realizada a atividade de Ideação de forma participativa com profissionais de tecnologia, conforme descrito no capítulo 0.

As principais atividades dessa etapa estão brevemente apresentadas no Apêndice III – Etapa 2: Mapeamento de Soluções IoT.

As próximas subseções apresentam o registro das principais atividades realizadas, análises e resultados gerados de acordo com a proposta do método IoT-PMHCS.

## 5.1 Relatório do *Workshop* de Ideação

Para a realização desse *Workshop* foi selecionado um grupo de 6 pessoas, com idades entre 19 e 49 anos de tecnologia, conforme descrito no capítulo 0.

Os voluntários moram na cidade de Campinas, SP sendo que 5 trabalham na área de Tecnologia da Informação e um é estudante de Engenharia de Produção e entusiasta de tecnologias inovadoras.

### 5.1.1 Adaptação do Método a Disponibilidade dos Voluntários

O segundo *Workshop* proposto pelo método IoT-PMHCS é o *Workshop* de Personas, com o objetivo de completar as informações das Personas obtidas inicialmente no *Workshop* de Valores e Necessidades, através do uso do Mapa de Empatia e da Jornada do Usuário, tornando o processo de ideação mais rico. Porém, foi avaliado pelos pesquisadores que as informações obtidas no *Workshop* de Valores e Necessidades na Etapa 1 do estudo de caso eram suficientes para o processo de Ideação, e aliado à pouca disponibilidade dos voluntários, optou-se por suprimir do estudo de caso o *Workshop* de Personas.

Dessa forma, todas as informações consideradas como entrada para o *Workshop* de Prototipação estariam completas, faltando apenas a jornada do usuário, que foi elaborada durante o *Workshop* de Prototipação.

Após a execução das atividades conclui-se que suprimir o *Workshop* de Personas não gerou problemas ou gaps para o processo como um todo: portanto recomenda-se que esse seja opcional, e sua necessidade será definida em função de mais informações sobre os *stakeholders* finais do desafio de design, suas limitações e necessidades.

Outra adaptação realizada, foi que os idosos não foram envolvidos nessa etapa junto com os profissionais de tecnologia por uma questão geográfica, uma vez que os idosos vivem em Minas Gerais e os profissionais de TI. Para suprir essa dificuldade e possibilitar a multisciplinaridade no método foi realizado um processo de Co-Criação em 2 grandes etapas: a primeira com os profissionais de TI e a segunda com os idosos. O uso de Personas para a primeira etapa demonstrou-se eficiente e suficiente para a execução dessa atividade, porém os profissionais de TI reportaram que a presença dos idosos nesse *Workshop* teria tornado o processo mais rico.

Importante também ressaltar que todas as lições aprendidas no *Workshop de Valores e Necessidades* foram incorporadas à organização e execução do *Workshop* de Ideação, o que tornou o processo mais eficiente e completo.

### 5.1.2 Organização do *Workshop*

A Tabela 20 descreve a organização do *Workshop* de Ideação, tais como material utilizado, participantes selecionados, duração média, entre outros a proposta do método IoT-PMHCS, contendo uma análise de eficácia do método para cada item, conforme a seguir:

**Tabela 20. *Workshop* de Ideação: recomendações para a organização**

| Organização | | Análise |
|---|---|---|
| **Material** | Artigos lúdicos (*e.g., post-its*, canetas coloridas, cartolina, música), vídeos para contextualização e motivação.<br>Material para adequação do ambiente de modo a facilitar a execução dessa atividade, tais como: *coffee break*, música ambiente e, materiais para tornar o local aconchegante etc. | Todo material foi considerado adequado e suficiente para a execução dessa atividade. |
| **Participan-tes** | Foram convidados 7 participantes, sendo que apenas 6 compareceram. Composição do grupo:<br>-Cientista da Computação: 1 mulheres e 1 homens<br>-Engenheiro da Computação: 1 homem<br>-Analista de Sistemas: 1 homem<br>-Engenheiro de Produção (estudante): 1 homem<br>Todos os voluntários vivem em Campinas, SP e não convivem diretamente com idosos.<br>Eles foram selecionados pela proximidade com a mediadora e por atuarem na área de tecnologia e serem entusiastas de tecnologias inovadoras.<br>Grande parte dos convidados já se conheciam previamente, o que facilitou a sinergia do grupo.<br>Foi convidado também um assistente para apoio as atividades, filho da mediadora. Porém, como um convidado faltou ele se candidatou à voluntário. | O convite foi feito pessoalmente e não houve resistência na aceitação.<br>O trabalho foi realizado sem assistente e não houve impacto na execução das atividades. |
| **Local** | A atividade foi realizada em Campinas, SP no apartamento da mediadora, em ambiente aconchegante e familiar.<br>A reunião inicial foi realizada na sala do apartamento e as atividades em grupo distribuídas no apartamento: 1 grupo ficou na sala, outro na cozinha e outro em um dos quartos. Os grupos eram formados por 2 pessoas cada. | O local foi adequado e todas as atividades foram realizadas com sucesso. Por ser um apartamento, todos os participantes ficaram próximos, o que facilitou à condução das atividades pela mediadora. |
| **Registro** | O registro das atividades foi feito por de forma textual e por fotos.<br>Os cartazes com os resultados foram armazenados para análise. | Os participantes não se sentiram à vontade em serem filmados.<br>Mas os registros realizados foram considerados satisfatórios para essa atividade. |
| **Duração** | A duração recomendada foi de 2 a 3 horas<br>Porém, a duração real da atividade foi de 3 horas, com 2 pausas para café.<br>Ao final foi feita uma confraternização entre os participantes. | Esse tempo foi adequado para os participantes, que não se demonstraram cansados ou desmotivados. |
| **Resultados Esperados** | Como resultado desse *Workshop* foi realizada a priorização de ideias mapeadas de acordo com as categorias assistivas e os critérios norteadores para cada Persona, através da criação da Matriz de Posicionamento. | Foi possível produzir os resultados esperados com a condução desse *Workshop* e completar o Mapa de Persona. |

### 5.1.3 Atividades Propostas no *Workshop*

A Tabela 21 apresenta o planejamento de atividades para esse *Workshop*.

**Tabela 21. *Workshop* de Ideação: planejamento de atividades**

| Objetivo | Atividades Sugeridas | Tempo |
|---|---|---|
| **1-Contextualizar** | - Contextualizar sobre o projeto (pode ser usado um pôster ou slides)<br>- Apresentar os resultados obtidos na fase anterior | 15 min |
| **2-Apresentação dos participantes** | - Pedir para que cada participante se apresente, respondendo às seguintes perguntas:<br>1-Quem eu sou (nome e idade);<br>2-Estilo de Vida (onde vivem, família, atividades);<br>3-Familiaridade com tecnologia. | 15 min |
| **3-Aquecimento** | - Realizar atividade de aquecimento, para a liberação da criatividade dos participantes.<br>- A sugestão é o Exercício da Vaca como sugerido por Osterwalder e Pigneur (2010), onde cada participante deve pensar em modelos inovadores com a vaca. As 3 melhores ideias são priorizadas por votação. | 25 min |
| **4-Vídeo Motivacional** | - Apresentar 2 vídeos motivacionais para estimular a criatividade dos participantes:<br>1- Explicar o conceito de IoT brevemente<br>2- Sugestões de soluções inovadoras com IoT | 5min |
| **Sugestão: pausa de 10min** | | |
| ***5-Workshop* de Brainstorm e Co-Criação** | - Organizar grupos de 3-4 pessoas para *Brainstorm e Co-Criação* com o objetivo de levantar o maior número de ideias possível no que se refere a tecnologias assistivas IoT.<br>- Cada grupo seleciona uma Persona que tenha mais afinidade. Essa sugestão pode ser feita pelo mediador ou selecionada pelo grupo.<br>- Cada grupo seleciona 1 ou 2 categorias de *Home Care* descritas no Diagrama de Afinidade do Etapa 1. Os grupos devem selecionar categoria diferentes.<br>- O grupo discute sobre alternativas de tecnologias assistivas para atender as necessidades da Persona selecionada dentro das categorias selecionadas.<br>- Anotar todos resultados em Post-Its e posicionar numa parede ou lousa debaixo da categoria selecionada. Importante: o grupo deve incluir nesse *Brainstorm* também necessidades pessoais, dentro da categoria selecionada, sendo que nesse caso é importante anotar em um *Post-It* de cor diferente;<br>- A cada 15 minutos, o grupo troca de Persona e continua o p processo de ideação dentro da categoria selecionada. Importante deixar claro que a troca é apenas de Personas, mas as categorias selecionadas pelo grupo continuam. Anotar os resultados em Post-Its e posicionar numa parede, lousa ou cartolina debaixo da categoria selecionada;<br>- Essa troca deve ser feita por 3 vezes.<br>- OBS: cada Persona deve ter a mesma cor de Post-it para as suas ideias, para que seja possível identificar as soluções que atendam cada necessidade. | 45 min |
| **Sugestão: pausa de 10min** | | |
| **6-Análise do Resultado** | - O mediador faz a organização das ideias geradas de acordo com as categorias, de forma visual numa parede ou lousa, com apoio do grupo;<br>- Devem ser eliminadas redundâncias e agrupadas semelhanças;<br>- O grupo atribui uma relevância a cada categoria, sendo a de maior valor a mais importante, e a de menor valor a menos importante;<br>- Gerar a Matriz de Posicionamento, relacionando Personas, ideias numa lousa, parede ou cartolina;<br>- Deve ser realizada a priorização por pesos se possível. | 20 min |
| **7-Apresentação dos resultados** | - O mediador apresenta o resultado ao grupo;<br>- O grupo realiza uma votação das 3 melhores ideias para cada Persona, priorizando as de maior abrangência e relevância. Se necessário priorizar novamente. | 15 min |
| **8-Encerramento** | - Apresentação dos resultados.<br>- Agradecimentos.<br>- Coletar feedback dos participantes. | 10 min |

### 5.1.4 Análise das Atividades Realizadas

A Tabela 22 apresenta uma breve análise das atividades realizadas para se atingir o objetivo desse *Workshop.*

**Tabela 22. *Workshop* de Ideação: análise das atividades realizadas**

| Atividade Planejada | Análise |
|---|---|
| **1-Contextualização** | A contextualização foi realizada com sucesso, com uma apresentação em *Power Point* projetada na TV da sala e todos confortavelmente acomodados. Durante a apresentação foram informados aos participantes os objetivos da pesquisa, as etapas do método, os principais conceitos e resultados esperados.<br>Foi enfatizada a importância da participação de todos, no sentido de ajudar o futuro da humanidade através deste projeto de pesquisa: isso ajudou no engajamento e melhor participação, sem a preocupação de serem julgados durante as atividades. |
| **2-Apresentação dos participantes** | Esta atividade foi muito bem-sucedida. |
| **3-Aquecimento** | Esta atividade foi bem-sucedida e atingiu seu objetivo de liberar a criatividade dos participantes e engajar o time.<br>Foi utilizado o exercício da Vaca, como sugerido por Osterwalder e Pigneur (2010). |
| **4-Vídeo Motivacional** | Foram apresentados 3 vídeos para despertar o interesse e criatividade dos idosos:<br>1-Drone ambulância:<br>https://www.youtube.com/watch?v=QAx6uRWfkSU<br>2- Grenny:<br>https://www.youtube.com/watch?v=BCK-v0cWE3k<br>3- GreatCall<br>https://www.youtube.com/watch?v=1S8ug0Xl1pU<br>4- Alexa:<br>https://www.youtube.com/watch?v=dpQueG8oxtI<br>1 (luzes), 2 (segurança), 5 (luzes e segurança) |
| **5-Divisão do Grupo por Personas** | A organização dos grupos por Personas foi feita pela mediadora e demonstrou-se satisfatória, permitindo uma boa execução dos trabalhos. Segue abaixo a distribuição proposta:<br>**Grupo 1 - Antônio**: 2 homens, de 29 e 49 anos.<br>**Grupo 2 - Maria de Lourdes**: 1 homem, 34 anos, e 1 mulher, 22 anos.<br>**Grupo 3 - Nair:** 2 homens, de 19 e 33 anos.<br>A escolha foi feita em função do nível de maturidade e capacidade de engajamento dos participantes. |
| ***6-Brainstorm* e Co-Criação** | Esta atividade foi bem-sucedida e o grupo trabalhou de forma atenta e participativa.<br>O resultado foi satisfatório e todos os grupos conseguiram chegar ao resultado esperado.<br>Para motivar os participantes, foi oferecido uma caixa de chocolates para o grupo que tivesse o maior número de ideias relevantes ao final do processo.<br>**Importante**: o exercício foi feito pensando na Persona, seus valores e necessidades. Cada grupo analisou por 10 minutos a categoria prioritária de *Home Care* para a Persona selecionada e mais uma categoria que permanecia no grupo mesmo com a mudança da Persona. |
| **7-Análise dos resultados *Brainstorm*** | Esta atividade exigiu mais proximidade e orientação da mediadora, pois os resultados foram analisados com base no Critérios Norteadores (Capítulo 4.4) Foram selecionadas as 10 ideias mais relevantes para serem pontuadas. E depois as 5 mais relevantes foram submetidas à votação de todos. O objetivo era selecionar as 3 ideias mais relevantes por Persona.<br>O resultado foi satisfatório e os objetivos atingidos. |
| **8-Encerramento** | A duração total da atividade foi de 3 horas.<br>O *Workshop* atingiu seus objetivos e com feedback positivo dos participantes.<br>Foi enviada pesquisa de participação aos participantes. |

### 5.1.5 Aquecimento

O objetivo da atividade 3 de Aquecimento foi ativar a capacidade criativa dos participantes. Para sua execução foi usado uma adaptação de exercício proposto por Osterwalder e Pigneur (2010), chamado "Exercício da Vaca", onde os participantes são convidados a identificar ideias inovadoras usando uma vaca.

A duração do exercício foi de 20 minutos e composta dos seguintes passos:

1. Definir as características da vaca
2. Pensar em modelos de negócio para a vaca
3. Escrever uma ideia/ post-its
4. Apresentar resultados e selecionar por votação as 3 melhores ideias

Num primeiro momento o grupo ficou inseguro com o exercício. Mas o resultado foi muito positivo, divertido, gerou engajamento, ativando a criatividade de todos (Figura 24).

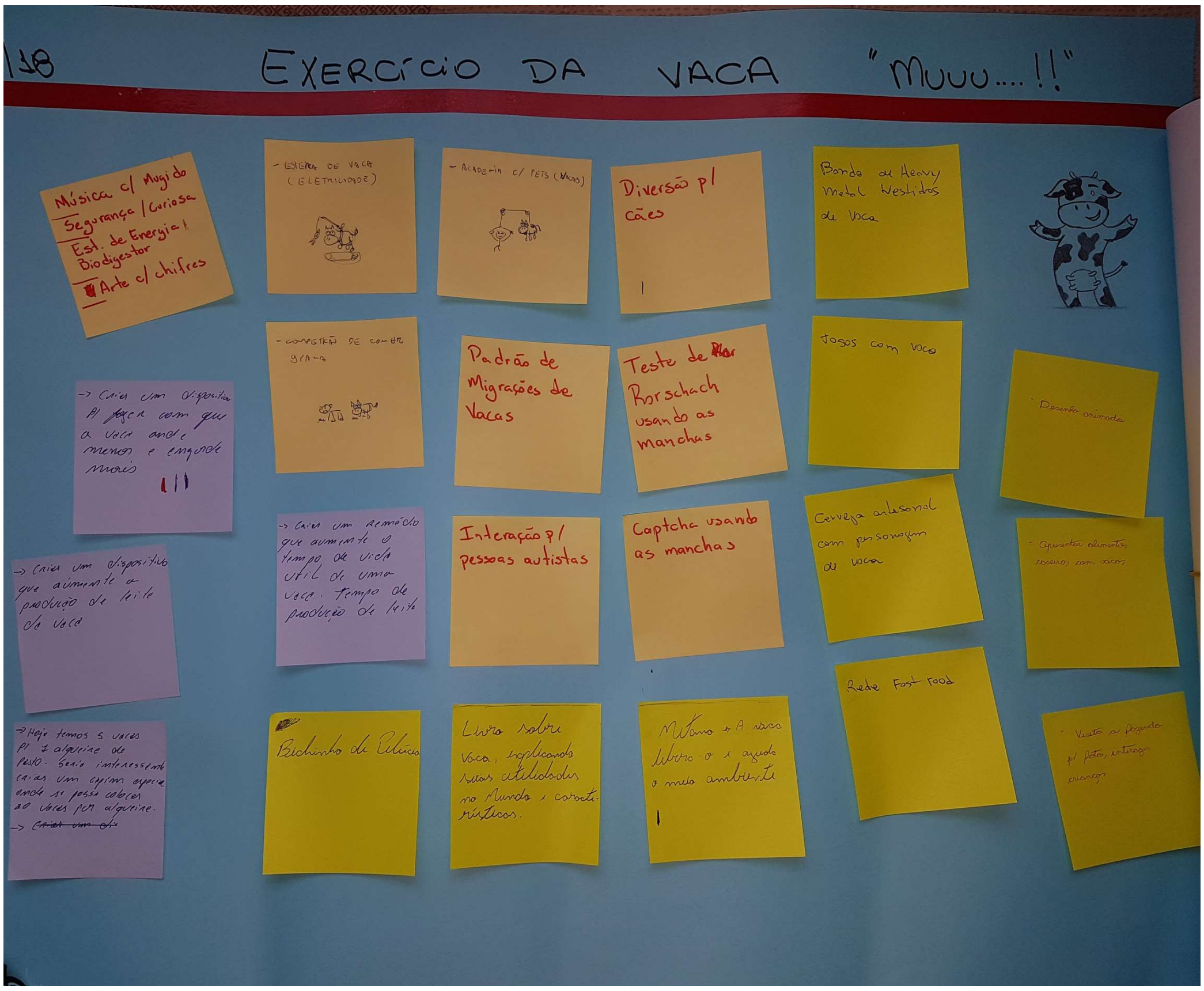

**Figura 24.** ***Workshop*** **Ideação: Aquecimento - Exercício da Vaca**

### 5.1.6 Brainstorm e Co-Criação

Para realizar a atividade 6 de Brainstorm e Co-Criação foram oferecidas as seguintes orientações aos participantes:

1. Cada equipe seleciona 2 Categorias de *Home Care*: uma deve ser prioritária para Persona selecionada e outra é sorteada;
2. O grupo discute por 10 minutos alternativa de tecnologias assistiva baseada em IoT para atender as necessidades da Persona dentro das categorias priorizadas. O objetivo é gerar o maior número de ideias nessas categorias. Mas o grupo não é impedido de idear em outras categorias: nesse caso sugere-se apenas identificar com uma cor de *Post-It* diferente (ROSA);
    a. Anotar uma ideia por *Post-It* e colocar na cartolina abaixo da categoria
3. Trocar a Persona e continuar o processo de ideação para a nova Persona, com as novas categorias;
4. Realizar a troca por 3 vezes, de tal forma que todos os grupos avaliem todas as

Personas e todas as categorias. O objetivo é não repetir ideias;

a. A troca é apenas de Persona. A categoria sorteada permanece no grupo

A seguir os resultados gerados por essa atividade organizado por Personas e categorias de *Home Care*

**Antônio**

Foram geradas as seguintes ideias para a Persona Antônio durante o *Workshop* de Co-Criação (Figura 25).

**Saúde**

- Dispositivo (relógio) para monitorar saúde (colesterol, pressão, batimentos) com alarmes para familiares em caso de emergência.
- Dispositivo que mede o grau de colesterol e ajuda no ajuste da dosagem e lembra a pessoa caso esqueça
- Sensores para boa postura com sistema de notificação.
- Identificação da qualidade da postura para prevenir a piora da hérnia de disco.
- Luzes que abaixam para melhorar o sono
- Dispositivo para diminuir o ritmo de agitação antes de dormir
- Reconhecimento facial para detecção de problemas de saúde
- Travesseiro com sensor de glicose no calor

**Mobilidade**

- Cadeado inteligente para abrir portas por proximidade, aviso para família quando chega em casa
- Luzes inteligentes nas bordas das portas para facilitar locomoção
- Sensores de queda na casa e sensor de tropeços
- Luzes inteligentes que ascendem quando acorda, com led automático em escadas, corredores e degraus.

**Comunicação**

- Não analisado

**Segurança**

- Sistema de segurança baseado em reconhecimento facial para pessoas que entrarem em casa fora do horário permitido

**Facilitar Dia a Dia**

- Dispositivo para identificar objetos e apoiar no esquecimento. Com GPS e luzes para localização.

**Figura 25. *Workshop* Ideação: sugestões para Persona Antônio por categoria**

**Maria de Lourdes**

Foram geradas as seguintes ideias para a Persona Maria de Lourdes durante o *Workshop* de Co-Criação (Figura 26).

**Saúde**

- Balança online, com aviso de peso acima do esperado e cálculo de massa corporal
- Balança de comida online, com acompanhamento familiar. Reporte semanal para nutricionista.
- Pulseira de avaliação alimentar com lembretes (água, comida, etc) quando ficar muito tempo sem ingerir alimentos.
- Caixinha de remédios online, com aviso de remédios faltantes e compra automática via Internet.
- Dispositivo que detecta humor e sugere distrações ou remédio certo para evitar depressão. Avisa a família em situações de perigo.

**Comunicação**

- Aviso inteligente caso ocorra algum acidente na casa ou com a idosa.

**Segurança**

- Detector de quedas que avisa a família.
- Detector de presença que avise familiares que a pessoa chegou bem ou não em casa.
- Dispositivo que tranque a casa automaticamente caso não detecte a presença de ninguém a aciona alarme.

**Facilitar Dia a Dia**

- Sistema inteligente para a cozinha com sensores e temporizadores.
- Sistema de reconhecimento de voz para comando de tarefas em geral.
- Botão de emergência em aplicativo no celular ou portátil.

**Mobilidade**

- Robô-Drone para auxílio de quedas com câmera para a família, suporte a movimentação e integração com telefone.

**Figura 26. *Workshop* Ideação: sugestões para Persona Maria de Lourdes por categoria**

## Nair

Foram geradas as seguintes ideias para a Persona Nair durante o *Workshop* de Co-Criação (Figura 27).

**Comunicação**

- Sistema de Comunicação com à família com sensores e alarmes de perigo.
- Dispositivo que controla sua alimentação, lembrando de alimentos importantes e medicamentos, com alarmes para cuidadores.
- Botão que faz ligação automática para pessoas cadastradas e integra com Smart Home para identificar situações de perigo.
- Kit com câmera e display (com quadro) que fique sempre disponível na sala e na cozinha do idosa e na casa dos filhos. Dispara alarmes em caso de emergência.

**Saúde**

- Dispositivo para monitoramento de pressão, batimentos e colesterol, com alarmes em caso de alterações bruscas.
- Dispositivo que lembre de tomar remédios nas horas e dosagem correta, com alarmes para cuidadores.

**Facilitar Dia a Dia**

- Geladeira inteligente para compras automáticas com post no Facebook.
- Cachorro-robô com lembretes de voz e comunicação via WhatsApp para familiares.
- Robô para limpeza.
- Televisão inteligente com detecção de face. Coloca o programa preferido para que a idosa fique feliz.

**Mobilidade**

- Cadeira de rodas automática com sensores e GPS.
- Corrimão inteligente que sai da parede com detecção de proximidade.
- Muleta inteligente que acende luzes no ambiente atual, com alarme de queda à cuidadores.

**Segurança**

- Dispositivo que faz barulho de cachorro e emite som de sirene da polícia caso sua casa seja invadida, com alarme para família e polícia.
- Alerta de invasão no celular para pessoas com problemas auditivos.

**Figura 27. Workshop Ideação: sugestões para Persona Nair por categoria**

### 5.1.7 Análise dos Resultados *Brainstorm*

Para realizar a atividade 7 de Análise dos resultados do *brainstorm* foram oferecidas as seguintes orientações:

1. Cada grupo avalia as ideias apresentadas para sua Persona:
    a. Eliminar redundâncias;
    b. Eliminar soluções que não sejam IoT;
    c. Agrupadas semelhanças;
2. Por votação o grupo seleciona as 10 ideias mais interessantes;
3. Classificar essas 10 ideias de acordo com seguintes Critérios Norteadores definidos na Tabela 19;
4. Os grupos selecionam as 5 ideias mais relevantes, ou seja, de maior peso, para cada Persona;
5. Todos os participantes selecionam por votação as 3 ideias mais relevantes para cada Persona.

A seguir os resultados gerados por essa atividade organizado por Personas e categorias de *Home Care*.

**Antônio**

Foram selecionadas pelo grupo representante da Persona Antônio as 10 ideias mais interessantes para serem avaliadas de acordo com os critérios norteadores. Essas ideias estão destacadas abaixo em vermelho (Figura 28).

**Saúde**

1. Dispositivo (relógio) para monitorar saúde (colesterol, pressão, batimentos) com alarmes para familiares em caso de emergência.
- ~~Dispositivo que mede o grau de colesterol e ajuda no ajuste da dosagem e lembra a pessoa caso esqueça~~
2. Sensores para boa postura com sistema de notificação.
- ~~Identificação da qualidade da postura para prevenir a piora da hérnia de disco.~~
3. Luzes que abaixam para melhorar o sono
4. Dispositivo para diminuir o ritmo de agitação antes de dormir
5. Reconhecimento facial para detecção de problemas de saúde
- ~~Travesseiro com sensor de glicose no calor~~

**Mobilidade**

6. Cadeado inteligente para abrir portas por proximidade, aviso para família quando chega em casa
7. Luzes inteligentes nas bordas das portas para facilitar locomoção
8. Sensores de queda na casa e sensor de tropeços
- ~~Luzes inteligentes que ascendem quando acorda, com led automático em escadas, corredores e degraus.~~

**Comunicação**

- Não analisado

**Segurança**

9. Sistema de segurança baseado em reconhecimento facial para pessoas que entrarem em casa fora do horário permitido

**Facilitar Dia a Dia**

10. Dispositivo para identificar objetos e apoiar no esquecimento. Com GPS e luzes para localização.

**Figura 28. *Workshop* Ideação: ideias preliminares selecionadas para a Persona Antônio**

Para cada ideia selecionada acima foram analisados os critérios norteadores e aplicados os respectivos pesos. Dessa forma foi criada uma Matriz de Posicionamento da Persona (Figura 29).

| Matriz de Posicionamento - Antônio | | Ideias | | | | | | | | | |
|---|---|---|---|---|---|---|---|---|---|---|---|
| **Matriz de Posicionamento** | **Peso** | **1** | **2** | **3** | **4** | **5** | **6** | **7** | **8** | **9** | **10** |
| **Atende ao valor da Persona?** | 1 a 4 | 4 | 4 | 4 | 4 | 4 | 0 | 4 | 4 | 0 | 4 |
| **Categoria de Home Care prioritária?** | 1 a 5 | 5 | 5 | 5 | 5 | 5 | 4 | 4 | 4 | 2 | 1 |
| **Facilidade de Uso pelo idoso ?** | 1 a 2 | 2 | 2 | 2 | 2 | 2 | 2 | 2 | 2 | 2 | 1 |
| **Viabilidade Técnica** | 0/ 1 | 0 | 1 | 1 | 1 | 0 | 1 | 1 | 1 | 1 | 1 |
| Total | **0 a 12** | **11** | **12** | **12** | **12** | **11** | **7** | **11** | **11** | **5** | **7** |

**Figura 29. *Workshop* Ideação: Matriz de Posicionamento Persona Antônio**

Foi solicitado ao grupo de trabalho selecionar as 5 ideias mais relevantes (maiores pesos) para votação final. Foram selecionadas as ideias 2, 3, 4, 7 e 8.

Ao serem submetidas para votação final de todos os voluntários, as ideias selecionadas como mais relevantes para a Persona Antônio foram:

- [3]: Luzes que abaixam para melhorar o sono, identificam automaticamente o horário do dia e usam à cromoterapia para reduzir stress
- [4]: Dispositivo para diminuir o ritmo de agitação antes de dormir
- [8]: Sensores de queda na casa e de tropeços

Tais ideias serão apresentadas aos idosos para escolha da mais relevante para à Persona Antônio no próximo *Workshop*.

**Maria de Lourdes**

Foram selecionadas pelo grupo representante da Persona Maria de Lourdes as 10 ideias mais interessantes para serem avaliadas de acordo com os critérios norteadores. Essas ideias estão destacadas abaixo em vermelho (Figura 30).

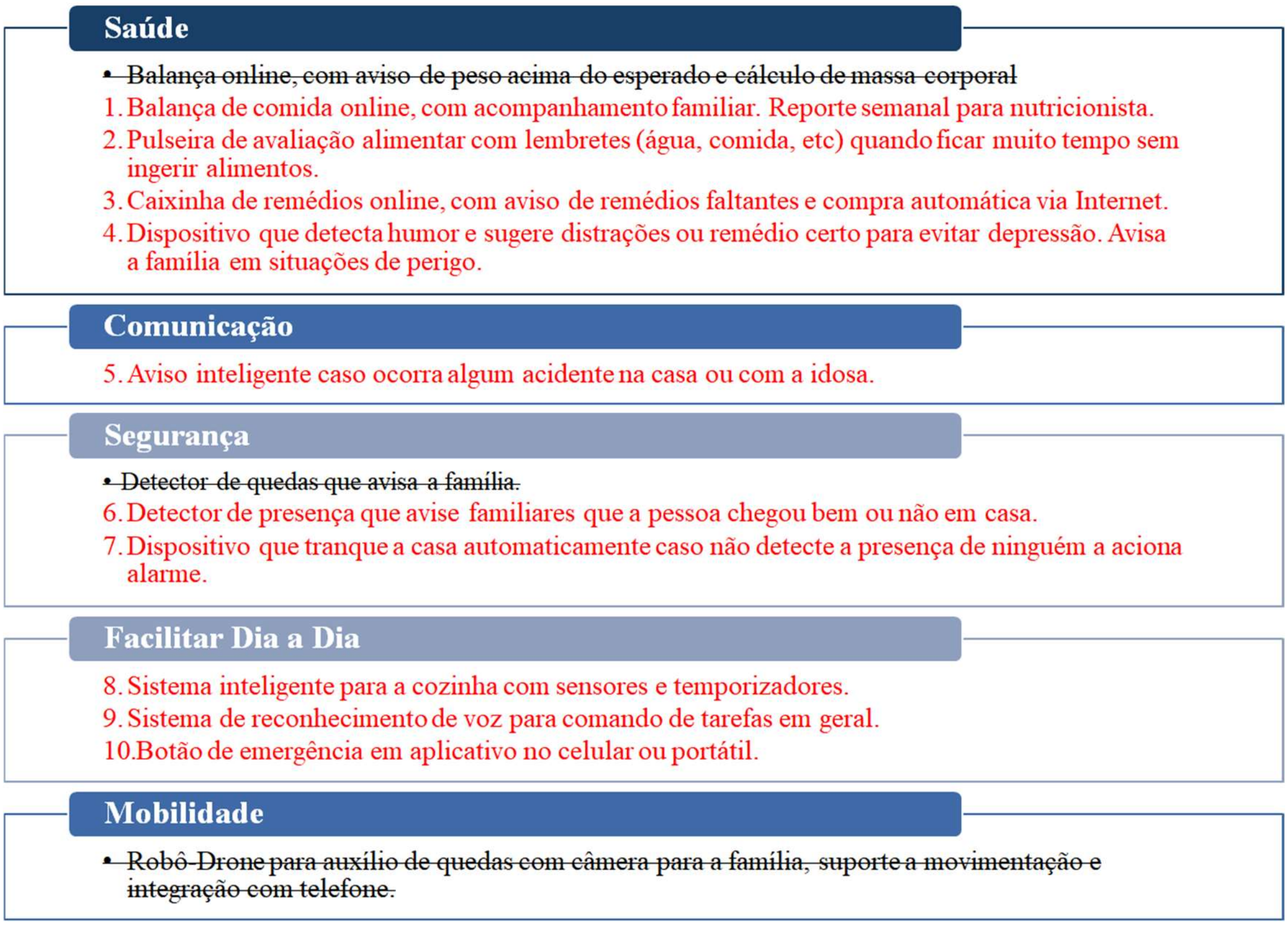

**Figura 30. *Workshop* Ideação: ideias preliminares selecionadas para Persona Maria de Lourdes**

Para cada ideia selecionada acima foram analisados os critérios norteadores e aplicados os respectivos pesos. Dessa forma foi criada uma Matriz de Posicionamento da Persona (Figura 31).

| Matriz de Posicionamento - Maria de Lourdes | | Ideias | | | | | | | | | |
|---|---|---|---|---|---|---|---|---|---|---|---|
| Matriz de Posicionamento | Peso | 1 | 2 | 3 | 4 | 5 | 6 | 7 | 8 | 9 | 10 |
| Atende ao valor da Persona? | 1 a 4 | 4 | 4 | 4 | 0 | 2 | 0 | 0 | 4 | 4 | 4 |
| Categoria de Home Care prioritária? | 1 a 5 | 5 | 5 | 5 | 5 | 4 | 3 | 3 | 2 | 2 | 2 |
| Facilidade de Uso pelo idoso ? | 1 a 2 | 1 | 2 | 2 | 2 | 2 | 2 | 2 | 2 | 0 | 2 |
| Viabilidade Técnica | 0/ 1 | 1 | 1 | 1 | 0 | 1 | 1 | 1 | 1 | 0 | 1 |
| Total | 0 a 12 | 11 | 12 | 12 | 7 | 9 | 6 | 6 | 9 | 6 | 9 |

**Figura 31.** ***Workshop*** **Ideação: Matriz de Posicionamento Persona Maria de Lourdes**

Foi solicitado ao grupo de trabalho selecionar as 5 ideias mais relevantes (maiores pesos) para votação final. Foram selecionadas as ideias 1,2,3,5,8.

Ao serem submetidas para votação final de todos os voluntários, as ideias selecionadas como mais relevantes para a Persona Maria de Lourdes foram:

- [3]: Caixinha de remédios online, com aviso de remédios faltantes e compra automática via Internet.
- [5]: Aviso inteligente caso ocorra algum acidente na casa ou com a idosa
- [8]: Sistema inteligente para a cozinha com sensores e temporizadores.

**Nair**

Foram selecionadas pelo grupo representante da Persona Nair as 10 ideias mais interessantes para serem avaliadas de acordo com os critérios norteadores. Essas ideias estão destacadas abaixo em vermelho (Figura 32).

**Comunicação**

1. Sistema de Comunicação com à família com sensores e alarmes de perigo.
2. Dispositivo que controla sua alimentação, lembrando de alimentos importantes e medicamentos, com alarmes para cuidadores.
3. Botão que faz ligação automática para pessoas cadastradas e integra com Smart Home para identificar situações de perigo.
4. Kit com câmera e display (com quadro) que fique sempre disponível na sala e na cozinha do idosa e na casa dos filhos. Dispara alarmes em caso de emergência.

**Saúde**

5. Dispositivo para monitoramento de pressão, batimentos e colesterol, com alarmes em caso de alterações bruscas.
6. Dispositivo que lembre de tomar remédios nas horas e dosagem correta, com alarmes para cuidadores.

**Facilitar Dia a Dia**

- ~~Geladeira inteligente para compras automáticas com post no Facebook.~~
- ~~Cachorro-robô com lembretes de voz e comunicação via WhatApp para familiares em caso de emergência.~~

7. Robô para limpeza.
8. Televisão inteligente com detecção de face. Coloca o programa preferido para que a idosa fique feliz.

**Mobilidade**

- ~~Cadeira de rodas automática com sensores e GPS.~~
- ~~Corrimão inteligente que sai da parede com detecção de proximidade.~~

9. Muleta inteligente que acende luzes no ambiente atual, com alarme de queda à cuidadores.

**Segurança**

10. Dispositivo que faz barulho de cachorro e emite som de sirene da polícia caso sua casa seja invadida, com alarme para família e polícia.

- ~~Alerta de invasão no celular para pessoas com problemas auditivos~~

**Figura 32. *Workshop* Ideação: ideias preliminares selecionadas para Persona Nair**

Para cada umas das 10 ideias selecionadas foram analisados os critérios norteadores e aplicados os respectivos pesos. Dessa forma foi criada uma Matriz de Posicionamento da Persona (Figura 33).

| Matriz de Posicionamento - Nair | | Ideias | | | | | | | | | |
|---|---|---|---|---|---|---|---|---|---|---|---|
| **Critérios Norteadores** | **Peso** | **1** | **2** | **3** | **4** | **5** | **6** | **7** | **8** | **9** | **10** |
| **Atende ao valor da Persona?** | 1 a 4 | 4 | 2 | 4 | 4 | 3 | 3 | 3 | 2 | 3 | 2 |
| **Categoria de Home Care prioritária?** | 1 a 5 | 5 | 5 | 5 | 5 | 4 | 4 | 3 | 3 | 2 | 1 |
| **Facilidade de Uso pelo idoso ?** | 1 a 2 | 2 | 2 | 2 | 1 | 2 | 2 | 2 | 2 | 2 | 2 |
| **Viabilidade Técnica** | 0/ 1 | 1 | 1 | 1 | 1 | 1 | 1 | 1 | 0 | 1 | 1 |
| **Total** | **0 a 12** | **12** | **10** | **12** | **11** | **10** | **10** | **9** | **7** | **8** | **6** |

**Figura 33. *Workshop* Ideação: Matriz de Posicionamento Persona Nair**

Foi solicitado ao grupo selecionar as 5 ideias mais relevantes, ou seja, as de maiores pesos, para votação final. Foram selecionadas as ideias 1, 2, 3, 4, 5.

Ao serem submetidas para votação final de todos os voluntários, as ideias selecionadas como mais relevantes para a Persona Nair foram:

- [2]: Dispositivo que controla sua alimentação, lembrando de alimentos importantes e medicamentos, com alarmes para cuidadores;
- [4]: Kit com câmera e display (com quadro) que fique sempre disponível na sala e na cozinha do idosa e na casa dos filhos. Dispara alarmes em caso de emergência;
- [5]: Dispositivo para monitoramento de pressão, batimentos e colesterol, com alarmes em caso de alterações bruscas.

Tais ideias serão apresentadas aos idosos para escolha da mais relevante para a Persona Nair no próximo *Workshop*.

## 5.2 Resultado Consolidado da Ideação

Resultado consolidado da Ideação durante o *Workshop (*Tabela 23).

**Tabela 23. *Workshop* de Ideação: resultado por Persona**

| | |
|---|---|
| **Nair** | [2]:Dispositivo que controla sua alimentação, lembrando de alimentos importantes e medicamentos, com alarmes para cuidadores;<br>[4]: Kit com câmera e display (com quadro) que fique sempre disponível na sala e na cozinha do idosa e na casa dos filhos. Dispara alarmes em caso de emergência;<br>[5]: Dispositivo para monitoramento de pressão, batimentos e colesterol, com alarmes em caso de alterações bruscas. |
| **Antônio** | [3]: Luzes que abaixam para melhorar o sono, identificam automaticamente o horário do dia e usam à cromoterapia para reduzir stress.<br>[4]: Dispositivo para diminuir o ritmo de agitação antes de dormir.<br>[8]: Sensores de queda na casa e de tropeços. |
| **Maria de Lourdes** | [3]: Caixinha de remédios online, com aviso de remédios faltantes e compra automática via Internet.<br>[5]: Aviso inteligente caso ocorra algum acidente na casa ou com a idosa<br>[8]: Sistema inteligente para a cozinha com sensores e temporizadores. |

Existe uma semelhança entre as ideias propostas para cada um dos idosos, agrupados acima por cores: ou seja, as cores iguais representam ideias semelhantes. Isso fornece uma forte indicação de que soluções nessa linha podem ser aceitas por mais de um grupo de Personas. No caso de uma solução personalizada por perfil de usuário, essas opções poderiam ser comuns a todos os perfis.

## 5.3 Mapa de Persona - Ideação

A partir dos resultados gerados por essa etapa é possível atualizar a versão do Mapa de Persona para cada Persona selecionadas para esse estudo de caso.

### 5.3.1 Antônio

A Figura 34 apresenta o Mapa da Persona Antônio agora atualizado a partir dos resultados gerados por esse *Workshop*.

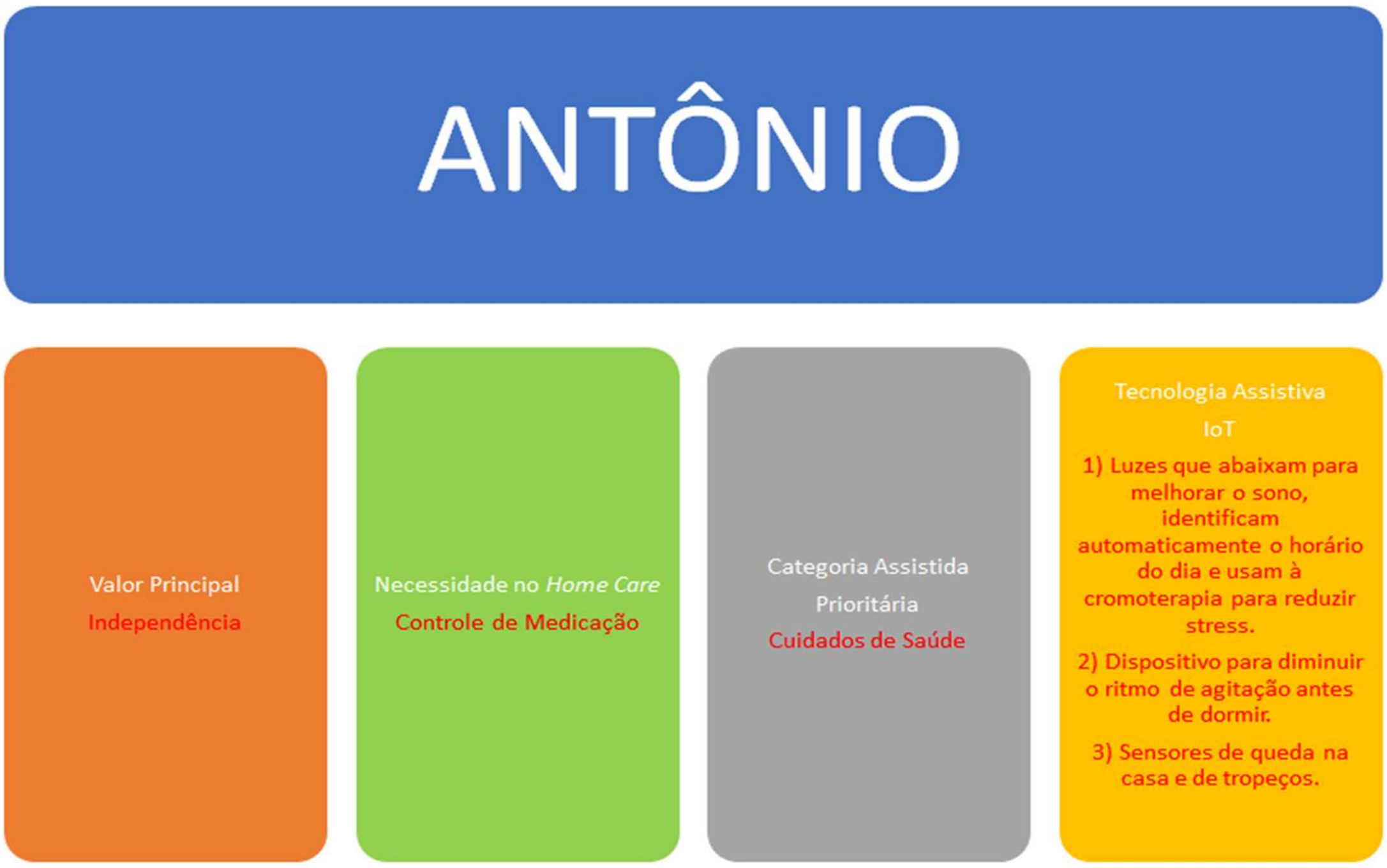

**Figura 34. Etapa 1 - Mapa de Persona atualizado com soluções IoT para Persona Antônio**

### 5.3.2 Nair

A Figura 35 apresenta o Mapa da Persona Nair agora atualizado a partir dos resultados gerados por esse *Workshop*.

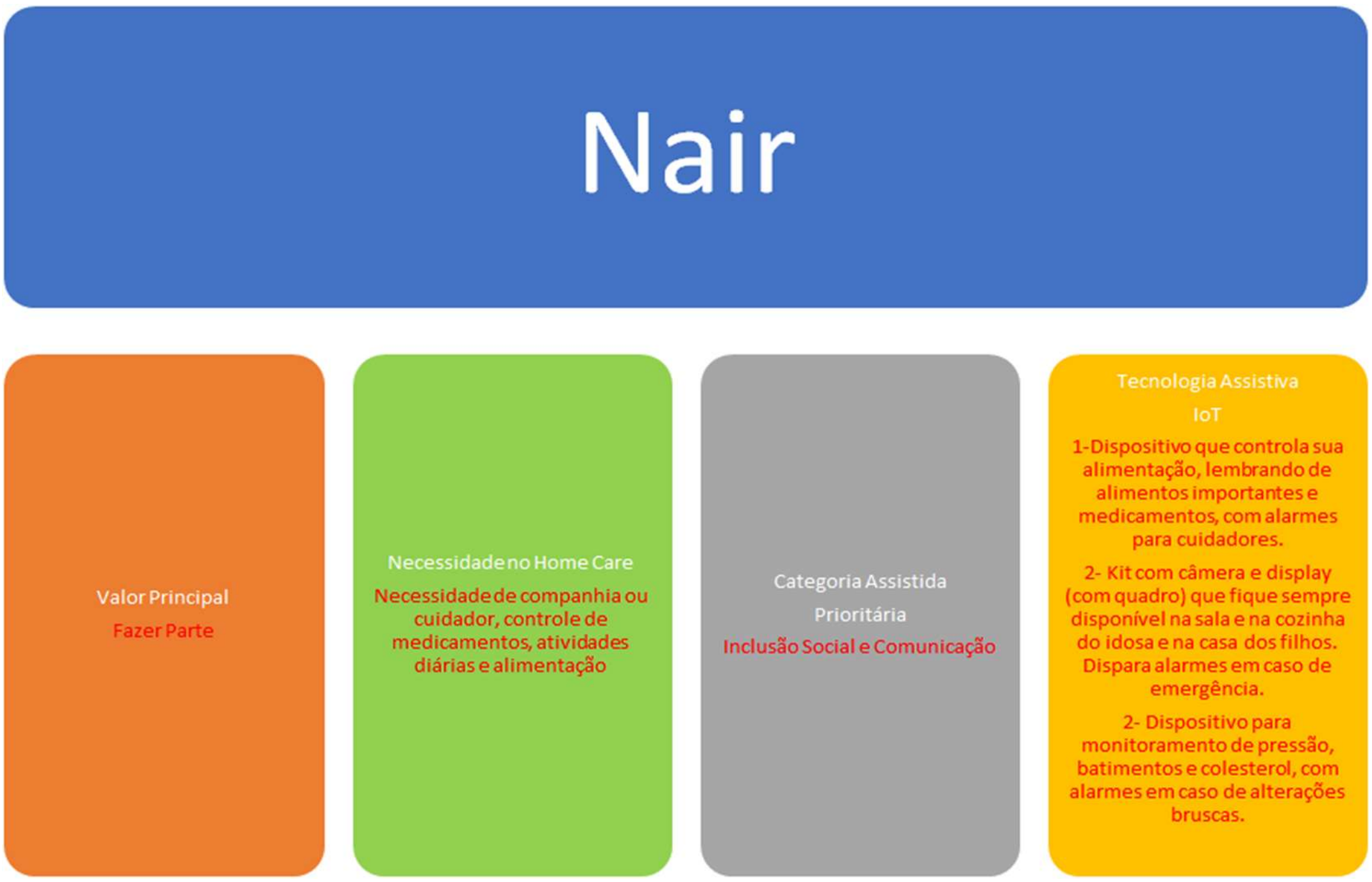

**Figura 35. Etapa 1 - Mapa de Persona atualizado com soluções IoT para Persona Nair**

### 5.3.3 Maria de Lourdes

A Figura 36 apresenta o Mapa da Persona Maria de Lourdes agora atualizado a partir dos resultados gerados por esse *Workshop*.

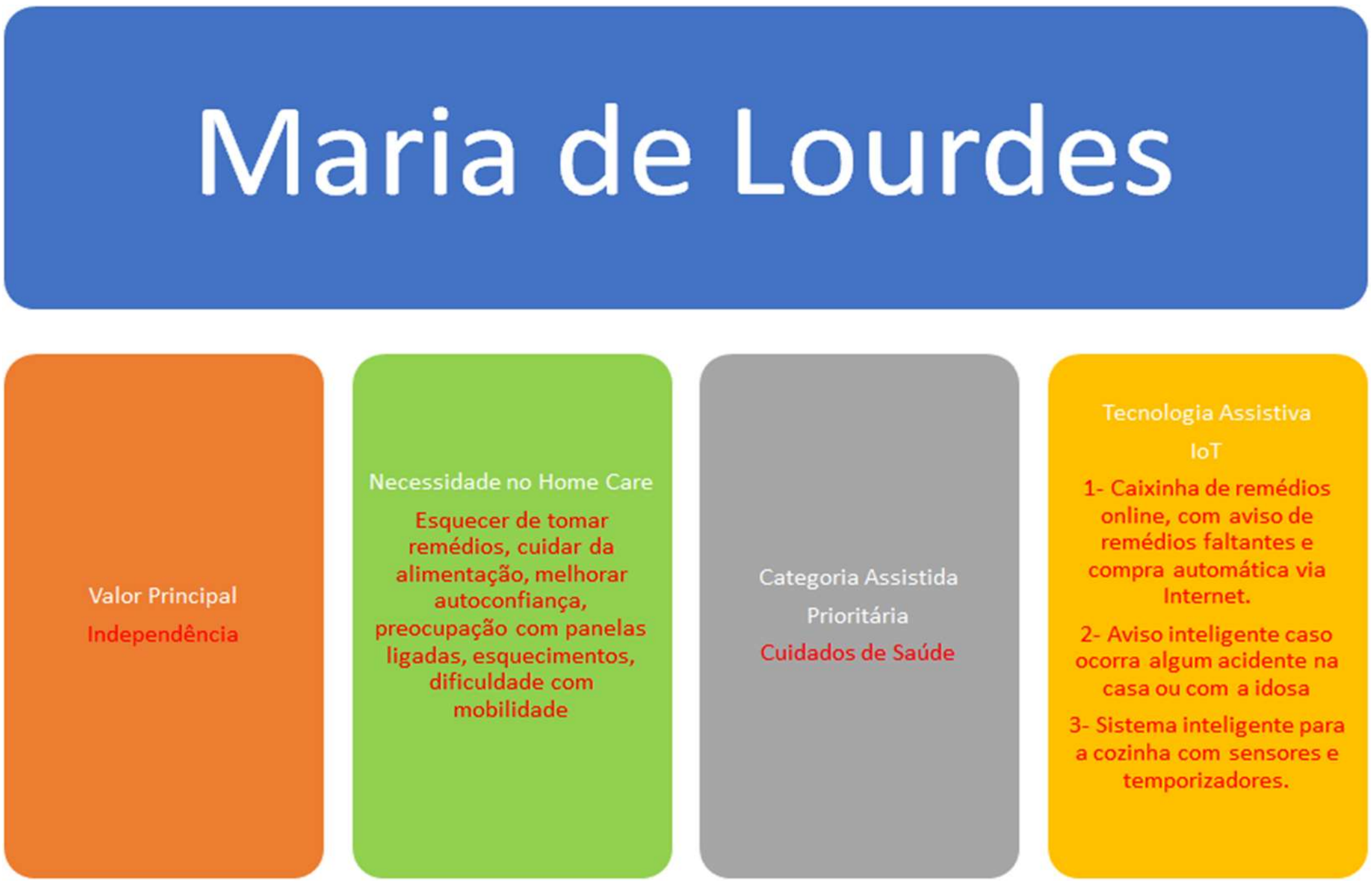

**Figura 36. Mapa de Persona atualizado com soluções IoT para Persona Maria de Lourdes**

## 5.4 Pesquisa de Participação

Os seguintes *feedbacks* foram coletados a após o encerramento das atividades:

- O resultado foi muito satisfatório e todos disseram que gostaram de participar e se disponibilizaram a voltar para o próximo *Workshop*. Seguem algumas frases e comentários que confirma essa impressão:
    - "Eu nunca havia participado de uma sessão de *Brainstorm* tão estruturada como essa.";
    - "Precisamos implantar esse processo de *Brainstorm* no nosso trabalho: os resultados serão muito melhores e as pessoas se sentirão mais motivadas a participar";
    - "À atividade de aquecimento (Exercício da Vaca) foi fundamental: sem ela ficaríamos muito travados durante a Ideação)".
- Os participantes não se demonstraram cansados ou desmotivados em nenhum momento;
- Sobre o local, foi sugerido manter o mesmo para as próximas atividades.

Além dessas impressões gerais, foi realizada uma coleta formal de feedback dos participantes através de um questionário *online.*

O resultado da pesquisa demonstra que os participantes ficaram satisfeitos com o processo e com a condução das dinâmicas. Porém, 67% dos participantes (ou seja, 4 em 6), acreditam que o processo teria sido mais rico se os idosos participassem do processo.

Como resultado dessa análise e solução de contorno, foi incluído um exercício no *Workshop* de Prototipação para que os idosos possam realizar um trabalho de Co-Criação a partir das ideias propostas pelo *Workshop* de Ideação, de tal forma que eles possam gerar novas ideias, explorar ideias já sugeridas e se sentirem participantes efetivos do processo de criação.

# 6 Etapa 2 - Mapeamento de Soluções IoT – Prototipação

Durante a Etapa 2 do método IoT-PMHCS foi realizada a atividade de Prototipação de forma participativa com os idosos, conforme descrito no capítulo 0.

As principais atividades dessa etapa estão brevemente apresentadas no Apêndice III – Etapa 2: Mapeamento de Soluções IoT.

Segue o registro das principais atividades realizadas, análises e resultados gerados de acordo com a proposta do método IoT-PMHCS.

## 6.1 Relatório *Workshop* de Prototipação

Para a realização deste primeiro *Workshop* foi selecionado um grupo de 9 idosos entre 60 e 80 anos da cidade de Muzambinho, MG, conforme descrito no capítulo 0.

### 6.1.1 Organização do *Workshop*

A Tabela 24 descreve a organização do *Workshop* de Prototipação, material utilizado, participantes selecionados, duração média, entre outros, de acordo com a proposta do método IoT-PMHCS, contendo uma análise de eficácia do método para cada item, conforme a seguir.

**Tabela 24. *Workshop* de Prototipação: análise de organização das atividades**

| Proposta de Organização | | Análise |
|---|---|---|
| **Material** | -Artigos lúdicos (*e.g., post-its*, canetas coloridas, cartolina, música), vídeos para contextualização e motivação;<br>-Material para adequação do ambiente de modo a facilitar a execução desta atividade, tais como: *coffee break*, música ambiente e, materiais para tornar o local aconchegante etc;<br>-Recomenda-se que todo material a ser utilizado como consulta pelos grupos para à realização das atividades seja impresso. | Todo material foi considerado adequado e suficiente para a execução desta atividade. |
| **Participantes** | Mandatório:<br>- *stakeholders* com envolvimento direto no sistema em foco, nesse caso, os idosos;<br>- a presença de um assistente para suporte e registro das atividades do evento;<br>Opcional: se possível, uma diversidade maior de *stakeholders*, como por exemplo, familiares, profissionais de saúde e profissionais de TI;<br>OBS: é recomendado nesta fase os mesmos | -Foram envolvidos os mesmos idosos da atividade anterior.<br>-Não houve assistente para esta atividade<br>-Como plano de contorno explicar as atividades antecipadamente para 2 voluntários, que foram pontos de apoio. |
| | participantes da fase anterior. | |
| **Local** | Recomenda-se um local apropriado para a execução desta atividade, contendo mesas, cadeiras e lousa para a realização desta atividade.<br>Importante existir uma TV ou projetor para a apresentação de slides e vídeos.<br>O ideal seria uma sala de reunião ou sala de aula, com espaço adequado para a atividade. | Foi usado o mesmo local do *Workshop* de Valores, mas agora com uma disposição melhor para a condução das atividades. |
| **Registro** | -Registrar os principais resultados de forma textual;<br>-Registro com fotos para análise posterior;<br>- É importante nesse *workshop* filmar o resultado da atividade de *storytelling* para análise posterior. | Os participantes não se sentiram à vontade em serem filmados.<br>Mas os registros realizados foram realizados utilizando fotos, anotações da moderadora e pesquisa final de participação, sendo considerado satisfatórios para esta atividade.<br>O *storytelling* final com um participante foi filmado. |
| **Duração** | Total de 2 a 3 horas, com pausas para café. | -O tempo total de 4 horas, sendo 2 horas em um dia 2 horas no outro.<br>-O tempo não foi suficiente para a realização de todas as atividades previstas no mesmo dia.<br>-Porém, as atividades foram realizadas dentro dos limites dos voluntários que não se sentiram cansados desta vez. |
| **Resultados Esperados** | Como resultado desse *Workshop* espera-se a prototipação das ideias mais relevantes e convergência do grupo da melhor alternativa de tecnologia inovadora a ser prototipadas pelo time de tecnologia. | Foi possível produzir os resultados esperados para apenas uma Persona, que foi seleciona para a continuidade das atividades do projeto. |

### 6.1.2 Atividades Propostas no *Workshop*

A Tabela 25 apresenta o planejamento de atividades para esse *Workshop*.

**Tabela 25. *Workshop* de Prototipação: planejamento de atividades**

| Objetivo | Atividades Sugeridas | Tempo |
|---|---|---|
| **1-Contextualizar** | -Contextualizar sobre o projeto (pode ser usado um pôster ou slides);<br>-Apresentar o conceito de IoT;<br>-Apresentar os resultados obtidos na fase anterior. | 15 min |
| **2-Apresentação dos participantes** | -Pedir para que cada participante se apresente, respondendo às seguintes perguntas:<br>1-Quem eu sou (nome e idade);<br>2-Estilo de Vida (onde vivem, família, atividades);<br>3-Familiaridade com TI.<br>(Caso os participantes sejam os mesmos, essa fase não precisa | 10 min |
| | ser executada) | |
| **3-Preparação do *Workshop*** | -Apresentar a proposta do *Workshop* e as principais ferramentas a serem utilizadas;<br>-Explicar sobre o material a ser utilizado, a organização dos grupos, os conceitos de *Sprint e* Jornada do Usuário;<br>-É recomendado sempre que possível a utilização de vídeos explicativos. | 10min |
| **4-Grupos de Trabalho** | -Organizar os participantes em grupos trabalho;<br>-Cada grupo deve selecionar uma Persona diferente, podendo essa escolha feita pelo mediador ou pelo grupo, por afinidade com a Persona selecionada;<br>-Se o grupo for o mesmo, pode ser mantido a mesma organização anterior; | NA |
| **5-Análise de Normas** | - Serão apresentados aos participantes um conjunto de Normas Cognitivas, Perceptuais e Avaliativas que irão apoiar na escolha da ideia mais relevante para cada Persona.<br>- Pode ser elaborado um quadro de critérios de escolha, com perguntas e respostas que facilitem a análise e escolha da melhor alternativa. | 30 min |
| **6-Escolha da Ideia prioritária para Prototipação** | -Uma vez avaliados os critérios de escolha, serão apresentados a cada grupo as ideias mais relevantes identificadas nas Etapas anteriores, incluindo as 3 ideias priorizadas durante o *Workshop* de Ideação;<br>-Cada grupo deve selecionar a mais relevante para a sua Persona, levando em consideração facilidade de uso, aderência aos valores, necessidades da Persona e os critérios de escolha selecionados;<br>-O grupo tem a liberdade de criar novas ideias com relação as já identificadas ou expandir os conceitos;<br>-Caso existam divergências entre a ideia prioritária para a Persona e para o grupo isso deve ser apontado, prevalecendo a escolha da ideia para a Persona. | 20min |
| **7-Jornada do Usuário*** | -Cada grupo deve selecionar uma situação de perigo dentro de casa para a Persona. Essa situação por ser sugerida pelo mediador;<br>-Neste momento a Jornada deve ser considerada sem o uso de tecnologia;<br>-O grupo constrói a Jornada do Usuário para a Persona selecionada, tendo como foco uma situação de perigo, e incluindo os seus próprios desafios e percepções.<br>- Essa jornada pode ser construída em papel, com desenhos ou linha do tempo;<br>- Recomenda-se que o moderador apresente um modelo para facilitar a condução das atividades pelo grupo;<br>* Caso o grupo tenha construído a Jornada em etapa anterior, essa pode ser usada como referência, tornando essa atividade opcional;<br>* O moderador pode substituir a elaboração da Jornada por um vídeo ou material complementar exemplificando a situação de perigo sem a tecnologia. | 30 min |
| **8-Apresentação dos resultados** | -Cada grupo apresenta a ideia selecionada para a Persona,a situação de perigo selecionada e o resultado da Jornada do Usuário (se houver). | 15 min |

| Sugestão: pausa de 10min | | |
|---|---|---|
| ***9*-Prototipação da Jornada do Usuário** | -O principal objetivo desta atividade é prototipar a solução IoT priorizada para cada Persona usando a Jornada do Usuário como referência;<br>-Cada grupo irá utilizar a Persona selecionada, a Jornada de Usuário criada e a solução IoT priorizada;<br>-A prototipação pode ser feita de 3 formas:<br>- Com LEGO® (prototipação em volume)<br>- Material e Desenhos (*Storyboard*)<br>- Encenação ou narrativa (*Storytelling*)<br>- Ou uma combinação de 2 ou mais formas<br>-Cada grupo pode selecionar a forma mais confortável ou o mediador pode sugerir uma forma diferente para cada grupo;<br>-Essa prototipação será feita de forma iterativa e interativa utilizando-se 2-3 *Sprints* de construção (15 min cada *Sprint*) para que seja possível criar a Jornada antes, durante e depois da situação de perigo;<br>-Sugere-se que uma pessoa do grupo seja o líder, para apoiar na condução das atividades;<br>-Ao final de cada *Sprint*, os idosos devem fazer uma demonstração do que já foi prototipado, obter feedbacks e se possível integrar com as ideias de outros grupos;<br>-Sugestão:<br>Deve ser usada o quadro de critérios de escolha (lista de Normas) como referência para validar a Jornada do Usuário e avaliar preferências na utilização dessa tecnologia com base nas expectativas do grupo. | 45-60 min |
| **Sugestão: pausa de 15min** | | |
| **10-Apresentação dos resultados** | -Cada grupo apresenta o resultado final da sua prototipação usando a técnica de *Storytelling* para materializar a ideia proposta;<br>-O grupo realiza uma votação para selecionar a ideia que mais agrada e atende as necessidades pessoais do grupo. Podem ser usados os Critérios Norteadores como referência. | 15 min |
| **11-Encerramento** | -Apresentação dos resultados finais<br>-Agradecimentos<br>-Coletar *feedback* dos participantes | 10 min |

### 6.1.3 Análise das Atividades Realizadas

A Tabela 26 apresenta uma breve análise das atividades realizadas para se atingir o objetivo desse *Workshop.*

**Tabela 26. *Workshop* de Prototipação: análise de atividades**

| Atividade Planejada | Análise |
|---|---|
| **1-Contextualização** | A contextualização foi realizada com sucesso, com uma apresentação em *Power Point* projetada numa TV da sala e todos confortavelmente acomodados. Durante a apresentação foram informados aos participantes os objetivos da pesquisa, as etapas do método, os principais conceitos, os resultados até o momento e as |
| | expectativas do *Workshop* proposto. |
| **2-Apresentação dos participantes** | Não foi necessário realizar novamente essa atividade pois o grupo foi praticamente o mesmo.<br>Todos estavam mais à vontade e descontraídos pois já havia sido criado um laço entre eles enquanto grupo de pesquisa. |
| **3-Preparação do *Workshop*** | -Todo material utilizado nas atividades foi impresso na forma de uma apostila de atividades contendo todos os exercícios a serem realizados, servindo de apoio aos voluntários.<br>- Todos os exercícios foram explicados imediatamente antes de sua execução, permitindo maior clareza na execução das atividades e menos dúvidas.<br>-Portanto essa atividade foi mais detalhada do que no *Workshop* de Valores, o que contribuiu para um maior entendimento dos exercícios propostos. |
| **4-Grupos de Trabalho** | Foram mantidos os grupos de trabalho do *Workshop* de Valores e as mesmas Personas: Antônio, Maria de Lourdes e Nair. |
| **5-Análise de Normas** | O resultado dessa atividade foi muito positivo e cada grupo conseguiu preencher o Quadro de Normas para a respectiva Persona, denominado nessa atividade de Quadro de Critérios de Escolha. |
| **6-Escolha da Ideia prioritária para Prototipação** | -A mediadora apresentou o conceito de soluções inovadoras para *Home care* e alguns dispositivos apresentados no *Workshop* de Ideação.<br>-O uso do Quadro de Critérios de Escolha foi fundamental para o direcionamento da escolha da ideia mais aderente a cada Persona.<br>-Foram apresentadas informações também para suporte a decisão (como vídeos e folders explicativos), considerados também importantes para os idosos materializassem a solução.<br>-Ao final dessa atividade foi realizada uma votação entre os participantes para decidir qual a solução de tecnologia iria seguir para a próxima etapa do método IoT-PMHCS. |
| **7-Jornada do Usuário*** | Esta atividade não foi realizada com todo grupo, pois o *Workshop* teve uma duração maior do que o previsto, e os voluntários optaram por encerrar as atividades.<br>Foi selecionado 1 grupo de 3 voluntários para continuar o *Workshop* no dia seguinte para a Persona Maria de Lourdes.<br>Na segunda etapa do *Workshop, o* grupo Maria de Lourdes não realizou a Jornada do Usuário sem a tecnologia. Optou-se por apresentar um vídeo para exemplificar a situação de perigo escolhida e o realizar a Jornada no Usuário com a tecnologia inovadora. |
| **8-Apresentação dos resultados** | Esta atividade não foi realizada como previsto.<br>Foi realizado o encerramento da primeira etapa do Workshop com todos os voluntários, onde todas as atividades foram consolidadas e todos informados dos próximos passos da pesquisa. |
| ***9*- Prototipação da Jornada do Usuário** | Esta atividade foi realizada apenas com o grupo Maria de Lourdes.<br>A situação de perigo escolhida foi esquecimento, seguido de queda e problema de saúde (aumento de pressão).<br>O grupo escolheu como ferramenta para a realização desta atividade o *Storytelling.*<br>A Jornada do Usuário para uso da tecnologia inovadora na situação de perigo foi realizada usando os seguintes recursos:<br>1- Elaboração da história e cenários de Antes, Durante e Depois usando *post-its e* uma cartolina, organizados em 3 *Sprints*;<br>2- Validação da Jornada usando o Quadro de Critérios de Escolha;<br>3- Não foram utilizados materiais adicionais como Lego, maquete ou desenhos, apesar de terem sido oferecidos;<br>4- A validação de cada *Sprint* foi realizada através da confirmação dos |
| | passos presentes nos *post-its e* uma cartolina por todos os presentes e ensaios parciais do *Storytelling.* |
| **10-Apresentação dos resultados** | Esta atividade foi realizada apenas com o grupo Maria de Lourdes.<br>O grupo optou pela escolha de uma das voluntárias para realizar o *Storytelling* no formato de uma entrevista. O seu resultado foi gravado em vídeo. |
| **11-Encerramento** | O encerramento foi muito positivo, e foram coletados feedbacks dos participantes. |

### 6.1.4 Quadro de Normas

Na atividade 5 – Análise de Normas, para apoiar os usuários na escolha da ideia mais adequada para cada Persona, foi usado a análise de Normas da Semiótica Organizacional.

Foi criado um artefato denominado Quadro de Normas, ou Quadro de Critérios de Escolha. Esse quadro é composto de Perguntas e Respostas de acordo com as categorias de Normas da Semiótica Organizacional priorizadas para essa atividade (Perceptuais, Cognitivas e Avaliativas).

A Tabela 27 a seguir apresenta o resultado consolidado dessa atividade para as Personas analisadas:

**Tabela 27. Quadro de Normas consolidado**

| **Questão** | **Respostas** | **Antônio** | **Maria L.** | **Nair** |
|---|---|---|---|---|
| 1- Qual seria a forma mais adequada da Persona acionar o socorro usando uma tecnologia? | A persona não tem familiaridade com tecnologia. Para que ela seja útil e fácil de usar esta tecnologia deve ser inteligente para detectar uma situação de perigo e acionar socorro, sem acionamento humano. | | | X |
| | O acionamento deve ser feito apenas pela própria Persona. Essa Persona não se sente confortável em ter acionamento automaticamente do socorro. | | | |
| | O acionamento pode ser feito pela própria Persona ou a tecnologia pode também acionar automaticamente o socorro. | X | X | |
| 2-Qual dessas alternativas seria mais adequado para a Persona perceber uma resposta ou ação de Tecnologia numa situação de perigo? | Alarme sonoro (um alarme ou buzina, por exemplo). | | | |
| | Alerta visual (através de imagem na tela de um celular, de um relógio ou TV). | | | |
| | Alerta luminosos (através de cores ou intensidade da iluminação, por exemplo). | | | |
| | Alarme através de mensagem de voz, informando situação de perigo e ações a serem tomadas. | X | X | X |
| 3-Qual a forma mais fácil da Persona compreender as informações emitidas pela Tecnologia? | Mensagens curtas e objetivas, apenas notificando perigo. | | X | X |
| | Mensagens detalhadas informando situação de perigo e ações a serem tomadas | X | | |
| | Alarmes sonoros (um alarme ou buzina, por exemplo) | | | |
| | Alarmes luminosos (através de cores ou intensidade da iluminação, por exemplo) | | | |
| 4-Quais informações a Persona gostaria de receber da tecnologia numa situação de perigo? | Não quero receber informações da tecnologia | | | |
| | Confirmação das ações (Ex: notificação de familiares) | X | X | X |
| | Avisar sobre proximidade do Socorro | X | X | X |
| 5-Quais ações a Persona gostaria que a tecnologia tomasse de forma automática numa situação de perigo? | Coletar informações do seu estado de saúde e manter a sua rede de apoio atualizada | | | X |
| | Filmar o local e enviar para sua rede de apoio | | X | |
| | Notificar prioritariamente os profissionais de saúde | | X | |
| | Notificar rede de apoio automaticamente por mensagem | X | X | |
| | Notificar rede de apoio automaticamente por ligação no celular | X | | |
| 6- Dado a familiaridade com tecnologia, a Persona teria dificuldade em interagir com tecnologia? | Como a Persona não gosta de tecnologia, seria muito difícil se adaptar a uma nova tecnologia. | | | |
| | Como a Persona tem grande facilidade com tecnologia, ela se adaptaria facilmente apenas assistindo um vídeo explicativo e lendo seu manual. | | | |
| | Deve haver um treinamento e/ou orientação de uso da nova tecnologia. Depois disso seria possível usá-la. | X | X | X |
| 7- Como a Persona prefere acionar a tecnologia numa situação de socorro? | Conversar com a tecnologia para acionar socorro (comando de voz). Exemplo: "Socorro queda", “Socorro fogo", "Ligar Médico". | | X | |
| | Apertar um botão de emergência numa pulseira ou colar | | X | |
| | Acionar o socorro através de um aplicativo no celular | | | |
| | A tecnologia deve ser inteligente para perceber o perigo e acionar o socorro automaticamente | X | X | X |
| 8- O que a Persona prioriza ao se deparar com uma nova tecnologia | A tecnologia deve ter uma aparência simples, mas bonita. | | | X |
| | A tecnologia deve ser imperceptível, não ver a tecnologia. | | | |
| | A tecnologia deve ter uma aparência inovadora, para se diferenciar de outros dispositivos eletrônicos. | | | |
| | A tecnologia pode ser usada num dispositivo que já existe e a Persona está acostumada a usar, como um celular, relógio etc. | X | X | |

Observa-se a partir dos resultados consolidados acima que a Persona Nair, como um idoso com maior dificuldade de interação com a tecnologia, optou pelo minimalismo e autonomia da tecnologia sempre que possível.

As demais Personas optam por acionamento por eles próprios, além das ações realizadas pela tecnologia de forma independente. Porém, se sente confortável em adotar um dispositivo de tecnologia que já conhece.

É fundamental que existam notificações, confirmação de ações, mantendo a rede de apoio atualizada.

### 6.1.5 Escolha da Ideia Prioritária para Prototipação

Na atividade 6 foram apresentados a cada grupo 6 ideias mais relevantes identificadas nas etapas anteriores: 3 ideias priorizadas durante o *Workshop* de Ideação e 3 ideias relacionadas à categoria de *Home Care* prioritária para cada Persona

Para a seleção das 3 ideias adicionais foi realizada uma análise de todas as sugestões de tecnologias para *Home Care* oferecidas até o momento, tanto em entrevistas, quanto questionários e *Workshop* de Ideação: foi um total de 56 ideias categorizadas em 16 grandes grupos, conforme apresentado na Figura 37. Essa informação foi apresentada aos idosos durante o *Workshop* de Prototipação.

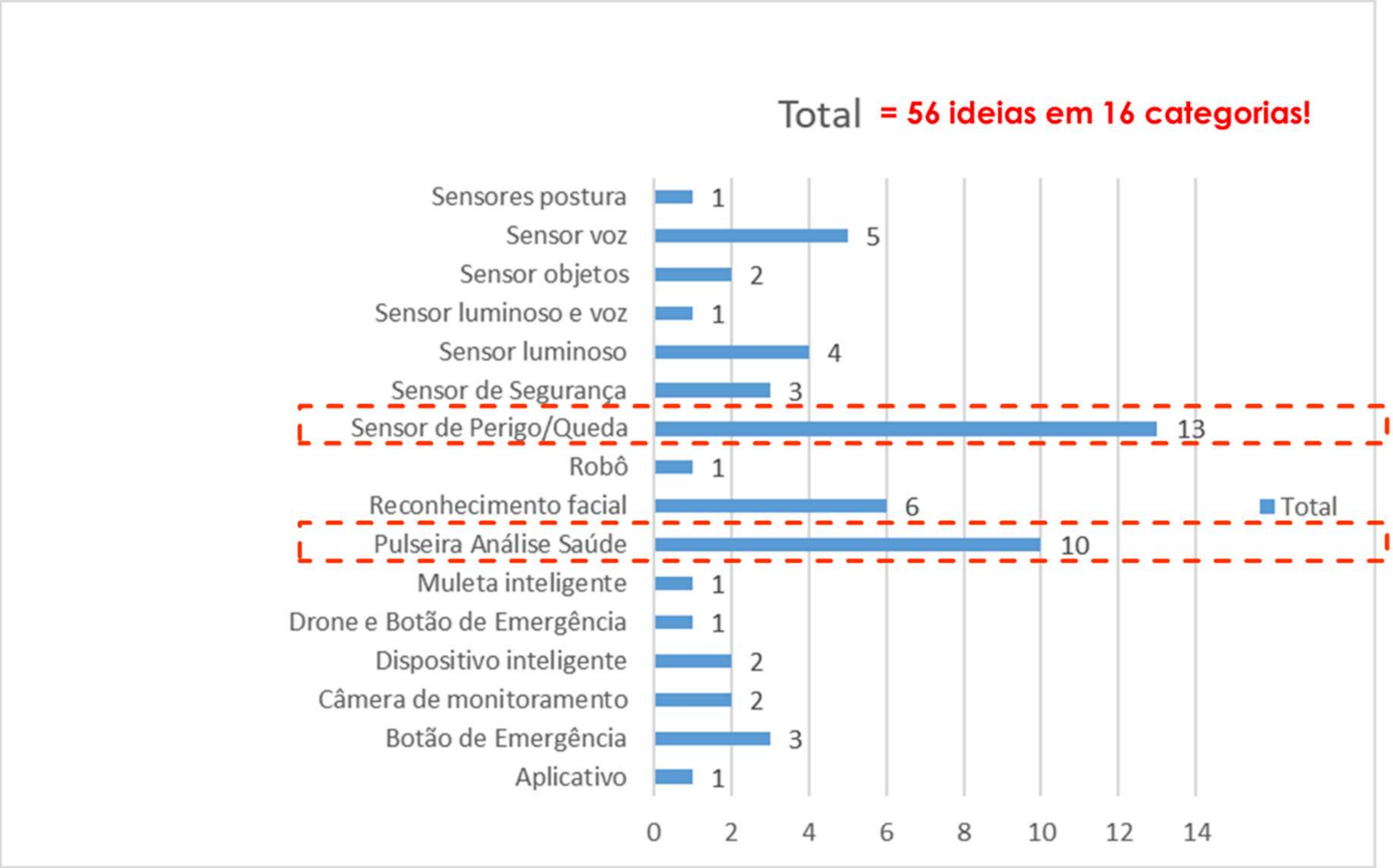

**Figura 37. Resultados das ideias sugeridas durante entrevistas, questionários e workshops, categorizados em 16 grupos de soluções**

A Tabela 28 apresenta as sugestões oferecidas para cada grupo de Persona. As 3 primeiras ideias de cada Persona foram selecionadas durante o *Workshop* de Ideação e as 3 seguintes escolhidas pela pesquisadora com base na categoria assistiva prioritária da Persona e nas análises apresentadas na Figura 37.

**Tabela 28. Ideias apresentadas para cada Persona**

| Grupo – Persona | Ideias |
|---|---|
| Grupo 1 – Antônio | 1. Luzes que abaixam automaticamente para melhorar o sono<br>2. Dispositivo para diminuir o ritmo de agitação antes de dormir<br>3. Sensores de queda na casa e sensor de tropeços<br>4. Pulseira para análise de sintomas de perigo (pulsação acelerada, pressão arterial alta ou alta): emissão de alerta via mensagem para o responsável ou grupo de familiares.<br>Acionamento do médico ou plano de saúde em caso de perigo. Nesse caso, o plano de saúde entra em contato com o cliente para verificar se existe algum problema de saúde.<br>O médico tem acesso ao histórico de saúde do paciente e informações atuais da situação de saúde do paciente na situação de perigo.<br>5. Reconhecimento facial para detecção de problemas de saúde e alarme para familiares em caso de problemas<br>6. Em caso de falta de energia, sensor aciona energia emergencial em toda a casa. O acionamento também poderia ser por voz. |
| Grupo 2 - Maria de Lourdes | 1. Caixinha de remédios online, com aviso de remédios faltantes e compra automática via Internet. |
| | 2. Aviso inteligente caso ocorra algum acidente na casa ou com a pessoa.<br>3. Sistema inteligente para a cozinha com sensores e temporizadores.<br>4. Dispositivo que detecta humor e sugere distrações ou remédio certo para evitar depressão. Avisa a família em situações de perigo.<br>5. Dispositivo que tranque a casa automaticamente caso não detecte a presença de ninguém a aciona alarme.<br>6. Sistema interligado nas casas com sistema de emergência, caso ocorra algum problema, um alerta seja acionado na central. |
| Grupo 3 – Nair | 1. Dispositivo que controla sua alimentação, lembrando de alimentos importantes e medicamentos, com alarmes para cuidadores.<br>2. Kit com câmera e display (com quadro) que fique sempre disponível na sala e na cozinha do idosa e na casa dos filhos. Dispara alarmes em caso de emergência.<br>3. Dispositivo para monitoramento de pressão, batimentos e colesterol, com alarmes em caso de alterações bruscas.<br>4. Pulseira para análise de sintomas de perigo (pulsação acelerada, pressão arterial alta ou alta): emissão de alerta via mensagem para o responsável ou grupo de familiares.<br>Acionamento do médico ou plano de saúde em caso de perigo. Nesse caso, o plano de saúde entra em contato com o cliente para verificar se existe algum problema de saúde.<br>O médico tem acesso ao histórico de saúde do paciente e informações atuais da situação de saúde do paciente na situação de perigo.<br>5. Pulseira que ao detectar perigo automaticamente filmaria o ambiente e a pessoa que está usando-a e aciona familiares e ou médico responsável.<br>6. "Drone" acionado pelo próprio usuário em situações de perigo, através de uma pulseira com botão de pane, conectado à uma Central e alerta pessoas cadastradas. |

O grupo tinha a liberdade também de acrescentar novas ideias ou complementar as ideias existentes.

Para realizar a escolha da ideia prioritária para cada Persona o grupo levou em consideração os seguintes critérios:

- Valor principal da Persona;
- Categoria de *Home Care* priorizada da Persona;
- Quadro de Normas com critérios de escolha preenchido para a Persona, segunda a Tabela 27;
- Vídeos[1] [2] com exemplos de tecnologias assistivas já disponíveis no mercado, para que eles pudessem visualizar uma solução ou parte, apresentados durante o *Workshop*:

[1] Drone ambulância: https://www.youtube.com/watch?v=QAx6uRWfkSU

[2] LinCare: https://www.youtube.com/watch?v=J1zLwYQpxSE&feature=youtu.be

A partir das escolhas, os Mapas de Persona de cada Persona foram atualizados com as ideias selecionadas e seus principais requisitos, como pode ser visto no capítulo 6.1.7 a seguir.

### 6.1.6 A escolha da Jornada do Usuário

Como o horário do *Workshop* já estava avançado e o grupo já estava cansado, a atividade 7 foi simplificada: foi sugerido ao grupo fazer uma votação pela ideia que mais gostaram e assim, seria realizada a prototipação de apenas uma solução, e não mais uma solução por Persona. As atividades do grupo foram encerradas e apenas um grupo de 3 voluntários retornou no dia seguinte para concluir as demais atividades previstas.

Ficou definido que a ideia mais interessante para o grupo era a pulseira com sensores de queda e saúde, com notificação da rede de apoio (familiares, médicos, vizinhos etc) e recursos IoT. A solução foi batizada pelo grupo de **Pulseira Inteligente**.

A Jornada do Usuário foi definida pelo grupo de 3 voluntários. Eles escolheram a persona Maria de Lourdes como objetivo dessa atividade e uma situação de perigo de episódio de queda seguido de um evento relacionado à saúde, como um aumento de pressão

Segue abaixo uma breve descrição da situação de perigo selecionada pelo grupo:

- Ao ir para seu quarto à noite, a idosa se recorda que esqueceu de tomar seu remédio de pressão
- Ao ir ao banheiro para tomar o remédio, ela sofre uma queda devido ao chão molhado
- Ocorre uma fratura decorrente a queda e a idosa não consegue se mover
- Em seguida um aumento de pressão devido ao nervosismo
- Como o celular não estava perto, não consegue chamar socorro
- Idosa mora sozinha e sua casa é recuada, o que impede vizinhos de ouvir seu pedido de socorro

Optou-se por uma questão de tempo não realizar a Jornada do Usuário sem o uso da tecnologia. Foram apenas realizados debates sobre quais as consequências desse acidente sem apoio de uma tecnologia, apresentado um vídeo explicando o que é a Jornada do Usuário e vídeo com estudo de caso de mercado com pulseiras assistivas[3].

### 6.1.7 Prototipação da Jornada do Usuário

Essa atividade teve por objetivo descrever o passo-a-passo da jornada do usuário da Persona Maria de Lourdes durante a situação de perigo escolhida pelos voluntários, considerando o uso da tecnologia assistiva Pulseira Inteligente.

Para facilitar a criação da Jornada do Usuário utilizando a tecnologia assistiva foram realizados os seguintes passos pelos voluntários segundo a Figura 38.

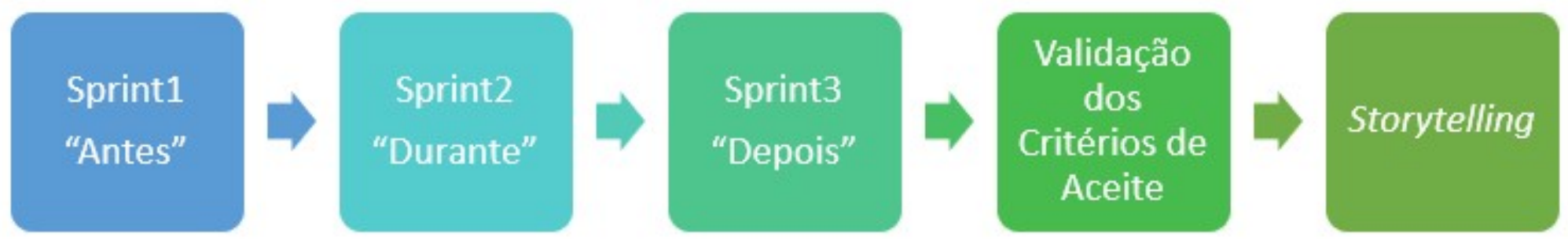

**Figura 38. Passos para a criação da Jornada do Usuário**

Para desenhar o "antes" da Jornada (*Sprint1*), o grupo fez a descrição da situação de perigo para situar a Persona no local e momento, respondendo as seguintes perguntas:

- Em qual cômodo da casa a persona estava?
- A Persona estava acompanhada ou só?
- Qual o estado físico e emocional da Persona no momento?
- Como ela aciona a tecnologia para pedir socorro?
- Quais são as premissas (configurações, treinamentos etc) precisam acontecer para que a tecnologia possa ser utilizada adequadamente pelo idoso?

Para desenhar o "durante" da Jornada (*Sprint2*), o grupo fez a descrição dos passos desde o acionamento do socorro até a chegada do socorro, descrevendo as ações da Persona e tecnologia assistiva também enquanto o socorro não chegava. Para tanto, foram criadas algumas orientações de apoio:

- Descreva quais os passos da situação de perigo, organizado numa linha do tempo:

[3] Jornada do Usuário TeleHelp: https://www.youtube.com/watch?v=xv9Zx-PkUQs

passo 1, passo 2, passo 3...

- Em cada passo descreva a ação da Persona, sendo uma ação por *post-it*
- Em casa passo descreva a ação da Tecnologia, sendo uma ação por *post-it*
- Usar cores diferentes para a ação da Persona e da Tecnologia

Para desenhar o “depois” da Jornada (*Sprint3*), o grupo fez uma descrição dos passos da chegada do socorro, respondendo as seguintes perguntas:

- Como a tecnologia saberá que a emergência já está controlada?
- Como os familiares serão notificados?
- Quais os próximos passos assim que o socorro chegar?
- Como será desativado o estado de emergência?

Além disso, durante essa atividade o grupo fez uma revisão de toda a Jornada do Usuário, avaliando os passos de acordo com o Quadro de Normas e as preferências da Persona. Mas observou-se também que as outras Personas poderiam ser atendidas de uma certa forma pela Jornada, realizando apenas pequenas algumas adaptações ou customizações em alguns passos.

Para apoiar a organização dessa atividade optou-se pelo uso de uma cartolina dividida de acordo com a Jornada do Usuários e post-its® para representar cada ação realizada. Foram usadas diferentes cores de post-its® para representar a atuação de cada *stakeholders* durante a Jornada.

O último passo dessa atividade foi a prototipação da Jornada do Usuário através da gravação de um vídeo de *Storytelling*, ou “Contação de Histórias”, dos passos dessa Jornada do Usuário.

O objetivo inicial era que cada grupo de voluntários utilizasse um conjunto pré-definido de materiais e técnicas para prototipação. Porém, como essa atividade foi realizada por amostragem, ou seja, com apenas um grupo e Persona, foi oferecido a eles todos os materiais disponíveis para escolha do que gerasse maior conforto e realidade na prototipação, dentre eles:

- Lego ®
- Maquete de uma casa
- Materiais para desenho

- *Storytelling*

O grupo optou por utilizar apenas o *Storytelling,* mas com um formato inédito, até então não pensado pela pesquisadora: eles optaram por realizar o *Storytelling* no formato de uma entrevista guiada pela mediadora com um dos voluntários.

O resultado foi a gravação de um vídeo de 10 minutos onde a Persona Maria de Lourdes, representada por um dos voluntários, narra sua saga durante a emergência usando a tecnologia.

## 6.2 Mapa de Persona - Prototipação

### 6.2.1 Antônio

A Figura 39 apresenta o mapeamento da Persona Antônio a partir dos resultados gerados pelo grupo, com a solução IoT escolhida e ampliada.

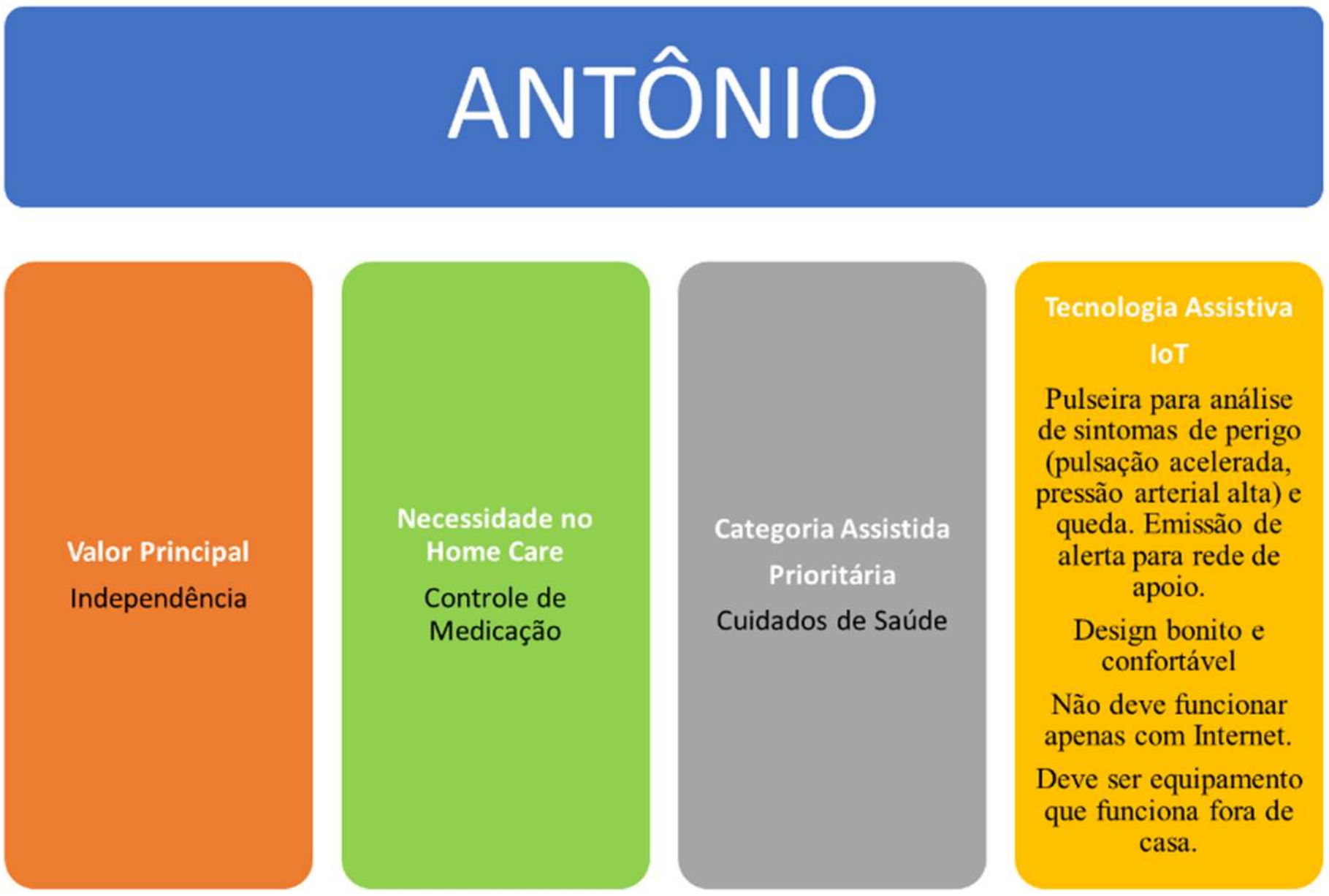

**Figura 39. Etapa 2-Prototipação: Mapa da Persona Antônio**

### 6.2.2 Maria de Lourdes

A Figura 40 apresenta o mapeamento da Persona Maria de Lourdes a partir dos resultados gerados pelo grupo, com a solução IoT escolhida e ampliada.

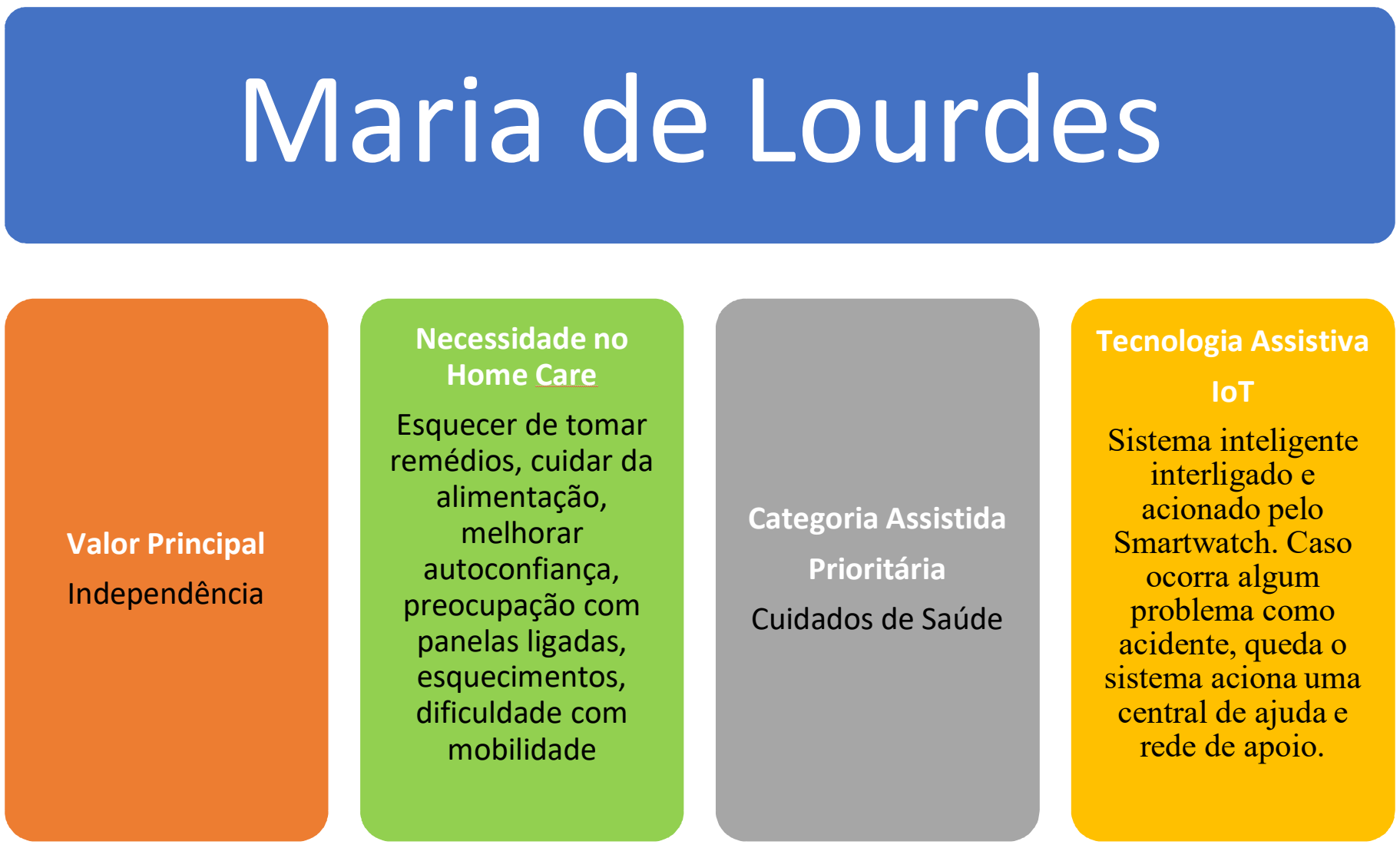

**Figura 40. Etapa 2-Prototipação: Mapa da Persona Maria de Lourdes**

### 6.2.3 Nair

A Figura 41 apresenta o mapeamento da Persona Nair a partir dos resultados gerados pelo grupo, com a solução IoT escolhida e ampliada.

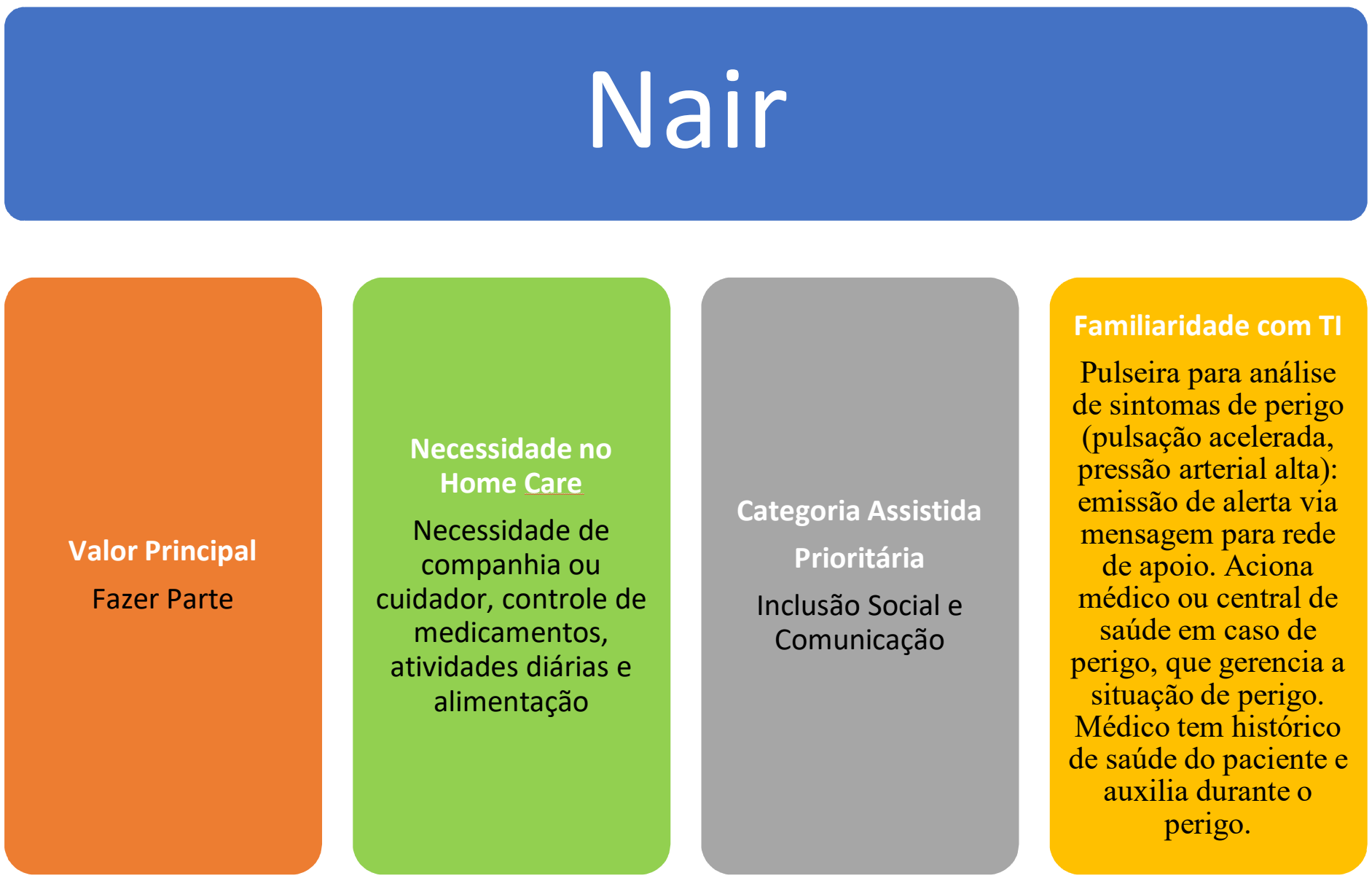

**Figura 41. Etapa 2 Etapa 2-Prototipação: Mapa da Persona Nair**

## 6.3 Jornada do Usuário

Durante a atividade de Prototipação, observou-se 4 grandes atores na Jornada do Usuário, descritos abaixo e representados na Figura 42:

- **Idoso**: ator principal, responsável por disparar ações mesmo de forma ubíqua;
- **Pulseira Inteligente**: dispositivo capaz de se comunicar com outros objetos e coletar informações do idoso relacionados a saúde e emergências. Composto de biosensores e sensores de queda;
- **Central de Controle**: composta por uma Central de Atendimento e controles automatizados entre os diferentes dispositivos e atores;
- **Aplicativo em celular para Rede de Apoio:** responsável por fornecer informações atualizadas a familiares, vizinhos, amigos sobre situação de saúde do idoso e informações em caso de emergência.

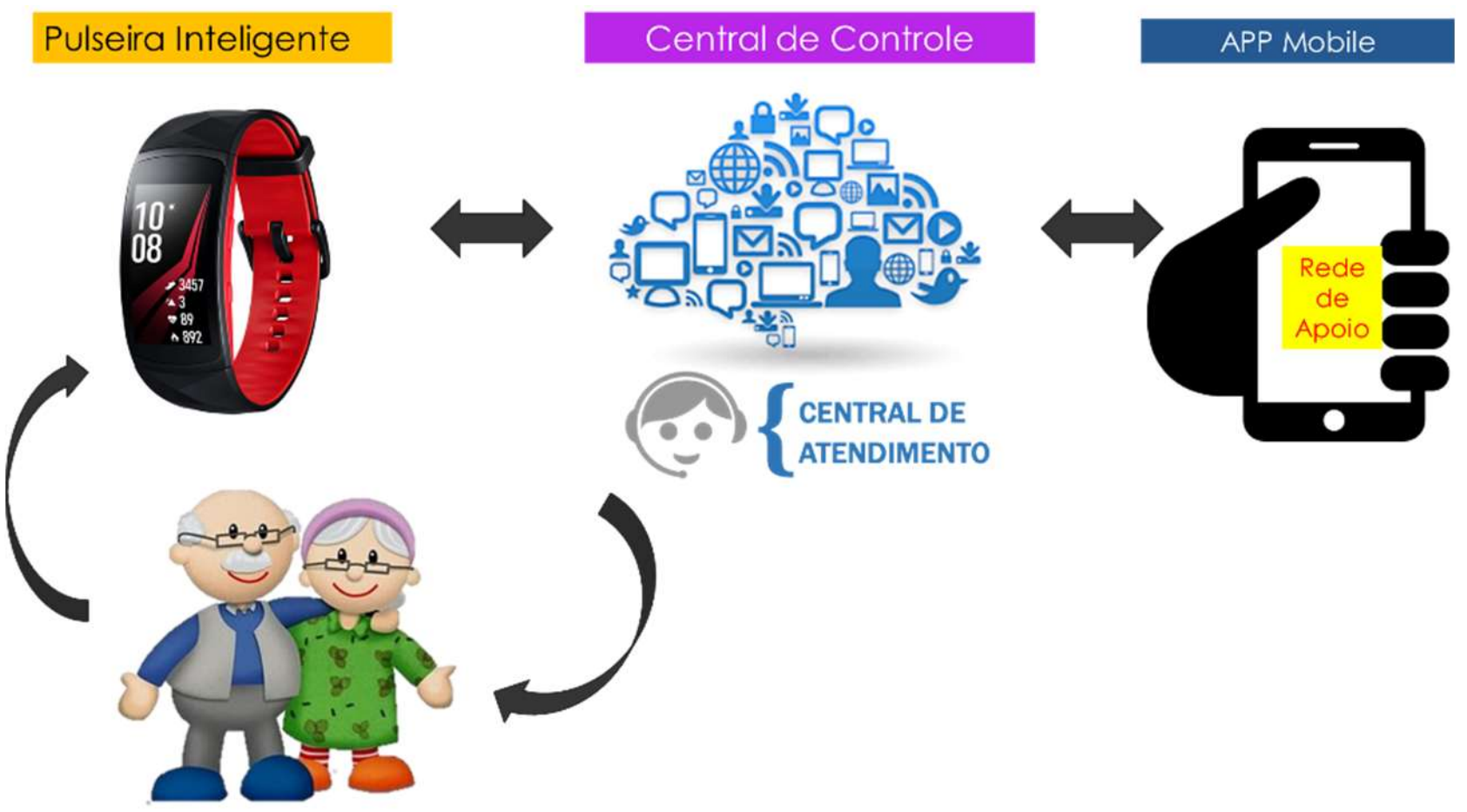

**Figura 42. Atores envolvidos na Solução IoT descrita pela Jornada do Usuário**

A Figura 43 e Figura 44 abaixo representam a Jornada do Usuário nos seus diferentes momentos:

- "Antes": passo 1
- "Durante": passos 2 e 3
- "Depois": passo 4

As cores indicam as ações dos diferentes atores durante a Jornada do Usuário:

- Idoso: Amarelo
- Central de Controle: Rosa
- Pulseira Inteligente: Laranja
- Aplicativo Celular: Azul

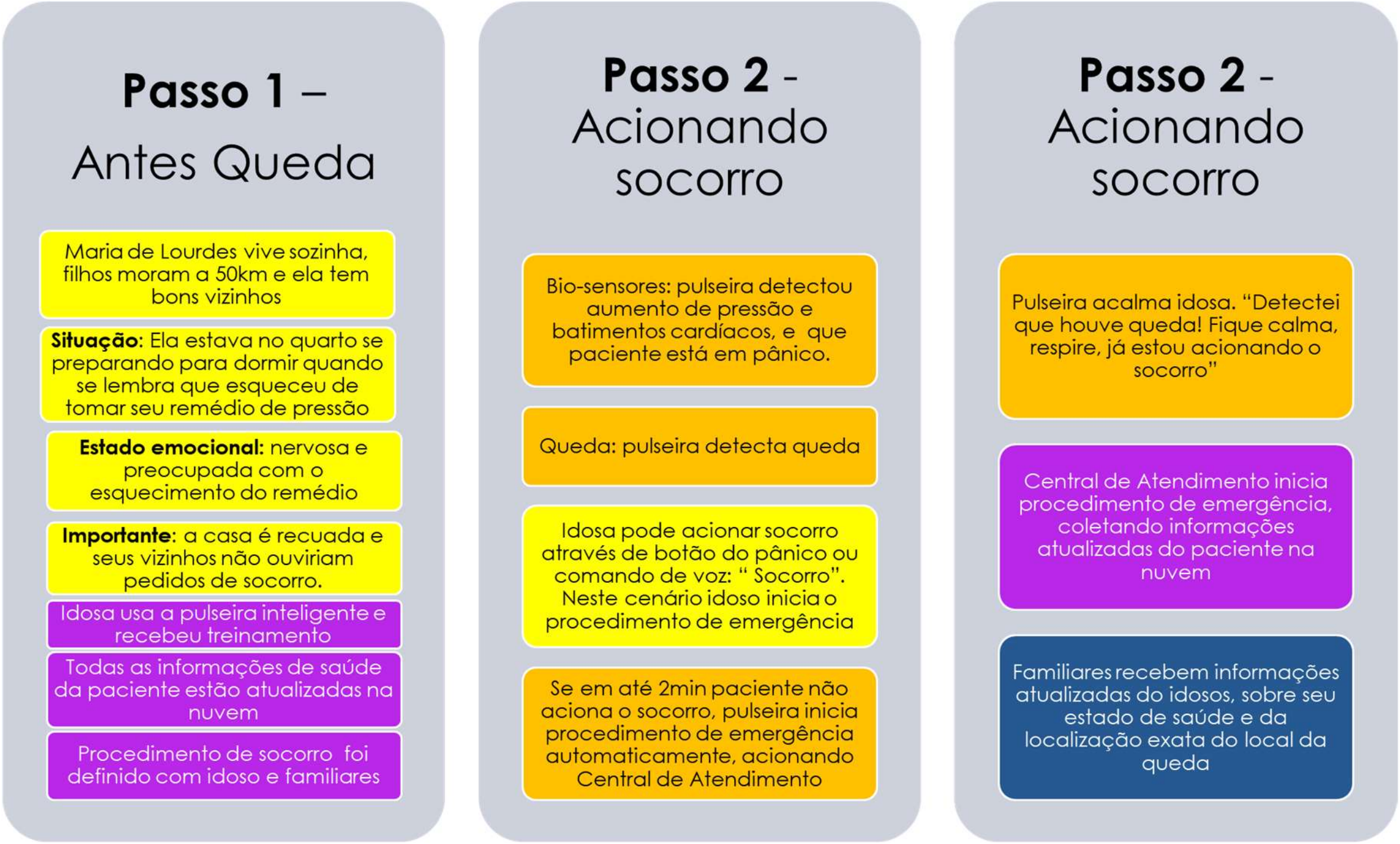

**Figura 43. Passos "Antes" e "Durante" da Jornada do Usuário**

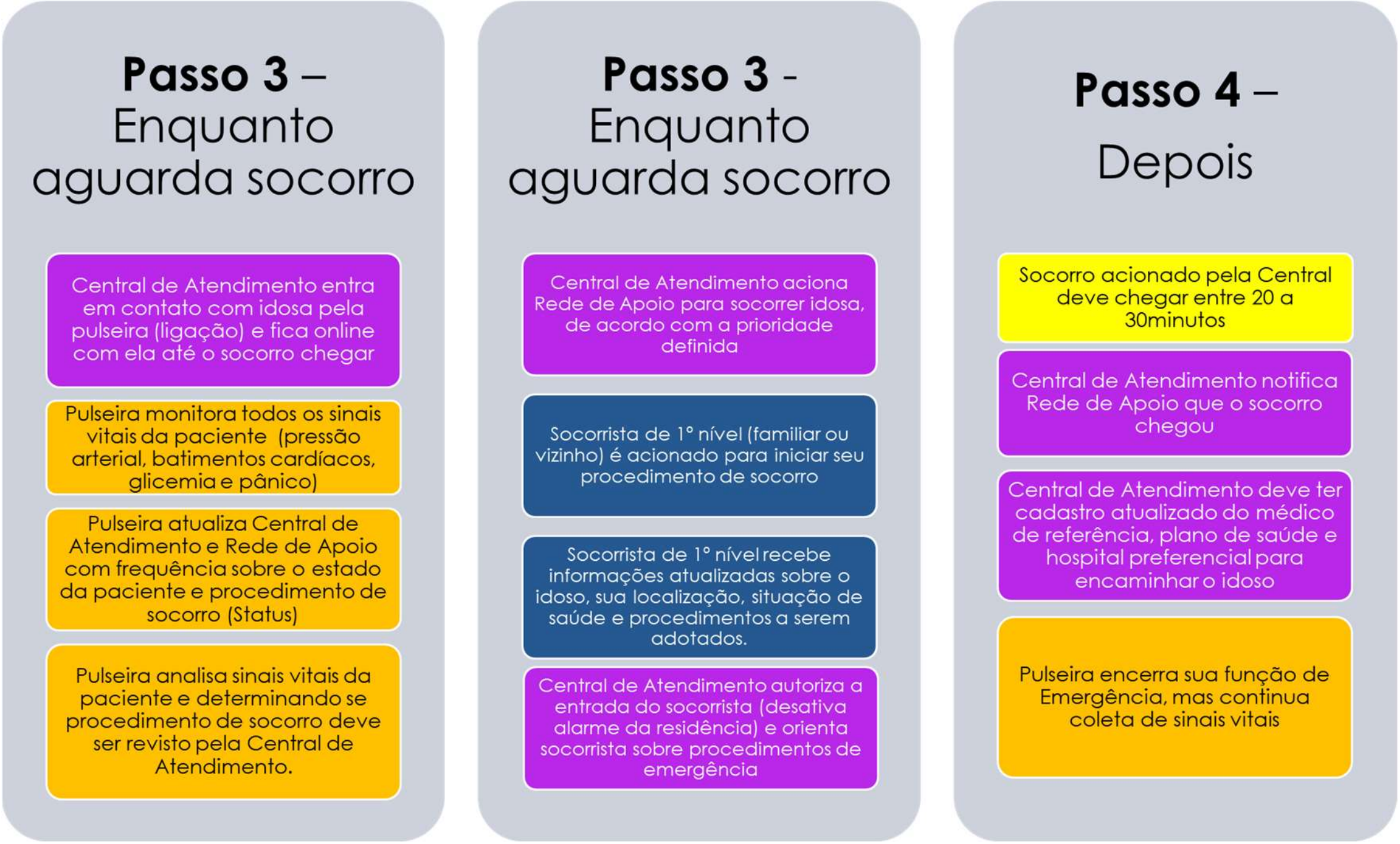

**Figura 44. Passos "Durante" e "Depois" da Jornada do Usuário**

## 6.4 Análise de Stakeholders

Após análise da Jornada do Usuário ficou claro os principais *stakeholder*, ou envolvidos, na solução a ser definida. São eles:

- *Stakeholder* primário: idosos
- *Stakeholder* secundário: rede de apoio formal
    - Central de Atendimento
    - Serviço Emergencial (SAMU)
- *Stakeholder* secundário: rede de apoio informal
    - Familiares
    - Vizinhos a amigos
- *Stakeholder* terciário
    - Médico
    - Hospital ou Clínica Médica
    - Plano de Saúde

## 6.5 Pesquisa de Participação

Após o *Workshop* de Prototipação na Etapa 2 do Método IoT-PMHCS foi solicitado aos participantes o preenchimento da Pesquisa de Participação para avaliação formal das atividades participativas ocorridas nos *Workshops* de Valores e Prototipação.

Seguem os principais resultados obtidos:

- 100% dos participantes consideram o tempo para execução das atividades adequado;
- 100% dos participantes consideram o material e as atividades adequadas para a realização das atividades;
- 100% dos participantes consideram as explicações suficientes para a condução das atividades;
- 100% dos participantes não consideram a presença dos profissionais de Tecnologia da Informação necessária e acreditam que iria até dificultar ou limitar os exercícios. O material fornecido foi suficiente para a realização das atividades;
- Cerca de 85% dos participantes não consideram a presença de profissionais de saúde necessária e acreditam que iriam limitar os exercícios. O material fornecido foi suficiente para a realização das atividades.

A participação dos idosos foi crescente e foi possível avaliar que no segundo *Workshop* o engajamento e envolvimento do grupo era maior e mais intensa do que na Etapa 1. Seguem algumas observações relevantes e que reforçam a efetividade do método:

- “Foi uma experiência ótima, abordando um assunto muito atual e necessário para a faixa etária proposta. Houve uma sintonia muito grande entre voluntários e pesquisadores”.
- “Achei ótimo! A mediadora conduziu muito bem as aulas e as explicações. O grupo estava bem homogêneo mostrando muito interesse pelo excelente trabalho, que era beneficiar a vida de muita gente no futuro.”

Os comentários reforçam que a preparação prévia dos *Workshops*, com material adequado e explicações claras para o público alvo cativam os participantes e geram maior engajamento.

Além disso, a utilização das Personas fez com os voluntários se sentissem mais à vontade ao falarem sobre situações de perigo e problemas indesejados, bem como preferências de interação com a tecnologia, sem se darem conta muitas vezes de que estavam falando deles próprios e não de outras pessoas.

Observou-se também que a presença de vídeos informativos, fotos e o Mapa de Personas permite que o processo participativo através da Co-Criação seja realizado à distância e em momentos diferentes entre idosos e profissionais de tecnologia da informação. Observa-se isto quando 100% dos idosos não considerem necessária a presença de profissionais de TI nos *Workshops*.

# 7 Etapa 3 – Design IoT

Durante a Etapa 3 do método IoT-PMHCS foi realizada a atividade de *Workshop* Técnico, de forma participativa com os profissionais de tecnologia, para definição dos requisitos de uma solução IoT, e *Workshop* de Validação com os idosos para validação da solução, conforme descrito no capítulo 0.

As principais atividades dessa etapa estão brevemente apresentadas no Apêndice IV – Etapa 3: Design IoT.

Segue o registro das principais atividades realizadas, análises e resultados gerados de acordo com a proposta do método IoT-PMHCS.

## 7.1 Relatório do *Workshop* Técnico

Para a realização desse *Workshop* foi selecionado um grupo de 10 pessoas, com idades entre 19 e 49 anos de tecnologia, da cidade de Campinas-SP e Belém-PA conforme descrito no capítulo 0.

### 7.1.1 Organização do *Workshop*

A Tabela 29 descreve a organização do *Workshop* de Prototipação, material utilizado, participantes selecionados, duração média, entre outros, de acordo com a proposta do método IoT-PMHCS, contendo uma análise de eficácia do método para cada item.

**Tabela 29. *Workshop* Técnico: análise de organização das atividades**

| Proposta de Organização | | Análise |
|---|---|---|
| **Material** | -P*ost-its*, canetas coloridas, cartolina, vídeos para contextualização e motivação;<br>- *Framework* IoT previamente analisado referente ao item "Semântica - Coisas", com a lista inicial de funcionalidades;<br>-Material para adequação do ambiente de modo a facilitar a execução dessa atividade, tais como: *coffee break*, música ambiente e, materiais para tornar o local aconchegante etc;<br>-Recomenda-se que todo material a ser utilizado como consulta pelos grupos para a realização das atividades seja impresso. | Todo material foi considerado adequado e suficiente para a execução dessa atividade. |
| **Participantes** | Mandatório: profissionais da área de TI (engenheiros, analistas, cientistas, pesquisadores ou entusiastas de tecnologia); | -Foram envolvidos os mesmos profissionais da atividade anterior e mais um grupo de estudantes da |
| | Opcional: idosos e familiares podem ser envolvidos nessa etapa, mas sua participação é opcional;<br>Recomendação 1: a presença de um assistente para suporte e registro das atividades do evento;<br>Recomendação 2: sempre que possível deve ser mantido o mesmo grupo de trabalho das etapas anteriores. | UFPA.<br>-Não houve assistente para essa atividade<br>-Não possível o envolvimento dos idosos nessa atividade por estarem em uma cidade diferente. Não foi considerada necessária essa presença no decorrer das por ser um *Workshop* mais técnico. |
| **Local** | Recomenda-se um local apropriado para a execução dessa atividade, contendo mesas, cadeiras e lousa.<br>Importante existir uma TV ou projetor para a apresentação de slides e vídeos.<br>O ideal seria uma sala de reunião ou sala de aula, com espaço adequado para a atividade. | Foi usado o mesmo local do *Workshop* de Ideação e como parte do grupo estava remoto foram usadas ferramentas de colaboração e comunicação a distância. |
| **Registro** | Registrar os principais resultados de forma textual.<br>Recomenda-se também registros com fotos e/ou vídeos para análise posterior. | Os participantes não se sentiram à vontade em serem filmados.<br>Mas os registros realizados foram realizados utilizando fotos, anotações da moderadora e pesquisa final de participação, sendo considerado satisfatórios para essa atividade. |
| **Duração** | Total de 2 a 3 horas, com pausas para café. | -O tempo total de 4 horas realizada presencial e 2 horas remotas por videoconferência.<br>-O tempo não foi suficiente para a realização de todas as atividades previstas no mesmo dia.<br>-Porém, as atividades foram realizadas dentro dos limites dos voluntários que não se sentiram cansados. |
| **Resultados Esperados** | Como resultado desse *Workshop* espera-se que o *Framework* IoT seja revisado e atualizado, bem como a lista de *User Stories*, critérios de aceite e normas referente ao item "Semântica - Coisas";<br>São analisadas também as "mensagens subliminares", ou seja, tudo que não foi dito explicitamente, mas se pode perceber através de comentários e observações durante o *Workshop*. | Foi possível produzir os resultados esperados para a solução priorizada, mas com um nível de detalhamento ainda preliminar. |

### 7.1.2 Atividades Propostas no *Workshop*

A Tabela 30 apresenta o planejamento de atividades para esse *Workshop*.

**Tabela 30. *Workshop* Técnico: planejamento de atividades**

| Objetivo | Atividades Sugeridas | Tempo |
|---|---|---|
| **1-Contextualizar** | -Contextualizar sobre o projeto (pode ser usado um pôster ou slides)<br>-Apresentar os resultados obtidos na fase anterior | 15 min |
| **2-Apresentação dos participantes** | -Pedir para que cada participante se apresente, respondendo às seguintes perguntas:<br>1-Quem eu sou (nome e idade);<br>2-Estilo de Vida (onde vivem, família, atividades);<br>3-Familiaridade com IoT.<br>Essa atividade é opcional caso os participantes sejam os mesmos de etapas anteriores. | 15 min |
| **3-Apresentar Informações da Solução** | -Apresentar a solução priorizada;<br>-Se for utilizada alguma solução de mercado ou pesquisa como referência, apresentar os requisitos da solução;<br>-Podem ser utilizados vídeos e materiais adicionais nessa etapa. | 10 min |
| **4-Apresentar *Framework* IoT** | -Apresentar o conceito do *Framework* IoT;<br>-Apresentar a proposta inicial criada. | 15 min |
| **5-*Workshop* de Brainstorm e Co-Criação** | - Organizar grupos de 3-4 pessoas para *Brainstorm e Co-Criação* com o objetivo de detalhar os requisitos da solução IoT priorizada;<br>- Cada grupo seleciona 1-2 funcionalidades diferentes da solução, de tal forma que todas as funcionalidades sejam avaliadas;<br>-Cada grupo realizar uma sessão de *Brainstorm* para detalhar as funcionalidades selecionadas, podendo nesse momento serem preenchidos outros itens do *Framework* IoT, além do priorizado “Semântica - Coisas”. O objetivo é principalmente validar a proposta do ponto de vista sistêmico e identificar requisitos de usabilidade usando-se *User Stories*;<br>-Cada grupo deve detalhar as funcionalidades utilizando *User Stories* contendo pelo menos 1 critérios de aceite cada uma;<br>-Anotar cada *User Stories* em um Post-Its e posicionar numa parede, lousa ou cartolina, abaixo da funcionalidade selecionada. Importante: o grupo deve incluir nesse *Brainstorm* também requisitos técnicos e de arquitetura identificados, sendo que nesse caso é importante anotar em um Post-It de cor diferente;<br>-A cada 20 minutos, o grupo troca de funcionalidades com outro grupo e continua o processo de *Brainstorm* dentro da nova funcionalidade. Anotar os resultados em *Post-Its* e posicionar numa parede, lousa ou cartolina debaixo da funcionalidade selecionada;<br>-Cada grupo deve complementar as *User Stories* da proposta do anterior e/ou identificar novas *User Stories;*<br>-Essa troca deve ser feita por 3 vezes;<br>-OBS-1: Cada grupo deve ter a mesma cor de Post-it, para que seja possível identificar as sugestões dadas por cada grupo.<br>OBS-2: Durante o processo de pesquisa será a avaliado a possibilidade de usar LEGO® para a prototipação; | 60 min |
| **Sugestão: pausa de 10min** | | |
| **6-Apresentação e Análise do Resultado** | -O mediador faz a organização das *User Stories* geradas de acordo com as funcionalidades, de forma visual numa parede, lousa ou cartolina, com apoio do grupo; | 50 min |
| | -Devem ser eliminadas redundâncias e agrupadas semelhanças;<br>-Se possível todos são lidos e o grupo discute sobre as especificações realizadas, incluindo novos critérios de aceite ou *User Stories* se necessário. | |
| **7-Encerramento** | -Apresentação dos resultados finais<br>-Agradecimentos<br>-Coletar feedback dos participantes | 10 min |

### 7.1.3 Análise das Atividades Realizadas

A Tabela 31 apresenta uma breve análise das atividades realizadas para se atingir o objetivo desse *Workshop.*

**Tabela 31. *Workshop* de Prototipação: análise de atividades**

| Objetivo | Análise |
|---|---|
| **1-Contextualizar** | A contextualização foi realizada com sucesso, com uma apresentação em *Power Point* projetada numa TV da sala e todos confortavelmente acomodados.<br>Durante a apresentação foram informados aos participantes os objetivos da pesquisa, as etapas do método, os principais conceitos, os resultados até o momento e as expectativas do *Workshop* proposto. |
| **2-Apresentação dos participantes** | Não foi necessário realizar novamente essa atividade pois o grupo foi praticamente o mesmo.<br>Todos estavam mais à vontade e descontraídos pois já havia sido criado um laço entre eles enquanto grupo de pesquisa. |
| **3-Apresentar Informações da Solução** | - Foram apresentadas as seguintes informações iniciais aos participantes:<br>• A solução priorizada na Etapa 2<br>• A Jornada do Usuário definida na Etapa 2.<br>• Vídeos e informações para reforçar o conceito de IoT.<br>• Vídeos sobre soluções já existentes no mercado, com foco no mercado brasileiro.<br>• Conceituação de User Story |
| **4-Apresentar *Framework* IoT** | -Foi apresentado o conceito do *Framework* IoT e a proposta inicial criada de requisitos da solução de acordo com a Jornada do Usuário. |
| **5-*Workshop* de Brainstorm e Co-Criação** | O grupo trabalhou na atividade de Co-Criação e gerou o maior número de *User Stories* dentro do tempo disponível.<br>A atividade foi extensa e para concluir os objetivos propostos foram realizadas 2 seções:<br>- Parte1: levantamento inicial das *User Stories* do “durante” da Jornada do Usuário. Cada grupo ficou responsável pela especificação de uma “coisa” de acordo com o Framework IoT.<br>- Parte2: Completar e detalhar as *User Stories* geradas na Etapa 1 e identificar as demais para o “antes” e “depois” da Jornada. Priorizar as “Histórias” mais relevantes.<br>Os resultados foram registrados na ferramenta *Trello* para permitir maior colaboração. |
| **6-Apresentação e Análise do Resultado** | Cada agrupo apresentou o seu resultado e foram incluídos novos critérios de aceite e *User Stories* quando necessário.<br>Observou-se que foram criados poucos critérios de aceite na Parte1, sendo esses |
| | melhor detalhados na Parte2. |
| **7-Encerramento** | O encerramento foi muito positivo, e foram coletados feedbacks dos participantes. |

### 7.1.4 Lista Inicial de Funcionalidades

Na atividade 4, para apoiar os voluntários na atividade de detalhamento foi apresentado a lista inicial de funcionalidades identificadas a partir da Jornada do Usuário e uso preliminar do *Framework* IoT, conforme mostra a Figura 45.

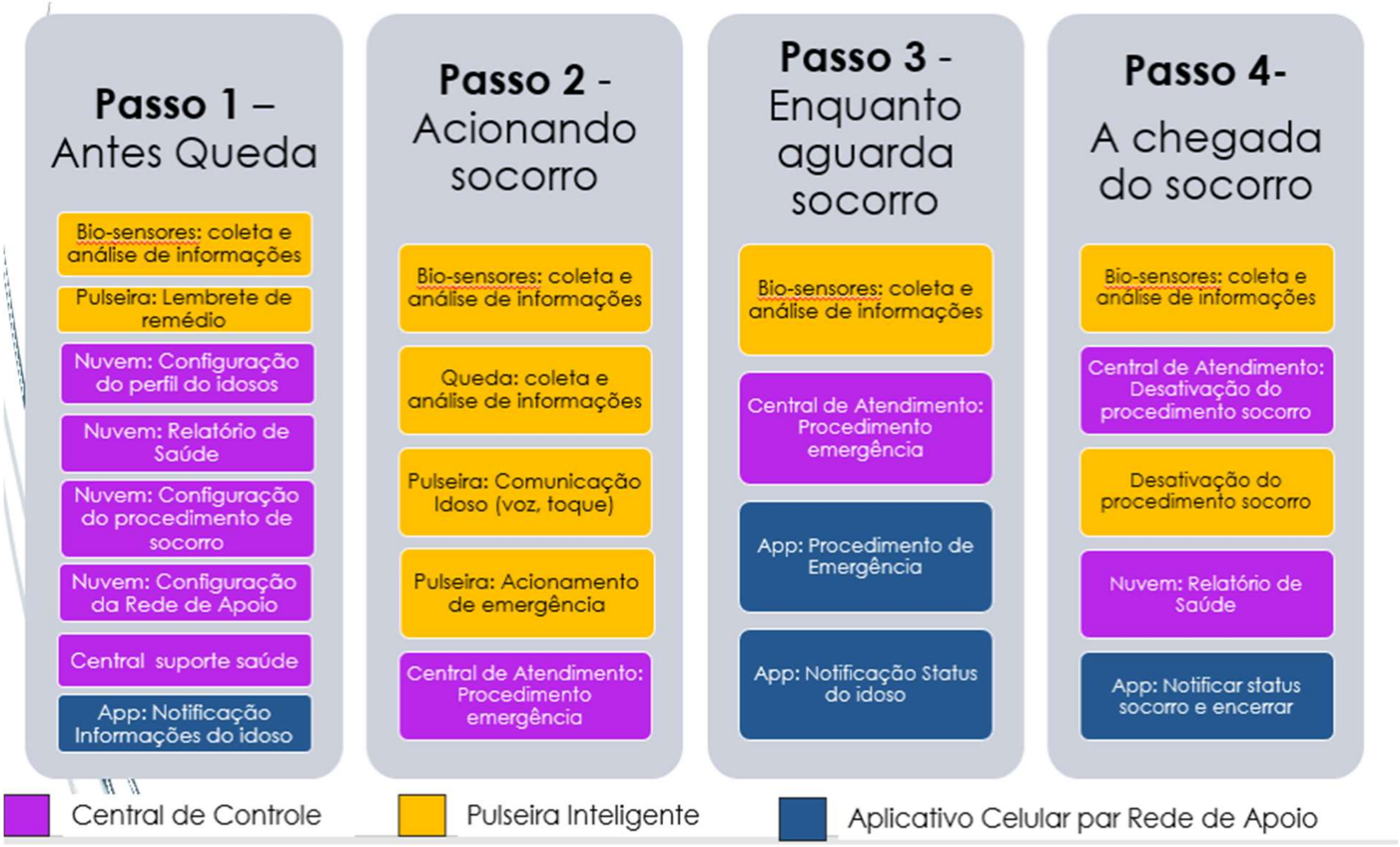

**Figura 45. Lista Inicial de Funcionalidades a Pulseira Inteligente**

### 7.1.5 Processo de Co-Criação

Para executar a atividade de Co-Criação, foi definido um processo que permitiu melhor organização das atividades.

De acordo com o *Framework* IoT (Koreshoff, Leong e Robertson, 2013) é importante para a especificação de uma solução IoT a identificação de todas as "coisas" que compreendem a solução, para assim realizar sua especificação. Para a solução "Pulseira Inteligente" foram identificadas 3 "coisas": pulseira, aplicativo mobile e a Central de Controle.

Aliados ao *Framework* IoT, foi combinada Jornada do Usuário para apoiar a especificação. Dessa forma, os voluntários deveriam identificar as *User Stories* de cada "coisa" nos diferentes momentos da Jornada: "antes", "durante" e "depois".

Seguem os passos utilizados para realizar essa atividade:

- Cada grupo fica responsável por avaliar as funcionalidades de uma "coisa" da Solução IoT: Pulseira, Central de Controle, Aplicativo;
- Será usado um Trello para organizar o detalhamento de cada "coisa". Todas as 3 pastas são compartilhadas;
- São realizadas 3 *Sprints* de 20minutos, um para cada etapa da Jornada do Usuário:
    - Na Parte 1, o principal objetivo é detalhar as *User Stories* do "Durante";
    - Na Parte 2, o principal objetivo é detalhar as *User Stories* do "Antes" e "Depois";
- Para cada etapa da Jornada, o grupo deve identificar as *User Stories* de cada funcionalidades. Caso existam novas funcionalidades essas devem ser identificadas
    - As Funcionalidades foram organizadas como "Etiquetas" no *Trello*
    - Colocar uma *User Stories* por *Card* e os critério de aceite como detalhamento
    - Classificar as *User Stories* por FUNCIONALIDADE
    - Se identificar mais funcionalidades, incluir NOVA ETIQUETA!
- **Co–Criação**: a cada *Sprint*, 1 pessoa do grupo é trocada de tal forma que:
    - O **Especialista** é fixo e responsável por manter a coesão das informações do seu dispositivo.
    - O **Integrador** não é fixo e a cada *Sprint* muda de time, sendo responsável por manter as informações dos dispositivos **integradas**

No início de cada *Sprint* cada pessoa explica o que já foi definido na etapa anterior para seu dispositivo e então discutem juntos as funcionalidades do dispositivo para a etapa da Jornada do Usuário representada pelo *Sprint*.

## 7.2 Solução IoT "Pulseira Inteligente"

Após o *Workshop* Técnico - Etapa 1 foi possível definir de forma macro a solução completa IoT, que como já mencionado é composta de 3 "coisas":

- **Pulseira Inteligente**: neste contexto é mais do que apenas um dispositivo eletrônico *wearable*. Ele é composto por uma solução onde a pulseira eletrônica com sensores capta as informações e é responsável pelo pré-processamento das informações e acionamento da Central de Controle. Esse pré-processamento pode ser feito na própria pulseira, no celular ou um *middleware* Arduino (Oliveira 2017), a ser definido durante o seu desenvolvimento.
- **Central de Controle**: é uma solução na nuvem responsável por receber informações da "Pulseira Inteligente" e disparar ações, dando suporte ao idoso e acionando a Rede de Apoio em caso de emergência. Ela pode ser acionada diretamente pelo idoso ou pela "Pulseira Inteligente" ao identificar automaticamente uma situação de perigo.
- **Outros Sensores**: na casa podem existir outros sensores que se comunicam com a "pulseira inteligente" para proteção do idoso e identificação de perigo. Como por exemplo um sensor de fumaça.
- **Aplicativo para Rede de Apoio (APP)**: A rede de apoio será acionada sobre a emergência através de um APP Mobile, que também gera relatórios diários sobre a saúde do idoso.

Após identificadas as funcionalidades de cada "coisa", foi possível gerar um diagrama simplificado da solução e suas integrações, conforme mostra a Figura 46.

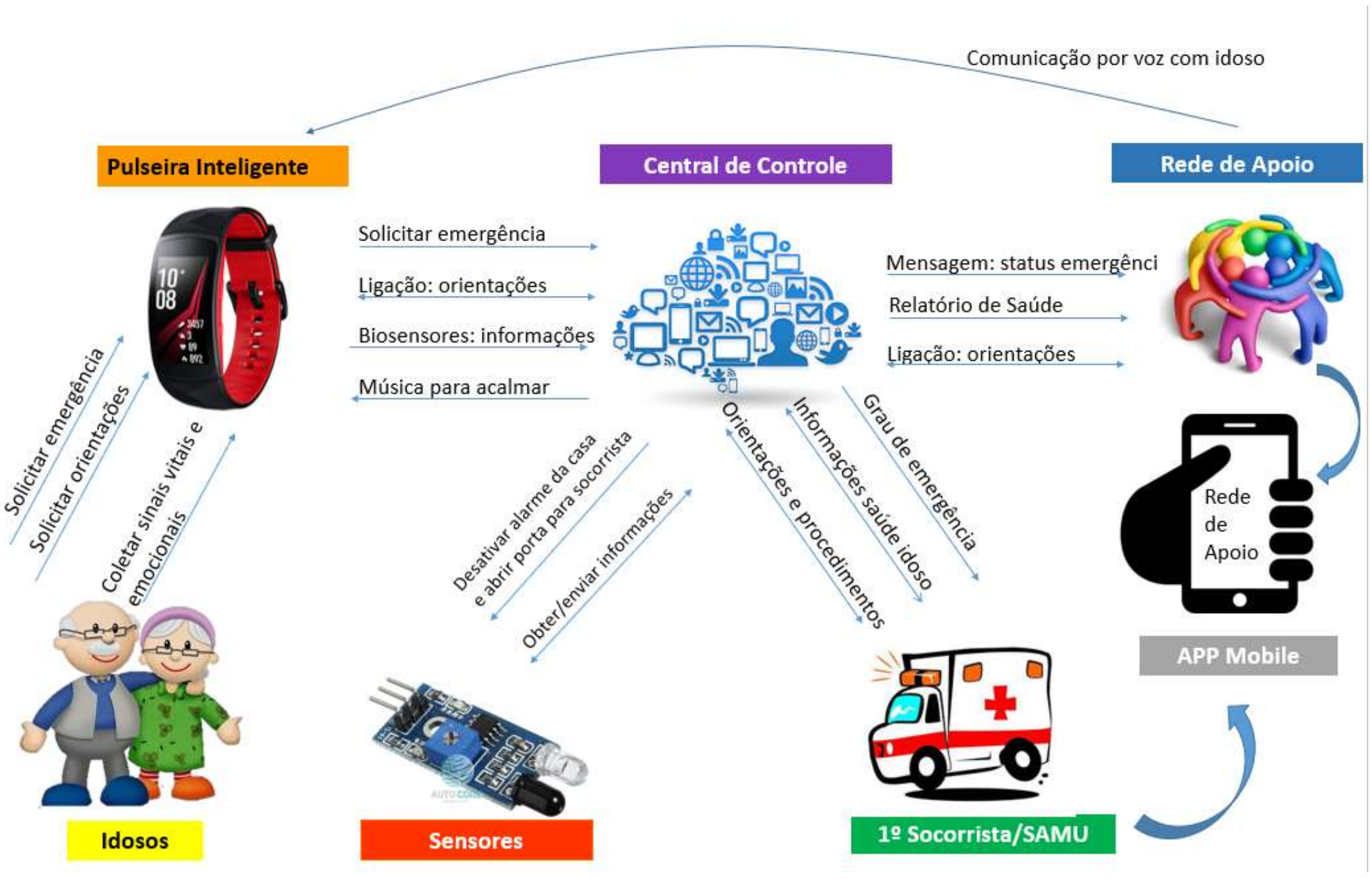

**Figura 46. Solução IoT Pulseira Inteligente**

Abaixo os principais requisitos funcionais e não funcionais macros da solução:

Lista de Funcionalidades, ou Requisitos Funcionais:

- Gestão de Acesso
- Suporte de Saúde
- Configurações: Rede de Apoio, Perfil do Idoso e procedimento de socorro
- Relatório de Saúde
- Procedimento de Emergência
- Encerramento da Emergência
- Comunicação Idoso
- Alerta Remédios
- Ação Humana
- Biosensores
- Outros Sensores

Lista de Requisitos Não Funcionais

- Integração
- Tratamento de Falhas

## 7.3 Lista de Funcionalidades para “Central de Controle”

Durante o *Workshop* Técnico na Etapa 3 do método IoT-PMHCS ficou definido que a Central de Controle será responsável por centralizar as informações do paciente e integrar a comunicação com os outros componentes da solução. Ela será formada por 3 componentes:

1- Central de Atendimento humano.

2- Um *middleware* para análise e controle de informações, elaboração de relatórios automáticos, tratamento de falhas, entre outros.

3- Um sistema com interface web que pode ser acessado pela Central de Atendimento para configurações e registros de atendimento.

A Central de Controle tem integração com o componente Pulseira Inteligente e Aplicativo para receber e enviar informações do idoso.

A Tabela 32 apresenta a lista de *User Stories* do componente Central de Controle organizada de acordo com a etapa da Jornada do Usuário (Antes, Durante - acionando o socorro e enquanto aguarda o socorro, Depois). É importante deixar claro que nesse *Workshop*, dado as restrições de tempo dos voluntários, as *User Stories* e os critérios de aceite foram criados de forma simplificada. As *User Stories* inclusive não atendem ao formato padrão recomendado, sendo usado apenas um texto simplificado para identificação.

**Tabela 32. Central de Controle: lista de User Stories**

| **Requisitos** | **Jornada** | ***User Storie*** | **Critérios de Aceite** |
|---|---|---|---|
| Suporte de Saúde | Antes | Atendimento de apoio ao idoso e familiares | Canal de comunicação aberto entre idosos e familiares. |
| | Antes | Receber uma ligação telefônica. | A Central ao receber uma ligação telefônica deve ser responsável por orientar os idosos nas mais diferentes situações.<br>Para isto ela deve acessar o cadastro do idosos com suas preferências e informações principais. |
| | Após | Monitorar idosos | Monitorar idosos após acidentes durante período pré-determinado. |
| Gestão de Acesso | Antes | Realizar login no sistema | -A senha deverá ser criptografada<br>-O login será o email do usuário<br>-Caso o usuário não estiver previamente cadastrado no sistema ele não terá acesso ao sistema e receberá uma mensagem informando ao tentar logar.<br>- Possui integração com Pulseira e Aplicativo |
| Configurações | Antes | Selecionar um atendente preferencial | -Selecionar o atendimento principal da Central de Controle:<br>-Selecionar pelo menos 2 atendentes principais/preferenciais |
| | Antes | Configurar o perfil do idoso | - Cadastro do idoso: informações sobre saúde, remédios, hospitais, convênios<br>- Preferência de atendimento |
| | Antes | Configurar procedimento de socorro | - Determinar um workflow padrão<br>- Avaliar alguns pontos de customização desse atendimento a serem definidos pelo idoso ou familiares |
| | Antes | Configurar o Rede de Apoio | - Cadastrar toda a rede de apoio<br>- Primeiro socorrista e atendimento padrão<br>- Definir uma lista de socorristas, por prioridade |
| Relatório de Saúde | Antes | Receber informações de batimento, respiração, pressão etc. | -Com qual frequência?<br>-Se o processamento foi feito no celular a Central poderia ser menor sobrecarregada<br>- Possui integração com Pulseira |
| | Antes | Gerar relatório de saúde | Analisar informações recebidas da pulseira, sensores e status da Central de Controle e preparar um relatório para envio periódico para Rede de Apoio |
| | Durante - Enquanto Aguarda Socorro | Manter Rede de Apoio informada | Em caso de emergência, enviar relatório periódico sobre situação de saúde do idosos e status do atendimento |
| | Durante - Enquanto Aguarda Socorro | Manter socorristas informados | Manter socorristas informados sobre situação do idoso até o socorro definitivo chegar no atendimento |
| Procedimento de Emergência | Durante - Acionando o socorro | Receber uma mensagem de emergência da pulseira | A Central recebe as informações da emergência e seleciona as possíveis ações de acordo com a gravidade da situação.<br>Iniciar um protocolo de verificação da chamada.<br>Diagnosticar o nível do problema. |
| | Durante - Acionando o socorro | Analisar Informações | Verificar informações vitais e compará-las com histórico para, possivelmente, disparar uma emergência com severidade alta<br><br>Processamento imediato no celular!<br>Se houver acionamento de perigo, então Central usa essas informações para processamento |
| | Durante - Acionando o socorro | Realizar uma ligação telefônica para a pulseira - Emergência | -Deve ser uma das etapas de verificação da emergência<br>-Deve ocorrer assim que receber a mensagem de emergência.<br>-Deve ocorrer dependendo do nível de bateria da pulseira. |
| | Durante - Acionando o socorro | Enviar uma mensagem para o primeiro socorrista | Essa mensagem é a de acionamento de emergência<br>-Ter uma lista de pessoas (parentes ou vizinhos) organizada por nível de prioridade.<br>-De preferência são pessoas que moram perto do idoso. |
| | Durante - Acionando o socorro | Enviar mensagens para a rede de apoio | Essa mensagem precisa ser enviada após uma rápida confirmação da emergência.<br>- Possui integração com Aplicativo |
| | Durante - Acionando o socorro | Enviar informações de saúde para médicos e primeiro socorrista | -Para o primeiro socorrista mandar um relatório mais sucinto possível e de fácil entendimento.<br>-Para os médicos pode ser enviado um relatório mais detalhado.<br>- Possui integração com Aplicativo |
| | Durante - Acionando o socorro | Tratar um lote de mensagens com informações coletadas pela pulseira enquanto a mesma estava desconectada | - Em situações de queda de rede.<br>- Em situações de envio periódico das informações Se a pulseira envia informações para a montagem do relatório de saúde, cabe à Central processar essas informações e gerar os relatórios nos formatos indicados (gráficos, relatórios textuais) |
| | Enquanto Aguarda Socorro | Enviar uma música de fundo enquanto o socorro não chega | opcional |
| | Enquanto Aguarda Socorro | Enviar mensagem para a pulseira destravar as portas e desarmar alarme | Isso deve ocorrer caso o primeiro socorrista estiver perto. |
| | Enquanto Aguarda Socorro | Realizar ligação para o primeiro socorrista para informar sobre procedimentos de emergência | -Ter uma lista de pessoas (parentes ou vizinhos)<br>-Caso o primeiro da lista não atenda - Ligar par o próximo.<br>-De preferência são pessoas que moram perto do idoso. |
| Tratamento de Falhas | Durante - Acionando o socorro | Realizar procedimentos de Verificação da Chamada de Emergência (Alarme Falso) | Verificar se a chamada não foi feita por engano no caso da chamada ter sido efetuada através de um botão.<br>-Isso pode ocorrer através do monitoramento dos sensores.<br>-Não tirar o fator humano da análise, (o idoso pode não estar bem emocionalmente) |
| | Durante - Acionando o socorro | Receber uma mensagem de alarme falso | |
| Encerramento da Emergência | Após | Desativar o status de emergência | Notificando toda a rede de apoio e dispositivos relacionados a situação de encerramento da emergência |

## 7.4 Funcionalidades para “Pulseira Inteligente”

Durante o *Workshop* Técnico na Etapa 3 do método IoT-PMHCS ficou definido que o componente Pulseira Inteligente será um *wearable* no formato de uma pulseira, que é responsável por monitorar a saúde do idoso. Tal pulseira tem integração com o componente Central de Controle para receber e enviar informações do idoso.

Seguem os principais requisitos não funcionais:

- Deve ser capaz de receber ligações da Rede de Apoio e da Central de Atendimento.
- Deve conter biosensores e ser capaz de identificar queda e localização do idoso.
- Deve ser capaz de se conectar via *bluetooth* ou *wifi*.
- Coleta, envia e recebe informações, mas não processa grande volume de informações, pois sua capacidade de processamento a princípio seria restrita. Portanto, processamento grande volume de informações seria feito num celular do idoso ou *middleware*.
- Ter tela para informar idosos sobre medicamentos, alertas.
- Aceitar comando de voz.

A Tabela 33 apresentam a lista de *User Stories* do componente Pulseira Inteligente organizada de acordo com a etapa da Jornada do Usuário (Antes, Durante - acionando o socorro e enquanto aguarda o socorro, Depois).

Importante deixar claro que, nesse *Workshop*, dado as restrições de tempo dos voluntários, as User Stories e os critérios de aceite foram criados de forma simplificada. As *User Stories,* inclusive, não atendem ao formato padrão recomendado, sendo usado apenas um texto simplificado para identificação.

**Tabela 33. Pulseira: lista de *User Stories***

| **Requisitos** | **Jornada** | ***User Storie*** | **Critérios de Aceite** |
|---|---|---|---|
| Gestão de Acesso | Antes | Realizar login | -A senha deverá ser criptografada<br>-O login será o email ou ID de rede social do usuário<br>-Caso o usuário não estiver previamente cadastrado no sistema ele não terá acesso ao sistema e receberá uma mensagem informando ao tentar logar. |
| Lembrete de Remédios | Antes | Alerta de Medicamento | -Idoso recebe um alerta minutos antes do horário programado para uso do medicamento.<br>-Alerta deve se manifestar em forma de notificação, com reprodução de um áudio que informe o medicamento, quantidade e entre outros detalhas anteriormente cadastrados. |
| Comunicação Idoso | Antes | Acionar a Central de Controle por comando de voz | Acionamento pode acontecer para:<br>-Acionamento de Socorro: ex. "Socorro"<br>-Solicitar comunicação com Central de Atendimento para orientações: ex. "Ligar Central"<br>-Solicitar informações sobre Saúde: ex. “Verificar Saúde” (Relatório de Saúde) |
| | Durante – Enquanto aguarda o socorro | Comunicar por Voz com o Usuário. | A Pulseira deve oferecer o recurso de acionamento do idoso pela Central de Controle para monitoramento e coleta de informações:<br>- Central de Atendimento entra em contato para avaliar ou confirmar a situação do idoso, mediante acionamento manual ou automático de socorro.<br>2.1: Se grave, Central permanece com o idoso até chegar o socorro.<br>2.2: Se não for grave, Central apenas coleta as informações necessárias, atualiza o Relatório de Saúde e envia para Rede de Apoio ou atualiza o *status* do Procedimento de Emergência. (notificação de *status* para a Rede de Apoio dentro da situação de emergência). |
| Biosensores | Antes | Medir Batimentos Cardíacos | -Será feita a coleta dos batimentos cardíacos através de sensores<br>-Detectar variações de acordo com os parâmetros cadastrados pelo idoso<br>-Esses dados serão enviados para a central. |
| | Antes | Medir Variação de Pressão | -Será feita a coleta da pressão arterial através de sensores |
| | | | - Detectar variações de acordo com os parâmetros cadastrados pelo idoso<br>-Esses dados serão enviados para a central. |
| | Antes | Identificar Queda | -Detectar alguma alteração brusca (através dos sensores) que indique uma possível queda.<br>-Avaliar possibilidade de realizar essa análise utilizando giroscópio e acelerômetro.<br>-Enviar para central um aviso de uma possível queda do idoso.<br>-Esses dados serão enviados para a central. |
| Outros Sensores | Antes | Monitoramento da postura | Classifica a postura humana básica – sentado, em pé ou deitado – bem como identifica situações que fogem a esse padrão. Pode ser usado para indicar que o idoso está muito tempo ocioso, ou confirmar uma queda, por exemplo. |
| | Antes | Monitorar fumaça | Identificar sinal de fumaça dentro de casa. |
| Ação Humana | Durante – Acionando o socorro | Detectar Entonação da Voz | -A atendente estará em comunicação direta com o idoso durante um procedimento de emergência ou monitoramento<br>-Deve analisar o estado de espírito da pessoa e em função da sua situação disparar o Procedimento de Emergência. |
| Procedimento de Emergência | Durante – Acionando o socorro | Integrar Dados para Identificar o Tipo de Problema | -A pulseira realiza um pré-processamento das informações coletados dos sensores para identificar o tipo de problema antes de notificar a Central.<br>- Ao notificar a Central esse pré-processamento e os dados coletados é enviado.<br>-A Central de Controle deve confirmar essas informações antes de acionar o procedimento de emergência.<br>-Deve ser avaliado se o pré-processamento será realizado na pulseira, no celular ou em um Arduino. |
| | Durante – Acionando o socorro | Acionar de Socorro Mediante Queda | - Uma vez a identificada a queda enviar sinal de "emergência" para central de controle notificando sobre uma possível queda do idoso.<br>-A Central de Controle deve confirmar essas informações antes de acionar o procedimento de emergência. |
| | Durante – Enquanto aguarda o socorro | Atualizar Informações – Biosensores | Durante o procedimento de emergência, a pulseira deve coletar as informações dos sensores e atualizar a Central frequentemente para que essa avalie se houve mudança na situação do idoso e alterar o procedimento, se necessário. |
| Encerramento da Emergência | Depois | Finalizar Aleta de Socorro | -Finalizar o procedimento de emergência (coleta de biosensores e informações)<br>-O desativa o alerta de socorro assim que toda a situação estiver resolvida. |

## 7.5 Lista de Funcionalidades para “Aplicativo”

Durante o *Workshop* Técnico na Etapa 3 do método IoT-PMHCS ficou definido que haverá um aplicativo no celular responsável por informar a Rede de Apoio sobre a situação de saúde do idosos e apoiar nos procedimentos de emergência.

O Aplicativo tem integração com o componente Pulseira Inteligente e Central de Controle para receber informações do idoso.

A Tabela 40 apresentam a lista de *User Stories* do componente Aplicativo organizada de acordo com a etapa da Jornada do Usuário (Antes, Durante - acionando o socorro e enquanto aguarda o socorro, Depois).

Importante deixar claro que nesse *Workshop*, dado as restrições de tempo dos voluntários, as *User Stories* e os critérios de aceite foram criados de forma simplificada. As *User Stories,* inclusive, não atendem ao formato padrão recomendado, sendo usado apenas um texto simplificado para identificação.

**Tabela 34. Aplicativo: lista de User Stories**

| Requisitos | Jornada | *User Storie* | Critérios de Aceite |
|---|---|---|---|
| Gestão de Acesso | Antes | Realizar Login | -A senha deverá ser criptografada.<br>-O login será o email ou ID de rede social do usuário.<br>-Caso o usuário não estiver previamente cadastrado no sistema ele não terá acesso ao sistema e receberá uma mensagem informando ao tentar logar.<br>-Ao se *logar* o usuário recebe informações do idoso (ou idosos) ao qual ele está relacionado.<br>- Realizar monitoramento individual ou integrado dos idosos relacionados. |
| Configurações | Antes | Cadastrar Plano de Ação - Preferências de Acionamento | -Permitir que a Rede de Apoio atualize o Plano de Ação, contendo a ordem de acionamento e procedimentos.<br>- O primeiro acionamento seria para contato de emergência informado pelo usuário<br>- Caso o contato de emergência não atenda, é feito contato com a central para avaliação da emergência e contato com socorristas |
| | Antes | Preferência da frequência de recebimento de informações | Cada usuário do APP definiria com que frequência receberia os relatórios de saúde. |
| Relatório de Saúde | Antes | Receber Relatório de Saúde Atualizado | Este relatório pode ser recebido de acordo com as preferências do familiar. Sugestões:<br>- Situação de saúde do idoso<br>- Situação medicamentos<br>- Qualidade do sono<br>- Emergências do dia etc |
| | Durante-Enquanto Aguarda o Socorro | Receber Relatório de Emergência Atualizado | Em casos de emergência, rede de apoio pode receber relatório de saúde atualizado.<br>Este relatório pode ser recebido de acordo com as preferências do familiar. Sugestões:<br>- Situação de saúde do idoso.<br>- Status do atendimento etc. |
| Procedimento de Emergência | Durante-Acionando o Socorro | Emitir Alertas Relativos ao Acidente | São sugeridos os seguintes tipos de alertas:<br>- P*ush notification* nos celulares dos contatos de emergência.<br>- Conectar ao *home inteligente* do contato de emergência (se houver) e realizar alertas (mudança de luz, aparelhos eletrônicos) para aviso da emergência.<br>- Chamar contato de emergência do celular do contato ativado se o mesmo não atender ao celular.<br>- Se ninguém atender a chamada, informar a Central de Controle para que faça uma ação emergencial. |
| | Durante-Acionando o Socorro | Enviar Informações ao Socorrista | - O contato de emergência ativado deve receber da Central de Controle um guia de primeiros socorros e plano de ação cadastrado.<br>- Esta informação será apresentada no aplicativo sempre que houver uma emergência.<br>- Serão apresentadas informações atualizadas do idoso ao socorrista:<br>- Informações de saúde coletadas dos biosensores<br>-Informações coletadas de outros sensores<br>- Localizações<br>- Outras informações do idoso devem ser exibidos<br>- Atualizar as informações a cada hora |
| | Durante-Enquanto Aguarda o Socorro | Comunicação de Voz Aberta com a Pulseira Durante Emergência | A comunicação de voz com o idoso será prioritária pela Central de Controle e essa gerencia a comunicação com outros familiares. |
| | Durante-Enquanto Aguarda o Socorro | Realizar *Chat* Coletivo | Abrir *chat* coletivo com todos os integrantes cadastrados na Rede de Apoio durante procedimento de emergência para troca de informações. |
| Tratamento de Falhas | Durante-Acionando o Socorro | Gerenciar Situações de Perda de Conexão | Realizar tratamento de falha mediante perda de conexão ou falta de contato. Sugestão:<br>- Se o aplicativo perder a conexão, notificar os outros usuários contatos de emergência da perda da pessoa<br>-Se não conseguir contato com idoso através da pulseira ou perder conexão ou não conseguir coletar dados atualizados do idoso, ativar plano de ação de emergência<br>-Se todas as pessoas perderem a conexão, ativar emergência médica/ambulância<br>-Em caso de perda de conexão com a Central, conectar diretamente aos contatos de emergência cadastrados. |
| Encerramento da Emergência | Depois | Desativar Procedimento de Emergência | -Finalizar o procedimento de emergência (coleta de biosensores e informações)<br>-O desativa o alerta de socorro assim que toda a situação estiver resolvida. |

## 7.6 Análise da Solução IoT

Para esta solução, foram identificados os seguintes cenários:

- Análise de Perigo.
- Acionamento do Socorro.
- Monitoramento Idoso Durante Emergência.
- Monitoramento de Rotina no Dia a Dia.

A “Análise de Perigo” foi apontado pela empresa IrisSenior[4], parceira do projeto, como o maior desafio que existe hoje em termos de tecnologia assistiva, pois os dispositivos atuais são reativos e não preditivos. Portanto, foi realizado um refinamento após o *Workshop* Técnico desse cenário. O cenário de “Análise de Perigo” pode ser dividido em 4 subcenários:

- Acionamento de emergência automático mediante problemas de saúde.
- Acionamento de emergência automático mediante queda.
- Acionamento de emergência manual.
- Monitoramento de Rotina do Idoso.

[4] https://irissenior.com.br/

### 7.6.1 Procedimento de Emergência – Acionamento Automático de Saúde

O acionamento de emergência pode ser realizado de forma automática, uma vez identificados problemas de saúde a partir da leitura de biosensores. Foi proposta a seguinte sequência de *User Stories*, segundo a Figura 47:

- Coletar informações de saúde (Biosensores).
- Verificar veracidade do Alarme.
- Se alarme for falso, cancela procedimento de emergência.
- Se alarme for verdadeiro:
  - Notificar Rede de Apoio.
  - Notificar Primeiro Socorrista.

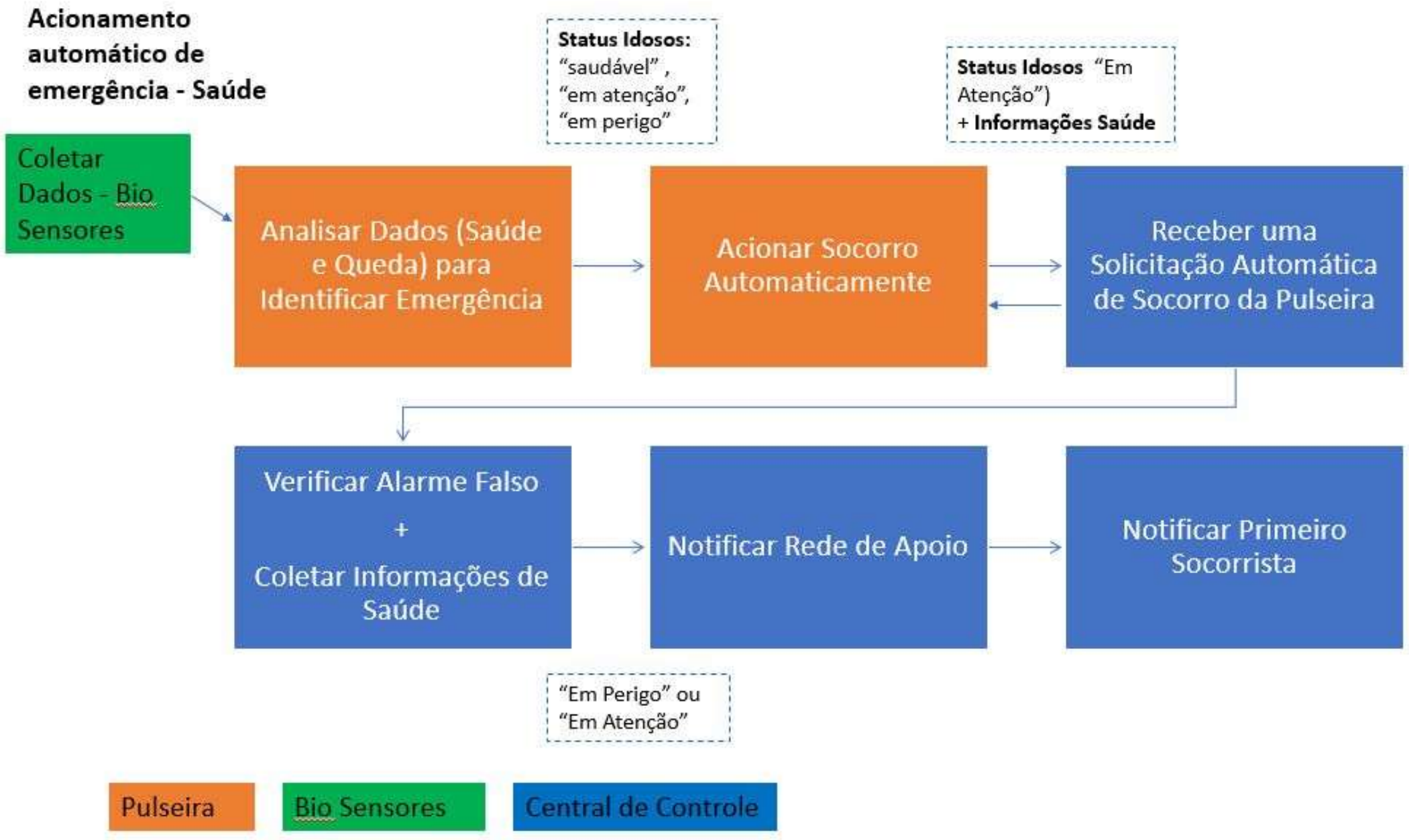

**Figura 47. Procedimento Automático de Saúde**

### 7.6.2 Procedimento de Emergência – Acionamento Automático Queda

O acionamento de emergência pode ser realizado de forma automática, uma vez identificados problemas de queda a partir da leitura de sensores. Foi proposta a seguinte sequência de *User Stories*, segundo a Figura 48:

- Coletar informações de queda.
- Verificar veracidade do Alarme.

- Se alarme for falso, cancela procedimento de emergência.
- Se alarme for verdadeiro:
    - Notificar Rede de Apoio.
    - Notificar Primeiro Socorrista.

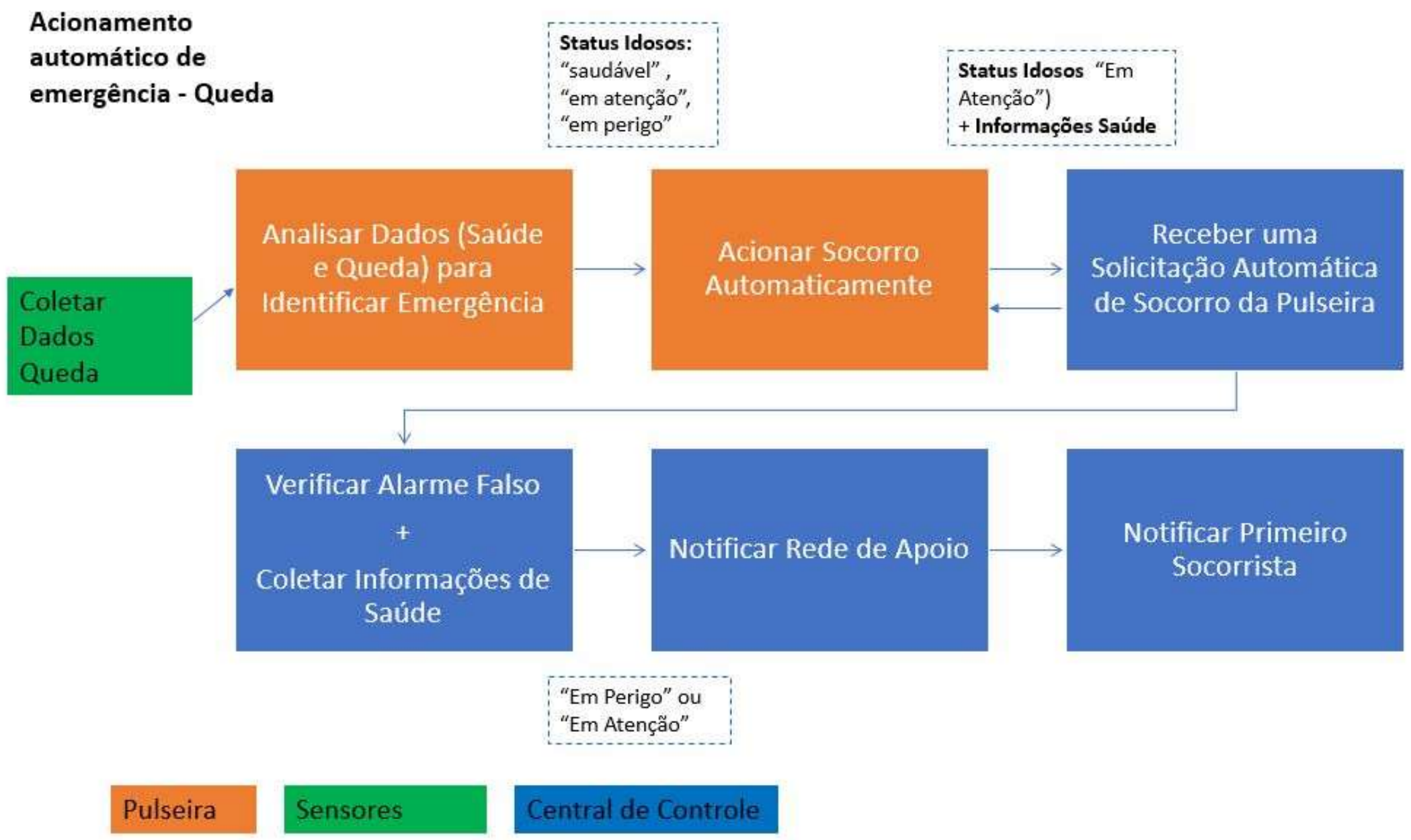

**Figura 48. Procedimento Automático de Queda**

### 7.6.3 Procedimento de Emergência – Acionamento Manual

O acionamento de emergência pode ser realizado de forma manual pelo idoso. Foi proposta a seguinte sequência de *User Stories*, segundo a Figura 49:

- Central de Controle é acionada pelo idosos:
    - Comando de voz.
    - Botão do Pânico.
- Verificar veracidade do Alarme.
- Se alarme for falso, cancela procedimento de emergência.
- Se alarme for verdadeiro:
    - Notificar Rede de Apoio.
    - Notificar Primeiro Socorrista.

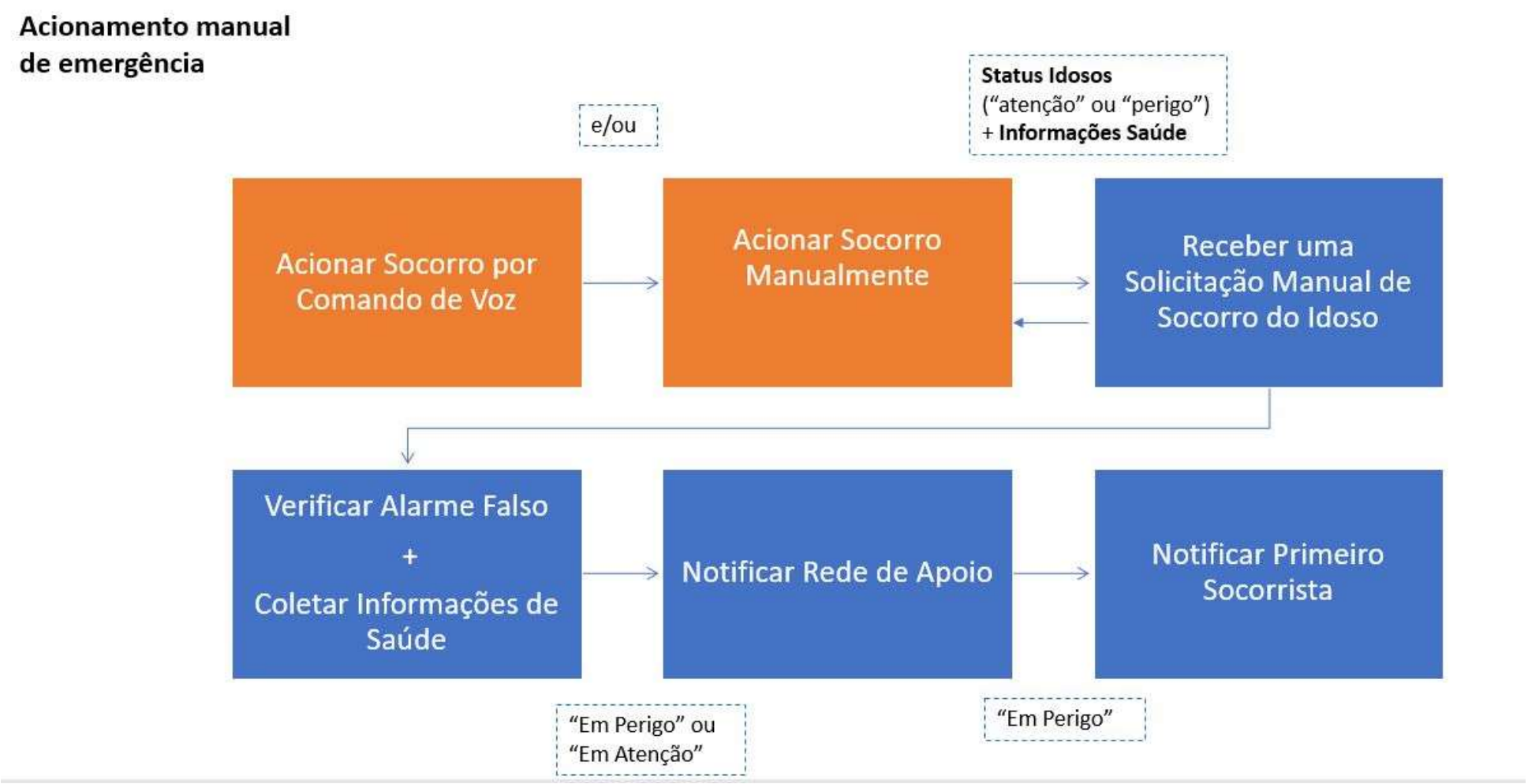

**Figura 49. Acionamento Manual**

### 7.6.4 Monitorar Rotina

Este cenário tem o objetivo de monitorar a rotina dos idosos para identificar desvios. Foi proposta a seguinte sequência de *User Stories*, segundo a Figura 50:

- Coletar informações de saúde do idoso – rotina.
- Verificar sensores de queda.
- Verificar outros sensores na casa (fumaça).
- Comparar com Dados Históricos.
- Verificar veracidade do Alarme.
- Se alarme for falso, cancela procedimento de emergência.
- Se alarme for verdadeiro:
    - Notificar Rede de Apoio.
    - Notificar Primeiro Socorrista.

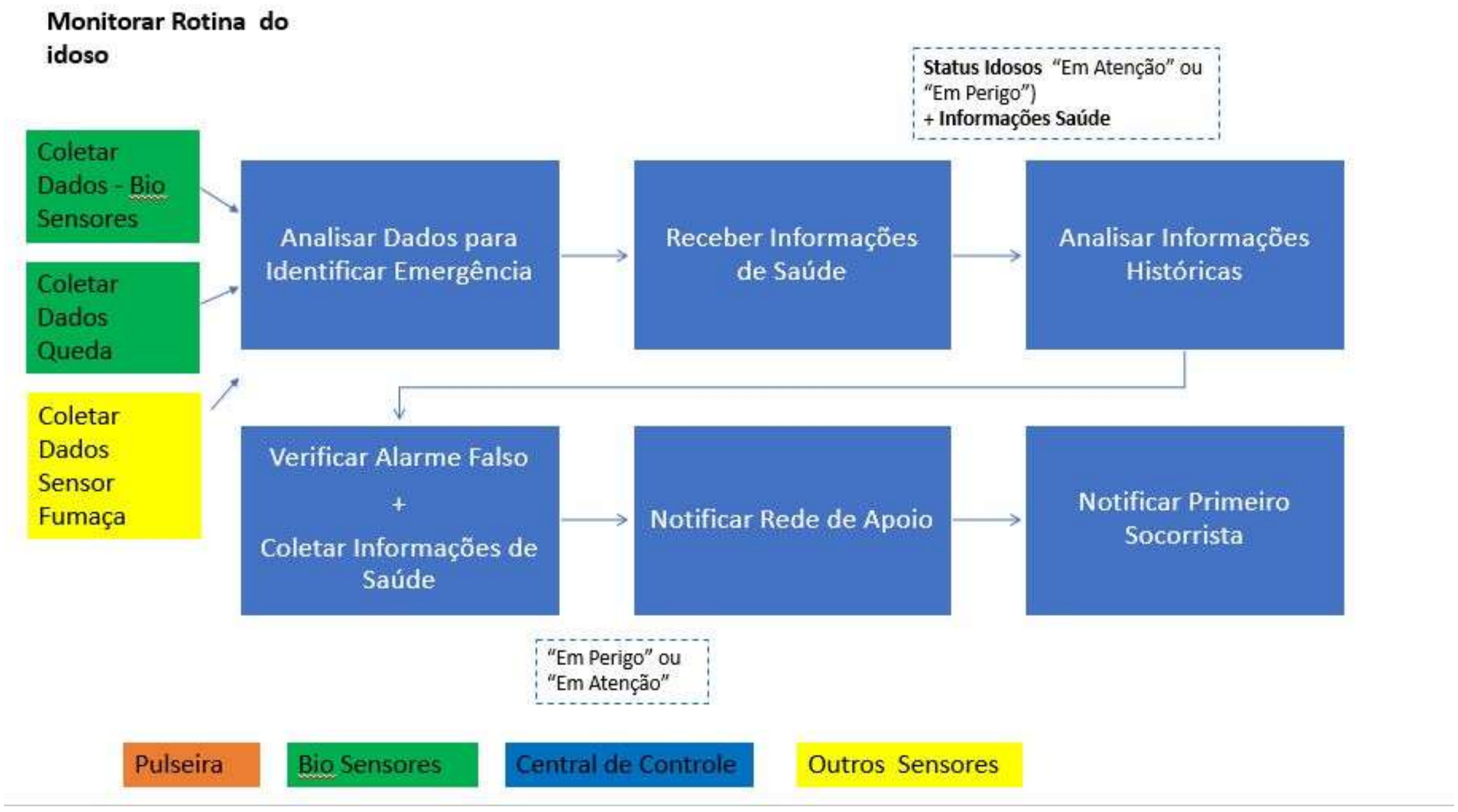

**Figura 50. Monitorar Rotina**

## 7.7 Aplicação do *Framework* IoT e Adaptações

Para apoiar o detalhamento da solução foram usadas duas ferramentas de apoio: a Jornada do Usuário e *Framework IoT*.

O cenário "Análise de Perigo" foi escolhido para utilizar uma adaptação do *Framework* IoT. Tal ferramenta foi aplicada nos subcenários de Procedimento de Emergência": Acionamento Automático de Saúde e Queda e Acionamento Manual.

### 7.7.1 Thing ou "Coisas"

Representa todos os componentes da solução, ou "coisas", que participam da solução nesse cenário, e seus requisitos não funcionais. São considerados "coisas" desse cenário:

- Central de Controle.
- Pulseira Inteligente.
- Outros Sensores.

A Tabela 35 representa os requisitos não funcionais de cada componente identificados de forma exemplificada para testar o *framework*.

**Tabela 35. Framework IoT: Coisas**

| Coisa | *Requisito não funcional* |
|---|---|
| Central de Controle | A Central de Controle será uma solução na nuvem e deve estar integrada aos demais componentes dessa solução. |
| Pulseira | A pulseira deve conter biosensores, giroscópio e acelerômetro para ser capaz de captar informações de saúde e queda do idoso. |
| | A pulseira deve ser capaz de identificar comando de voz de forma simples. |
| | A pulseira deve ter um design minimalista, simples, mas com design bonito e moderno. |
| Sensor de Fumaça | A solução deve ser composta por sensor capaz de identificar alto nível de $CO2$ em um determinado local. |

### 7.7.2 Semântica-Coisa

Descreve todas as análises que serão feitas a partir de dados coletados das "coisas". Para tanto serão realizados os seguintes passos:

- Listar as funcionalidades de cada "coisa" de acordo com a Jornada do usuário, pensando no "antes", "durante" e "depois".
- Identificar todos os cenários que compõe a solução e realizar o detalhamento usando *User Stories* para cada Cenário.
- Para cada cenário identificar como os dados de cada "coisa" são processados e suas integrações

A Tabela 36 representa os requisitos funcionais de cada componente representados na forma de *User Stories* de forma exemplificada para testar o *framework*.

**Tabela 36. Framework IoT: Semântica-Coisa**

| **Coisa** | **Funcionalidade** | ***User Stories (título)*** | ***User Stories (descrição)*** |
|---|---|---|---|
| Central de Controle | Procedimento Emergência | [CentralControle] Verificar Alarme Falso | *Como* Central de Controle<br>*Quero* confirmar informações de saúde do idoso e se existe emergência<br>*Para* acionar Procedimento de Emergência corretamente e evitar alarme falso |
| | Procedimento Emergência | [CentralControle] Coletar Informações de Saúde | *Como* Central de Controle<br>*Quero* coletar informações da Pulseira Inteligente sobre de saúde do idoso<br>*Para* confirmar Procedimento de Emergência |
| | Relatório Saúde | [CentralControle] Receber Informações de Saúde | *Como* Central de Controle<br>*Quero* receber informações da Pulseira Inteligente sobre de saúde do idoso (batimento cardíaco, pressão etc)<br>*Para* monitorar frequentemente a saúde do idoso e emitir alerta em caso de emergência |
| Pulseira | Biosensores | [Pulseira] Medir de Dados de Saúde | *Como* pulseira inteligente<br>*Quero* coletar dados de saúde (batimentos cardíacos, pressão etc) através de sensores<br>*Para* monitorar e informar a saúde do idoso |
| | Biosensores | [Pulseira] Identificar Queda | *Como* pulseira inteligente<br>*Quero* coletar através de sensores de giroscópio e acelerômetro alguma alteração brusca que indique uma possível queda.<br>*Para* monitorar e informar a saúde do idoso |
| | Procedimento Emergência | [Pulseira] Acionar Socorro por Comando de Voz | *Como* idoso<br>*Quero* acionar Pulseira Inteligente via comando de voz<br>*Para* acionar emergência na Central de Controle |
| | Procedimento Emergência | [Pulseira] Acionar Socorro Manualmente | *Como* idoso<br>*Quero* acionar Pulseira Inteligente via comando de voz<br>*Para* acionar emergência na Central de Controle |
| Sensor de Fumaça | Procedimento Emergência | Identificar Alto Nível de CO2 na Casa | *Como* Sensor de Fumaça<br>*Quero* monitorar o nível de CO2<br>*Para* notificar perigo |

### 7.7.3 Semântica

Descreve todas as funcionalidades que serão feitas dentro de cada cenário a partir de dados coletados e análises realizadas de cada "coisa". Para tanto serão realizados os seguintes passos:

- Listar as funcionalidades de cada "coisa" de acordo com a Jornada do usuário, pensando no "antes", "durante" e "depois".
- Identificar todos os cenários que compõe a solução.
- Identificar todas as funcionalidades e processamentos realizados por cada "coisa".
- Realizar o detalhamento usando *User Stories* para cada Funcionalidade, Cenário e "Coisa".

A Tabela 37 representa os requisitos funcionais de cada componente representados na forma de *User Stories* de forma exemplificada para testar o *framework*.

**Tabela 37. Framework IoT: Semântica**

| **Coisa** | **Funcionalidade** | ***User Stories (título)*** | ***User Stories (descrição)*** |
|---|---|---|---|
| Central de Controle | Relatório de Saúde | [CentralControle] Analisar Informações Históricas | *Como* Central de Controle<br>*Quero* analisar as informações de saúde do idoso (batimento cardíaco, pressão etc) comparando com seu histórico<br>*Para* monitorar saúde do idoso e emitir alerta em caso de emergência |
| | Gestão de Acesso | [Central de Controle] Realizar Acesso a Central de Controle | *Como* Usuário do sistema<br>*Quero* Realizar o acesso seguro ao sistema<br>*Para* Ter acesso as minhas funcionalidades |
| | Configurações | [Central de Controle] Configurar o Perfil do idoso | *Como* Central de Controle<br>*Quero* Realizar o cadastro das informações do idoso (dados cadastrais, plano de ação, histórico de saúde, perfil)<br>*Para* dar ao idoso uma identidade única e customizada na aplicação |
| | Procedimento Emergência | [CentralControle] Receber uma Solicitação Automática de Socorro da Pulseira | *Como* Central de Controle<br>*Quero* receber notificação de emergência automática da Pulseira Inteligente<br>*Para* identificar sua gravidade e realizar Procedimento de Emergência |
| | Procedimento Emergência | [CentralControle] Receber uma Solicitação Manual de Socorro do Idosos | *Como* Central de Controle<br>*Quero* receber notificação de emergência diretamente pelo idoso por comando de voz ou acionamento manual<br>*Para* identificar sua gravidade e executar Procedimento de Emergência |
| | Procedimento Emergência | [CentralControle] Notificar Primeiro Socorrista | *Como* Central de Controle<br>*Quero* Enviar mensagem ao socorrista<br>*Para* informar emergência do idosos |
| | Procedimento Emergência | [CentralControle] Notificar Rede de Apoio | Como Central de Controle<br>*Quero* identificar o STATUS de saúde do idoso e se houve queda<br>*Para* informar Rede de Apoio |
| | Procedimento Emergência | [CentralControle] Manter Rede de Apoio Informada | Como Central de Controle<br>*Quero* coletar o STATUS de saúde do idoso periodicamente e STATUS de emergência durante o Procedimento de<br>*Para* informar Rede de Apoio |
| Pulseira | Gestão de Acesso | [Pulseira] Realizar Acesso Personalizado a Pulseira Inteligente | *Como* usuário idoso<br>*Quero* me cadastrar e acionar a Pulseira Inteligente<br>*Para* ter acesso aos recursos de forma personalizada |
| | Procedimento Emergência | [Pulseira] Analisar Dados de Saúde para Identificar Emergência (Pré-Processamento) | *Como* pulseira inteligente<br>*Quero* integrar e analisar os dados de saúde<br>*Para* monitorar a saúde do idoso e identificar seu STATUS de emergência |
| | Procedimento Emergência | [Pulseira] Analisar Dados Queda para Identificar Emergência | *Como* pulseira inteligente<br>*Quero* integrar e analisar os dados de posição do idoso<br>*Para* monitorar a saúde do idoso e identificar seu STATUS de emergência |
| | Procedimento Emergência | [Pulseira] Analisar Dados dos Sensores para Identificar Emergência | *Como* pulseira inteligente<br>*Quero* integrar e analisar os dados de outros sensores aos dados de saúde e queda do idoso<br>*Para* monitorar o idoso e identificar seu STATUS de emergência |
| | Procedimento Emergência | [Pulseira] Acionar Socorro Automaticamente | *Como* pulseira inteligente<br>*Quero* identificar o STATUS do idoso<br>*Para* acionar emergência na Central de Controle |

### 7.7.4 Atores

Observou-se uma necessidade também de deixar claro todos os atores que participam da solução, bem como as interações com cada "coisa" e a semântica relacionada.

Portanto, sugere-se adaptação ao *Framework IoT* para representar todos os atores da solução, conforme apresentado na Figura 46. São eles:

- Idosos.
- Rede de Apoio.
- Central de Atendimento.
- Equipe de Socorro.

### 7.7.5 Restrições

Como o principal objetivo dessa pesquisa era um método de apoio a especificação funcional da solução IoT, o *Framework* IoT foi usado como método de apoio com maior foco em "*Things*" e "Semântica". Não foi realizado detalhamento de Internet, apenas realizado um levantamento preliminar.

A utilização do *Framework IoT* como referência aliada a Jornada do Usuário demonstrou-se como ferramenta importante e eficiente de apoio a especificação funcional da solução "Pulseira Inteligente".

Mesmo a especificação sendo feita por amostragem permitiu detalhamento inicial dos requisitos de forma orientada e organizada, inclusive sendo utilizada como referência inicial para o desenvolvimento do protótipo dessa solução.

Porém, tal protótipo não pode ser construído durante o tempo de projeto de pesquisa devido a indisponibilidade dos voluntários, optando-se por utilizar uma solução de mercado para os próximos passos de validação desse método.

### 7.7.6 Pesquisa de Mercado – Solução

Uma vez que não foi possível realizar a criação de um protótipo, optou-se por avaliar as soluções de mercado existentes hoje no Brasil. A Tabela 38 apresenta as soluções avaliadas no mercado brasileiro.

**Tabela 38. Soluções Avaliadas no Mercado Brasileiro**

| Empresa | Principais recursos | Apoia Pesquisa |
|---|---|---|
| IrisSenior[5] | -Botão de emergência em pulseira simples<br>-Sensor de queda em colar<br>-Sensor de fumaça<br>-Viva-voz de alta potência<br>-Central de Serviços 24h<br>-Não é uma solução IoT | Sim |
| Cuidador Digital[6] | -Botão de emergência em pulseira simples<br>-Viva-voz de alta potência | Não |

[5] https://irissenior.com.br/
[6] https://www.cuidador.digital/

| | | |
|---|---|---|
| | -Não possui Central de Serviços<br>-Não é uma solução IoT | |
| HelpCare[7] | -Botão de emergência em pulseira simples<br>-Sensor de queda em colar<br>-Sensor de fumaça<br>-Viva-voz de alta potência<br>-Central de Serviços 24h<br>-Não é uma solução IoT | Não |
| TeleHelp[8] | -Botão de emergência em pulseira simples<br>-Sensor de queda em colar<br>-Sensor de fumaça<br>-Viva-voz de alta potência<br>-Central de Serviços 24h<br>-Não é uma solução IoT | Não |
| TecnoSenior[9] | -Botão de emergência em pulseira simples<br>-Sensor de queda em colar<br>-Sensor de fumaça<br>-Viva-voz de alta potência<br>-Central de Serviços 24h<br>-Não é uma solução IoT | Não |
| Lincare[10] | -Botão de emergência em pulseira com recursos *smartwatch*<br>-Sensor de queda na pulseira<br>-Sensor de fumaça<br>-Viva-voz de alta potência<br>-Central de Serviços 24h<br>-Solução na Nuvem<br>-Não é uma solução IoT, mas é a solução que mais se aproxima da Pulseira Inteligente | Não |

Para a escolha da solução d mercado a ser usada como protótipo da solução, foram usados os seguintes critérios:

- Acionamento manual pelo idoso.
- Acionamento automático mediante queda ou biosensores.
- Central de Atendimento para centralizar suporte a idoso e familiares.
- Plano de Ação previamente cadastrado.

7 http://www.helpcarebrasil.com.br/
8 http://www.telehelp.com.br/
9 https://tecnosenior.com/
10 http://lincare.com.br/

- Acionamento automático de familiares

A solução IrisSenior é uma solução típica de mercado, que se assemelha a outras soluções assistivas existentes hoje no mercado brasileiro. Essa solução foi selecionada para este projeto de pesquisa por atender aos requisitos acima estabelecidos, e por aceitar a parceria de pesquisa de forma voluntária, através do empréstimo dos equipamentos para teste e análises por um período de 2 meses.

A Tabela 35, Tabela 36 e Tabela 37 apresentam os principais requisitos identificados através da aplicação do *Framework IoT* e priorizados para o *Workshop* de Validação, através do uso da solução de mercado IrisSenior.

## 7.8 Relatório do *Workshop* Validação

Para a realização desse *Workshop* foi selecionado um grupo de 5 idosos, residentes na cidade de Muzambinho, SP, conforme descrito no capítulo 0.

### 7.8.1 Organização do *Workshop*

A Tabela 39 descreve a organização do *Workshop*, material utilizado, participantes selecionados, duração média, local, método de registro e resultados esperados. De acordo com a proposta do método IoT-PMHCS, esta tabela também contém uma análise de eficácia do método para cada item.

**Tabela 39. Workshop de Validação: Análise de organização de atividades**

| Proposta de Organização | | Análise |
|---|---|---|
| **Material** | -P*ost-its*, canetas coloridas, cartolina, vídeos para contextualização e motivação;<br>-Material para adequação do ambiente de modo a facilitar a execução dessa atividade, tais como: *coffee break*, música ambiente e, materiais para tornar o local aconchegante etc;<br>-Recomenda-se que todo material a ser utilizado como consulta pelos grupos para a realização das atividades seja impresso. | Foi usado como material desse *Workshop*:<br>- Aparelho IrisSenior<br>- Câmera para filmar<br>- Caderno de Apoio para preparação<br>- Avaliações |
| **Participantes** | Mandatório: idosos e familiares devem ser envolvidos nessa etapa.<br>Opcional: profissionais da área de TI (engenheiros, analistas, cientistas, pesquisadores ou entusiastas de tecnologia) e da saúde.<br>Recomendação 1: a presença de um assistente para suporte e registro das atividades do | -Foram convidados 5 idosos do grupo inicial para realizar essa atividade, sendo 1 representante de cada grupo de Personas.<br>-Houve assistente para apensa uma atividade. |
| | evento.<br>Recomendação 2: sempre que possível deve ser mantido o mesmo grupo de trabalho das etapas anteriores. | |
| **Local** | Recomenda-se um local apropriado para a execução dessa atividade, contendo mesas, cadeiras e lousa.<br>Importante existir uma TV ou projetor para a apresentação de slides e vídeos.<br>O ideal seria uma sala de reunião ou sala de aula, com espaço adequado para a atividade. | Foi usado para essa atividade a casa de cada um dos voluntários, para que eles vivenciassem a simulação num ambiente real. |
| **Registro** | Registrar os principais resultados de forma textual.<br>Recomenda-se também registros com fotos e/ou vídeos para análise posterior. | O registro dessa atividade foi feito na forma de gravação de vídeo e áudio, além de anotações sobre os resultados e avaliações. |
| **Duração** | Total de 2 a 3 horas, com pausas para café. | -O tempo foi de 1 hora por voluntário, incluindo a avaliação. |
| **Resultados Esperados** | Como resultado desse *Workshop* espera-se a validação do protótipo e de todo do método IoT-PMHCS;<br>São analisadas também as "mensagens subliminares", ou seja, tudo que não foi dito explicitamente, mas se pode perceber através de comentários e observações durante o *Workshop*. | Foi possível produzir os resultados esperados para a solução priorizada e foi possível aplicar todos os métodos selecionados |

### 7.8.2 Atividades Propostas no *Workshop*

A Tabela 40 apresenta o planejamento de atividades para esse *Workshop*, incluindo atividades sugeridas e pauta.

**Tabela 40. Workshop de Validação: planejamento de atividades**

| Objetivo | Atividades Sugeridas | Tempo |
|---|---|---|
| **Contextualizar** | -Contextualizar sobre o projeto (pode ser usado um pôster ou slides)<br>-Apresentar os resultados obtidos na fase anterior | 15 min |
| **Apresentação dos participantes** | -Pedir para que cada participante se apresente<br>-Essa atividade é opcional caso os participantes sejam os mesmos de etapas anteriores. | 15 min |
| **Apresentar Informações da Solução** | -Apresentar a solução priorizada e os cenário propostos para uso e avaliação;<br>-Se for utilizada alguma solução de mercado ou pesquisa como referência, apresentar os requisitos da solução;<br>-Apresentar aos participantes as técnicas de DP a serem utilizadas<br>-Podem ser utilizados vídeos e materiais adicionais nessa etapa. | 15 min |
| ***Workshop* de Cenários e Encenação** | -Reunir grupos de 2-3 idosos, sendo que cada grupo representa uma Persona selecionada para essa pesquisa (Nair, Maria de Lourdes ou Antônio);<br>- Caso exista a participação de profissionais da saúde ou TI, esses devem ser distribuídos nos grupos;<br>-Distribuir para cada grupo um cenário relevante, ou seja, uma *User Story* detalhada sobre o uso dessa tecnologia;<br>-Cada grupo deve ler os critérios de aceite, discutir e validar o entendimento com o mediador, caso seja necessário;<br>-O grupo deve planejar a encenação de cada cenário proposto de tal forma a ficar o mais real possível, distribuindo os papéis e ações;<br>-Pode ser usado como referência a Jornada do Usuário proposta no *Workshop* de Prototipação, caso seja aplicável;<br>OBS: Durante o processo de pesquisa será a avaliado a possibilidade de usar LEGO® para análise de cenários | 60 min |
| **Sugestão: pausa de 10min** | | |
| **Encenação e Análise do Resultado** | -Cada grupo deve realizar a sua encenação, com o apoio do mediador;<br>- Ao final de cada encenação, o mediador realiza uma sessão de *Brainstorm* para coletar *feedback* da solução para aquela Persona, respondendo as seguintes perguntas:<br>-O que eu gostei e quero manter<br>-O que eu não gostei e quero mudar<br>-O que não existe e quero criar<br>-Quais as limitações essa tecnologia oferece?<br>-Quais os benefícios? | 45 min |
| **Encerramento** | -Apresentação dos resultados finais<br>-Agradecimentos<br>-Coletar feedback dos participantes sobre o método e sobre a solução prototipada | 10 min |

### 7.8.3 Análise das Atividades Realizadas

A Tabela 41 apresenta uma breve análise das atividades realizadas para se atingir o objetivo desse *Workshop.*

**Tabela 41. Workshop de Validação: análise de atividades**

| Objetivo | Análise |
|---|---|
| **Contextualizar** | - Foi elaborada uma apostila contendo todas as informações sobre esse *Workshop* e entregue antecipadamente a cada voluntário. |
| **Apresentação dos participantes** | - Esse *Workshop* foi realizado de forma individual, portanto essa atividade não foi realizada. |
| **Apresentar Informações da Solução** | - Foi elaborada uma apostila contendo todas as informações sobre esse *Workshop* e entregue antecipadamente a cada voluntário.<br>- No dia do *Workshop* foram realizadas as seguintes atividades:<br>- Breve apresentação da solução.<br>- Treinamento da solução de mercado. |
| | - Apresentação dos métodos de avaliação. |
| ***Workshop* de Cenários e Encenação** | -Essa atividade não foi realizada conforme recomendação do método, por ter sido usada uma solução de mercado e não um protótipo da solução escolhida.<br>Para a encenação foram realizadas as seguintes atividades:<br>- Apresentação do cenário propostos para o voluntário.<br>- Apresentação dos métodos de avaliação. |
| **Encenação e Análise do Resultado** | - A encenação foi feita individualmente pelos voluntários, na forma de uma simulação de situação de perigo ou de uma entrevista.<br>- Cada voluntário realizou a encenação em sua residência e essa foi gravada.<br>- Ao final da encenação, os voluntários avaliam a solução respondendo um questionário (ver Apêndices I e II).<br>- 3 voluntários foram selecionados para o uso prolongado da solução e avaliação. |
| **Encerramento** | -Foram coletados *feedbacks* dos participantes sobre o método IoT-PMHCS. |

### 7.8.4 Encenação e Análise de Resultados

Para realizar a atividade de Encenação e Análise dos Resultados foi utilizada uma sequência estruturada de atividades. Seguem as principais atividades planejadas para essa etapa:

- Cada voluntário recebe um caderno para se recordar do trabalho e experimentos já realizados, antes do experimento.
- Agendar uma reunião de 2 horas com cada voluntário: planejar uma por dia
- Realizar piloto com um voluntário e fazer adaptações necessárias.
- Os cenários devem ser realizados por pelo menos 3 voluntários, sendo um representante de cada Persona (Antônio, Nair e Maria de Lourdes).
- Se possível envolva pelo menos um familiar em cada simulação.
- Usar a Jornada do Usuário da Etapa 2 (ver seção 6.3) como referência para a escolha do cenário de encenação. Mas cada voluntário deve ser incentivado para se comportar de forma natural e mais próxima da sua realidade, podendo esse cenário ser alterado sempre que desejado.
- Preparar o ambiente para a simulação:
    - A experiência deve ser realizada na casa do voluntário.
    - Idoso deve ter sido treinado sobre uso do dispositivo e processo de simulação.
    - Todos os cadastros na IrisSenior devem ter sido realizados.

- Dispositivos IrisSenior devem ter sido configurados (pulseira ou colar e viva-voz).
- Todo o processo deve ser registrado com vídeo e fotos para posterior transcrição.
- Após realizada a encenação, solicitar ao idosos que responda um questionário contendo as seguintes perguntas:
    - O que eu gostei e quero manter.
    - O que eu não gostei e quero mudar.
    - O que não existe e quero criar.
    - Quais as limitações essa tecnologia oferece?
    - Quais os benefícios?

## 7.9 Cenários de Simulação

Para realizar a simulação, foi considerado um cenário diferente para cada voluntário, organizado conforme a Tabela 42.

**Tabela 42. Organização das atividades de simulação**

| Nome | Idade | Resistência à Tecnologia | Facilidade com Tecnologia | Grupo Persona | Tipo de Cenário | Descrição | Uso Prolongado |
|---|---|---|---|---|---|---|---|
| Elizabete (Beta) | 68 | Não | Não | Maria de Lourdes | TO BE | -Simulação parcial.<br>-Acionamento automático por comando de voz.<br>-Cenário: problema de saúde e queda.<br>-Atendimento com sucesso. | Sim |
| Lúcia | 72 | Não | Sim | Maria de Lourdes | AS IS | -Simulação parcial.<br>-Acionamento manual.<br>-Cenário: queda.<br>-Atendimento com falha. | Não |
| Luiz | 67 | Não | Sim | Antônio | AS IS | -Simulação parcial.<br>-Acionamento manual.<br>-Cenário: problema de Saúde<br>-Atendimento com sucesso. | Sim |
| Maria José (Zezé) | 69 | Sim | Não | Nair | TO BE | -Simulação total.<br>-Acionamento automático por comando de voz.<br>-Cenário: problema de Saúde e fumaça em casa<br>-Atendimento moderado. | Não |
| Elga | 64 | Não | Sim | Nair | AS IS | -Simulação parcial. | Sim |
| | | | | | | -Acionamento automático por sensor de queda.<br>-Atendimento com falha. | |

Considera-se como tipo de cenário "AS IS" todas as encenações com as características atuais do equipamento da IrisSenior. Nesse cenário, a simulação acontece de forma parcial, ou seja, o voluntário simula uma situação de perigo, aciona o dispositivo e a Central IrisSenior realiza o atendimento como em uma situação real.

O cenário "TO BE" considera a existência de características adicionais ao equipamento IrisSenior, como por exemplo acionamento por comando de voz, existência de um *smartwatch*, biosensores e análise integrada de sensores. A simulação nesse caso pode ser de forma parcial, quando o voluntário simula uma situação de perigo e a Central IrisSenior é acionada, ou de forma integral, quando todo o cenário é simulado, sem a presença da Central Iris.

O critério de distribuição de cenários por voluntário não foi aleatório: ficou definido que todos os voluntários com baixa familiaridade com tecnologia iriam executar um cenário do tipo "TO BE", para que pudessem vivenciar a experiência de uso mais completa e ubíqua.

A Tabela 43 apresenta as *User Stories* selecionadas para simulação no Workshop de validação, bem como o tipo de cenário "AS IS" e/ou "TO BE".

**Tabela 43. Requisitos Priorizados para Workshop de Validação por cenário**

| Dispositivo | Funcionalidade | *User Stories* | AS IS | TO BE |
|---|---|---|---|---|
| Central de Controle | Procedimento Emergência | Verificar Alarme Falso | X | |
| | Procedimento Emergência | Coletar Informações de Saúde | | X |
| | Relatório Saúde | Receber Informações de Saúde | | X |
| | Relatório de Saúde | Analisar Informações Históricas | X | X |
| | Gestão de Acesso | Realizar Acesso a Central de Controle | X | X |
| | Configurações | Configurar o Perfil do idoso | X | X |
| | Procedimento Emergência | Receber uma Solicitação Automática de Socorro da Pulseira | X | X |
| | Procedimento Emergência | Receber uma Solicitação Manual de Socorro do Idosos | X | X |
| | Procedimento Emergência | Notificar Primeiro Socorrista | X | X |
| | Procedimento Emergência | Notificar Rede de Apoio | X | |
| | Procedimento Emergência | Manter Rede de Apoio Informada | | X |
| Pulseira | Biosensores | Medir de Dados de Saúde | | X |
| | Biosensores | Identificar Queda | X | X |
| | Procedimento Emergência | Acionar Socorro por Comando de Voz | | X |
| | Procedimento Emergência | Acionar Socorro Manualmente | X | |
| | Gestão de Acesso | Realizar Acesso Personalizado a Pulseira Inteligente | | X |
| | Procedimento Emergência | Analisar Dados de Saúde para Identificar Emergência (Pré-Processamento) | | X |
| | Procedimento Emergência | Analisar Dados Queda para Identificar Emergência | X | X |
| | Procedimento Emergência | Analisar Dados dos Sensores para Identificar Emergência | | X |
| | Procedimento Emergência | Acionar Socorro Automaticamente | X | X |
| Sensor de Fumaça | Procedimento Emergência | Identificar Alto Nível de $CO_2$ na Casa | | X |

## 7.10 Análise de Acessibilidade – Uso Prolongado

Para realizar essa atividade foram selecionados 3 voluntários, sendo um de cada grupo conforme apresenta a Tabela 42. Seguem as atividades planejadas:

- Duração proposta: **15 dias**
- Explicar para os usuários o objetivo da análise.
- Explicar o funcionamento do dispositivo.
- Voluntário deverá realizar uso prolongado do dispositivo dentro do tempo especificado.
- Ao final do período será realizada análise de Uso Prolongado, de forma orientada pelo pesquisador respondendo às perguntas:
    - O que eu gostei e quero manter.
    - O que eu não gostei e quero mudar.
    - O que não existe e quero criar.
    - Quais as limitações essa tecnologia oferece?
    - Quais os benefícios?

## 7.11 Resultados e Análises

Após realizada a atividade de simulação, foram coletados *feedbacks* dos voluntários sobre a experiência simulada. Observou-se resultados e relatos positivos com relação a facilidade de uso da solução *Home Care*. Isso se deve principalmente ao fato da solução IrisSenior ser por essência simples, fácil de usar, não causar ansiedade, e possuir um plano de ação e atendimento adequados. Não deve ser também desconsiderado o treinamento e orientação aos usuários antes do uso da solução como um fator fundamental para a percepção de facilidade de uso.

Com relação a utilidade percebida e motivação de uso de uma solução *Home Care*, os usuários apresentaram-se motivados ao uso principalmente após vivenciar uma situação de perigo, mesmo que simulada. Porém, essa motivação diminui de forma significativa o uso do equipamento por um período maior. Nesse cenário, os usuários ficam desmotivados ou indiferentes ao uso, e simplesmente abandonam a solução após 2 dias em média. Isso ocorre, pois, não há percepção de utilidade ou valorização do design do dispositivo (como é o caso dos grupos Antônio e Maria de Lourdes), ou ainda pois a resistência a tecnologia é mantida (como no caso do grupo Nair).

Acredita-se que isso ocorra principalmente devido a faixa etária do público em análise (entre 60 a 70 anos), suas condições de saúde ainda permitirem autonomia em casa e fora dela, e sua relação com tecnologia ser moderada. Para maior estímulo ao uso de uma solução *Home Care*, seria necessário oferecer funcionalidades de ganho indireto (como um relógio, alarme etc), sem perder a facilidade e simplicidade de uso, tornando o dispositivo mais atrativo ao uso diário e com um design mais motivador.

No caso de usuários resistentes, como é o grupo Nair, a sugestão é que a solução seja a mais ubíqua e simples possível, sem a necessidade de interação com os idosos. Nesse caso, fica claro durante todo o processo que para usuários mais resistentes desse grupo a aceitação ao uso só seria possível mediante recomendação médica ou dos familiares. Caso contrário, a solução não será aceita.

A Tabela 44 apresenta uma proposta de melhorias no Dispositivo IrisSenior dentro das análises realizadas após a simulação. A análise está organizada por grupo de Persona, considerando seu valor principal, a aceitação de tecnologia e se o requisito foi identificado para cada uma das soluções – IrisSenior e a solução IoT Pulseira Inteligente.

**Tabela 44 Análise por Dispositivo, considerando se atende ou não ao requisito solicitado após *Workshop* de Validação**

| Grupo | Valor principal | Aceita TI | Proposta | IrisSenior | Pulseira Inteligente |
|---|---|---|---|---|---|
| Todos | NA | NA | **Integração**: Dispositivo deve possuir acesso a Central integrado, para que ele possa ser usado em qualquer lugar, dentro e fora de casa; | Não | Sim |
| Todos | NA | NA | **Simplicidade**: Minimalista, prezar pela simplicidade. Se os recursos adicionais dificultarem o uso, será descartado; | Sim | Não |
| Todos | NA | NA | **Treinamento**: garantir treinamento do dispositivo e processo aos usuários; | Sim | Não |
| Todos | NA | NA | **Monitoramento**: realizar monitoramento do idosos aos familiares; | Não | Sim |
| Todos | NA | NA | **Plano de Ação**: permitir criação previa de um plano de ação para cada tipo de Persona e com atualização rápida pela Rede de Apoio | Parcial | Sim |
| Todos | NA | NA | **Customização:** permitir a customização da solução por tipo de usuário | Não | Sim |
| Antônio e Maria Lourdes | Independência | Médio a Alto | **Engajamento**: Oferecer mais recursos de benefício direto aos idosos para estimular o uso tais como relógio, despertador, entre outros; | Não | Sim |
| Antônio e Maria Lourdes | Independência | Médio a Alto | **Design**: prezar pelo design e oferecer opções de cores e tamanhos; | Não | Parcial |
| Antônio e Maria Lourdes | Independência | Médio a Alto | **Biosensores**: Dispositivo deve permitir medições que possam trazer maior controle aos idosos de sua saúde (batimentos cardíacos, pressão arterial, nível de ansiedade, queda etc). | Apenas queda | Sim |
| Antônio e Maria Lourdes | Independência | Médio a Alto | **Privacidade**: preservar a privacidade dos idosos, respeitando seu valor principal. Evitar câmeras e acessórios invasivos. | Sim | Sim |
| Nair | Fazer Parte | Baixo a Médio | **Ubíqua:** ser uma solução a mais ubíqua possível e integrada. | Parcial | Sim |
| Nair | Fazer Parte | Baixo a Médio | **Biosensores:** Dispositivo deve permitir medições que possam trazer maior controle aos familiares da saúde do idoso (batimentos cardíacos, pressão arterial, nível de ansiedade, queda etc). Idosos não deve ter acesso; | Parcial | Sim |
| Nair | Fazer Parte | Baixo a Médio | **Privacidade**: preservar a privacidade dos idosos, respeitando seu valor principal. Evitar câmeras e acessórios invasivos; | Sim | Sim |
| Nair | Fazer Parte | Baixo a Médio | **Central**: oferecer um atendimento personalizado e diferenciado a esses usuários, uma vez que seu valor principal é "Fazer parte". | Sim | Sim |

A solução Pulseira Inteligente atende aproximadamente a 80% dos requisitos identificados, enquanto que a solução Iris Senior atende a aproximadamente 50%. Isto demonstra que adaptações são necessárias na solução IrisSenior, mas não seria a reconstrução de uma nova solução e sim adaptações e ampliação em tecnologia e processos.

Além disso, a escolha de uma solução de mercado semelhante a solução especificada pelo método apoiou o processo de avaliação, mas não substitui a criação de um protótipo. Ela foi importante para uma coleta preliminar de informações sobre a solução antes de maiores investimentos para prototipação.

Acredita-se que com essa análise e resultados seja possível iniciar a prototipação da solução "Pulseira Inteligente" de forma mais assertiva e estruturada. Com isso, as etapas futuras teriam menos chance de retrabalho, uma vez que a solução já foi previamente avaliada por amostragem pelos futuros usuários. Desta forma, aumenta-se a probabilidade de ser uma solução melhor aceita pelos usuários dentro do perfil de Persona analisado, identificando inclusive pontos de customização e adaptações pertinentes a uma solução de mercado já disponível.

# 8 Conclusão

O estudo de caso do método IoT-PHMCS foi realizado com sucesso e todas as etapas do método foram cumpridas, entretanto foram necessárias adaptações.

Foram identificados os seguintes pontos de destaque no método IoT-PMHCS, adaptados durante a execução do estudo de caso, e que devem ser evidenciados e mantidos:

- Preparação preliminar de cada *Workshop*, com a reserva do local e material a ser utilizado;
- Sempre que possível manter o grupo de voluntários em todo o processo: isso aumenta a participação e entrosamento do grupo, com melhores resultados e redução do tempo de execução do processo;
- Uso de material lúdico durante os processos participativos traz maior engajamento e envolvimento dos idosos;
- Ter uma apostila impressa com um resumo das atividades a serem realizadas;
- Sempre que possível utilizar material de apoio como vídeos e material ilustrativos, pois enriquecem o trabalho e apoiam a materialização de conceitos inovadores, como por exemplo IoT e AAL;
- A forma de conduzir as atividades com o grupo de idosos deve ser mais lenta, com explicações mais detalhadas, para melhor compreensão.
- Não explicar todos os exercícios no início e sim na medida em que forem necessários;
- Realizar apresentações parciais dos resultados sempre que possível;
- Reforçar ao final de cada Etapa os resultados obtidos e os próximos passos;

- Se houver assistente, todas as atividades devem ser detalhadamente explicadas para uma condução adequada do trabalho;
- São recomendadas pequenas pausa durante o processo, não ultrapassando o tempo de 90 minutos consecutivos de atividades com os idosos durante o *Workshop*;
- A utilização das Personas faz com os voluntários se sintam mais à vontade ao falar sobre situações de perigo e problemas indesejados, bem como preferências de interação com a tecnologia, sem se darem conta muitas vezes de que estavam falando deles próprios e não de outras pessoas;
- Sempre que possível devem ser utilizadas equipes multidisciplinares em um mesmo processo participativo. Porém, quando não for viável ou possível, sugere-se a presença de artefatos que apoiem na comunicação entre os grupos, como foi o caso do canvas Mapa de Personas. Ele permitiu que o processo participativo através da Co-Criação fosse realizado à distância e em momentos diferentes entre idosos e profissionais de tecnologia da informação, sem trazer prejuízos para o estudo de caso;
- A materialização da solução através do uso simulado apresentou-se muito rica, pois trouxe a vivência de uma solução idealizada. Isso permite a criação de um processo iterativo, executado em etapas e com melhoria progressiva devido a coleta de *feedbacks* parciais;
- O processo IoT-PMHCS deve ser adaptado sempre que necessário para atender as necessidades e restrições do grupo em análise: tempos, artefatos e atividades podem ser adaptadas, desde que seus impactos sejam mensurados e não comprometam a execução do método e resultados;
- Quadro de Normas da Semiótica Organizacional foi fundamental para o processo de escolha da tecnologia e das preferências de cada Persona. A combinação do canvas de Mapa de Persona e o Quadro de Normas demonstrou-se como ferramenta poderosa, pois fez com os voluntários diminuíssem as barreiras contra a tecnologia e exercessem a empatia através das Personas. Pensar antecipadamente na tecnologia antes dela ser apresentada e escolhida foi primordial, o que poderia gerar resistência se fosse ao contrário;

Foram identificados os seguintes pontos de melhoria e limitações no método IoT-PMHCS, para pesquisas futuras:

- Não foi possível realizar processos participativos envolvendo todos os grupos de voluntários devido as restrições de local e limitações pessoais;
- Não foi possível materializar a solução *Home Care* através de um protótipo mais próximo a solução idealizada;
- Não foi possível maior envolvimento de profissional da saúde durante todo o processo, devido a limitações pessoais;
- Não foi possível coletar *feedbacks* dos usuários em todos os processos participativos, por limitação de tempo dos participantes;
- Não foi possível filmar todos os processos participativos por limitações dos participantes;
- Não foi possível realizar o trabalho com as 5 Personas segundo proposta de Gonçalves e Bonacin (2017), sendo selecionadas no escopo desse trabalho apenas 3 Personas;
- Não foi possível contar com a ajuda de um assistente em todos os processos, o que teria trazido maior apoio e suporte nas atividades;

- Com relação ao tempo de duração das atividades, a sugestão é não ultrapassar 2 horas por sessão de processo participativo. Quando a atividade for maior do que isso, quebrar em tarefas menores e realizadas em dias diferentes. Isto pois observou-se que os idosos após 2 horas de atividades ficam cansados e não produzem mais, reduzindo o engajamento;
- Evitar usar termos técnicos durante as atividades participativas. Por exemplo, para a execução do Workshop de Prototipação foi importante a contextualização dos voluntários sobre o termo "protótipo", que não era de conhecimento dos presentes. Ao se introduzir um novo termo, sempre apresentar exemplos, vídeos, usar analogias;
- O *Workshop* Técnico deve ser dividido em 2 ou mais etapas de detalhamento, de forma iterativa e interativa;

- A escolha de uma solução de mercado semelhante a solução especificada pelo método apoiou o processo de avaliação, mas não substitui a criação de um protótipo. Ela foi importante para uma coleta preliminar de informações sobre a solução antes de maiores investimentos para prototipação.

No Apêndice VI –Registro das Atividades Participativas pode ser visualizado o registro das atividades realizadas, dos voluntários e de alguns artefatos gerados que tanto enriqueceram este projeto de pesquisa.

Após esse estudo de caso, todas as adaptações pertinentes e possíveis foram realizadas na proposta do Método IoT-PMHCS. A versão final encontra-se no documento de Dissertação de Mestrado (Podestá, Bonacin e Gonçalves, 2018).

## 9 Referências

## Apêndice I – Etapa 0: Estudo Exploratório e Planejamento

Esta etapa tem por objetivo entender o problema de design, realizando uma imersão sobre os grandes temas que norteiam todo o trabalho a ser feito, além de realizar um plano macro de atividades de DP a serem executadas.

A Figura 51 apresenta a proposta inicial das atividades a serem realizadas na Etapa 0 do Método IoT-PMHCS. Tais atividades foram revisadas e adaptadas após aplicação nesse estudo de caso. A versão final encontra-se no documento de Dissertação de Mestrado (Podestá, Bonacin e Gonçalves, 2018).

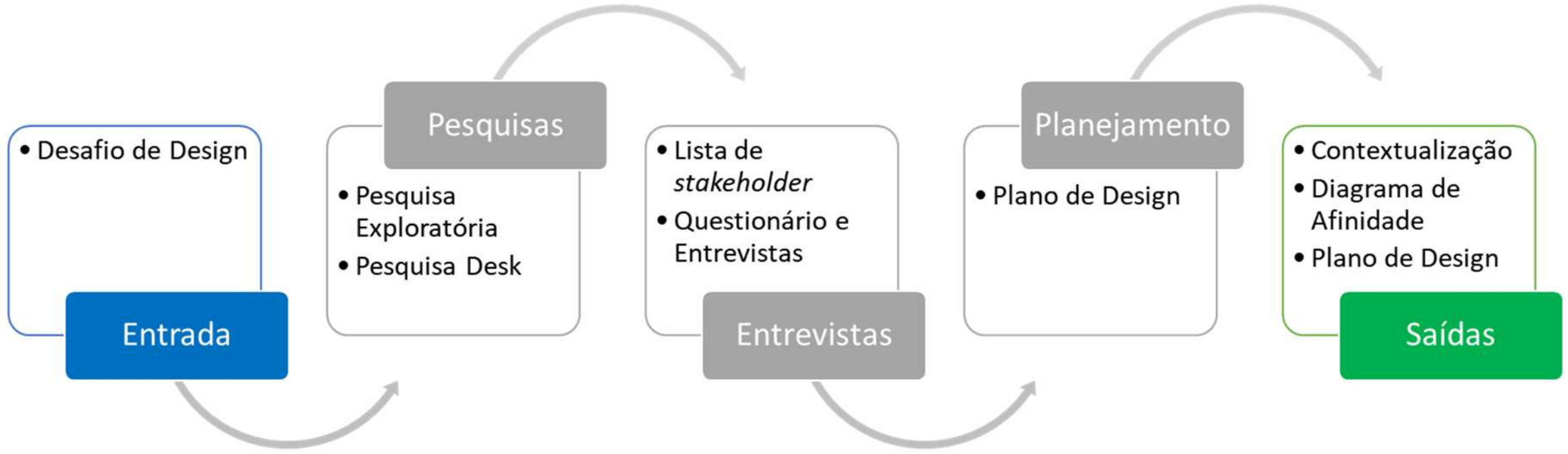

**Figura 51. Atividades da Etapa 0 - Estudo Exploratório e Planejamento**

## Apêndice II – Etapa 1: Mapeamento de Personas, Valores e Necessidades

Nessa etapa deve-se identificar valores e desafios, ou necessidades, dos *stakeholders* principais (nesse caso os idosos) por meio de atividades de DP, mapear as Personas a serem utilizadas, desenvolver o Mapa de Persona inicial e identificar os Critérios Norteadores.

A Figura 52 apresenta a proposta inicial das atividades a serem realizadas na Etapa 1 do Método IoT-PMHCS. Tais atividades foram revisadas e adaptadas após aplicação nesse estudo de caso. A versão final encontra-se no documento de Dissertação de Mestrado (Podestá, Bonacin e Gonçalves, 2018).

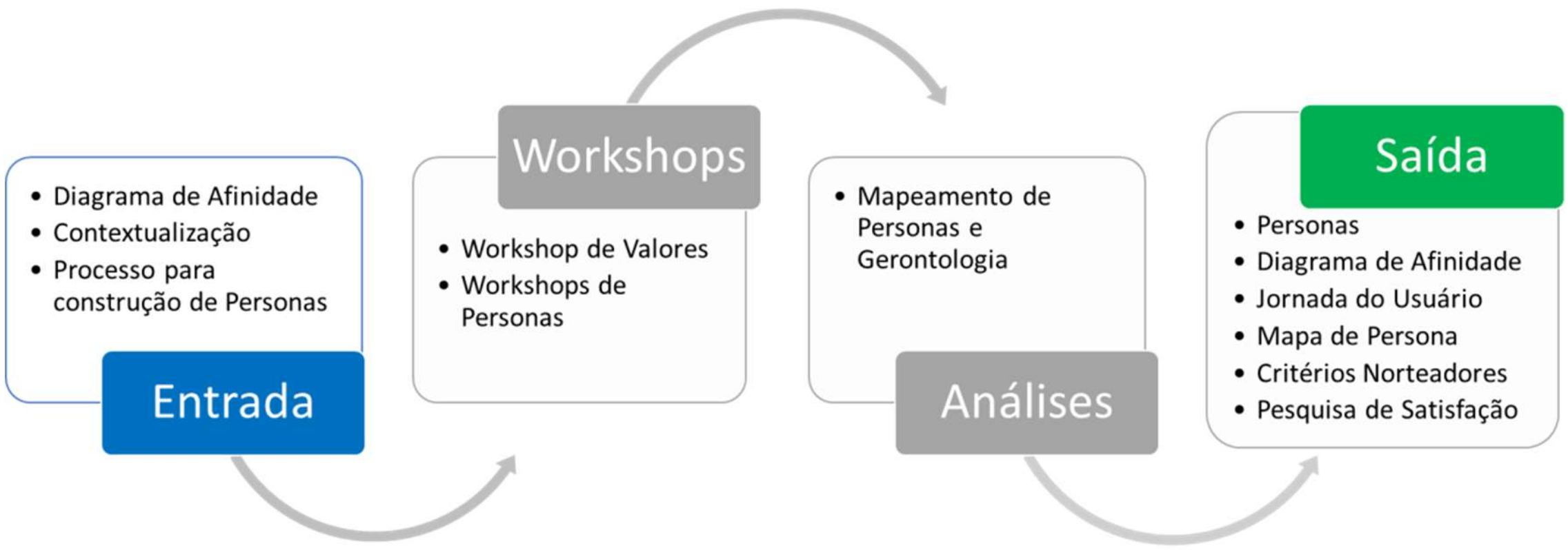

**Figura 52. Atividades da Etapa 1: Mapeamento de Personas, Valores e Necessidades**

## Apêndice III – Etapa 2: Mapeamento de Soluções IoT

Uma vez mapeadas as Personas, seus valores e diferentes necessidades, esta etapa tem por objetivo avançar no IoT-PMHCS para identificar alternativas de tecnologias baseada no conceito de IoT para atender as mais diferentes necessidades de cada uma das *Personas*.

A Figura 53 apresenta a proposta inicial das atividades a serem realizadas na etapa 2 do Método IoT-PMHCS. Tais atividades foram revisadas e adaptadas após aplicação nesse estudo de caso. A versão final encontra-se no documento de Dissertação de Mestrado (Podestá, Bonacin e Gonçalves, 2018).

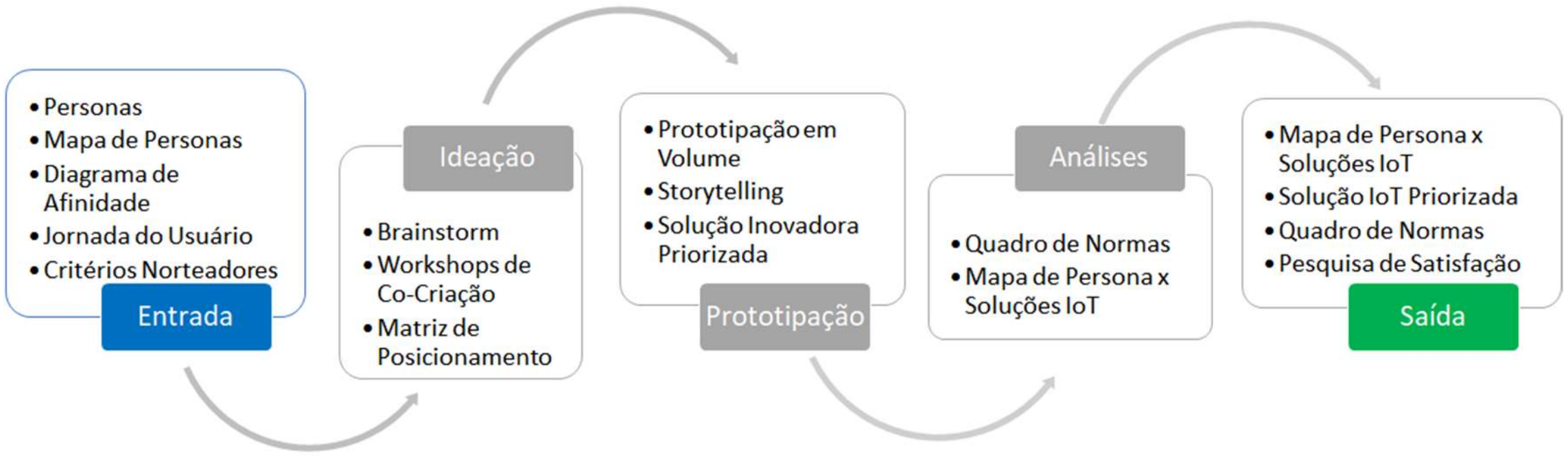

**Figura 53. Atividades da Etapa 2 - Mapeamento de Soluções IoT por Personas**

## Apêndice IV – Etapa 3: Design IoT

Finalmente, a última etapa do IoT-PMHCS tem por objetivo realizar o design de uma solução IoT identificada na Etapa 2, visando entregar solução de alto valor para os *stakeholders* e aprofundar na sua concepção na forma de requisitos funcionais.

A Figura 54 apresenta a proposta inicial das atividades a serem realizadas na Etapa 3 do Método IoT-PMHCS. Tais atividades foram revisadas e adaptadas após aplicação nesse estudo de caso. A versão final encontra-se no documento de Dissertação de Mestrado (Podestá, Bonacin e Gonçalves, 2018).

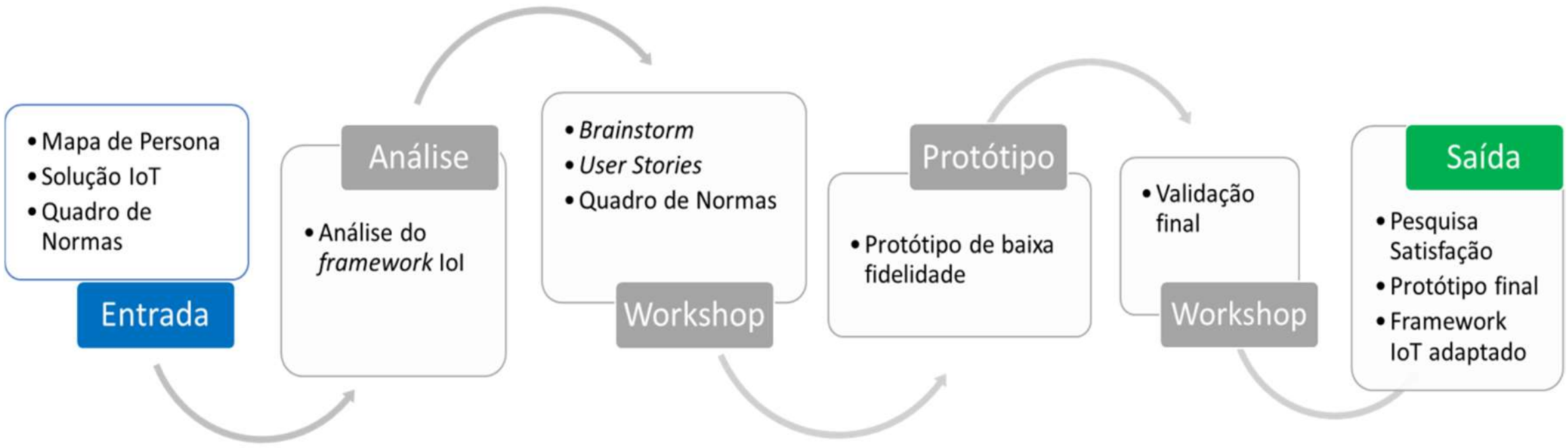

**Figura 54. Atividades da Etapa 3 – Design de Soluções IoT**

## Apêndice V – Diagrama de Afinidades de Soluções AAL Propostas

A partir do questionário e entrevistas foi possível criar o Diagrama de Afinidades dos desafios em casa segundo a Tabela 45, que será usado como referência nas próximas etapas do método IoT-PMHCS.

**Tabela 45. Etapa 0: Diagrama de Afinidade de Soluções AAL**

| Tipo Tecnologia | Categoria AAL | Sugestão de IoT | Categoria Solução |
|---|---|---|---|
| Sensores | Segurança | Sensor que pode identificar um estranho dentro de casa e acionar a polícia; | Reconhecimento facial |
| Sensores | Segurança | Reconhecimento de face ao entrar em casa, e acesso a base da polícia para identificar se é um ladrão já cadastrado; | Reconhecimento facial |
| Câmera | Segurança | Câmera para monitorar cuidadores, para evitar maus tratos; | Câmera de monitoramento |
| Sensores | Segurança | Acionamento de alarmes e câmeras por comando de voz | Sensor voz |
| Sensores | Facilitação atividades diárias | Sensores em determinados itens (chaves, celular, óculos, carteira, bolsa) com acionamento por voz ou acionamento por celular. Emite alerta no objeto esquecido para facilitar localização. Ex: "Localizar Celular", "Localizar Óculos" etc; | Sensor objetos |
| Sensores | Facilitação atividades diárias | Acionamento de energia e eletroeletrônicos por comando de voz. | Sensor voz |
| Sensores | Assistência mobilidade | Sensores luminosos nas escadas com acionamento por voz ou presença; | Sensor luminoso e voz |
| Sensores | Assistência mobilidade | Alarme monitorado com câmera em casa e comando de voz para acionar socorro em caso de queda | Sensor voz |
| Sensores | Assistência mobilidade | O sistema de alarme em casa poderia identificar acidentes em casa e enviar uma mensagem de alerta por celular de familiares; | Sensor de Perigo /Queda |
| Sensores | Assistência mobilidade | Em caso de falta de energia, sensor aciona energia emergencial em toda a casa. O acionamento também poderia ser por voz. | Sensor luminoso |
| Wearable | Cuidados de | Pulseira para análise de sintomas de perigo (pulsação | Pulseira |
| | saúde | acelerada, pressão arterial alta ou alta): emissão de alerta via mensagem para o responsável ou grupo de familiares.<br>Acionamento do médico ou plano de saúde em caso de perigo. Nesse caso, o plano de saúde entra em contato com o cliente para verificar se existe algum problema de saúde;<br>Histórico de saúde na nuvem, para integrar com a pulseira. Ajuda no registro de histórico e no resgate de informações para análise de riscos e problemas de saúde | Análise Saúde |
| Sensores | Segurança | Sensores para monitorar sinais de perigo com envio de alertas por celular | Sensor de Segurança |
| Câmeras | Segurança | Câmeras com monitoramento pelo celular por familiares | Câmera de monitorament o |
| Sensores | Assistência mobilidade | Iluminação automática com sensores para evitar quedas | Sensor luminoso |
| Sensores | Assistência mobilidade | Sensores com acionamento por voz no banheiro, quarto e próximo a escadas | Sensor voz |
| Wearable | Cuidados de saúde | Pulseira, quando aceitas pelo idosos, podem ajudar a monitorar temperatura, frequência cardíaca, pressão arterial, frequência respiratória. Emissão de alertas de perigo e de socorro para médicos e familiares. Pode ser combinada com controle de medições, para emitir alertas sobre horários dos remédios e alertas para familiares caso a medicação não seja tomada. Permite acionamento de socorro por comando de voz em caso de perigo, quedas, dores etc; | Pulseira Análise Saúde |
| Sensores | Assistência mobilidade | Sensor que identifica queda por som e dispara um SMS ou WhatsApp de alerta; | Sensor de Perigo /Queda |
| Sensores | Assistência mobilidade | Dispositivo que mediante movimento brusco de queda do aparelho celular notifica familiares; | Sensor de Perigo /Queda |
| Sensores | Segurança | Sistema interligado nas casas com sistema de emergência, caso ocorra algum problema, um alerta seja acionado na central; | Sensor de Perigo /Queda |
| Wearable | Cuidados de saúde | Pulseira que ao detectar perigo automaticamente filmaria o ambiente e a pessoa que está usando-a e aciona familiares e ou médico responsável; | Pulseira Análise Saúde |
| Wearable | Cuidados de saúde | Pulseira de comunicação, que pode ser acionada para chamar socorro com um clique ou aperto; | Botão de Emergência |
| Wearable | Cuidados de saúde | Pulseira que mostre todos os sinais vitais do paciente direto ao hospital com endereço; | Pulseira Análise Saúde |
| Aplicativo | Cuidados de saúde | Cadastramento dos idosos da cidade, com doenças, medicamentos e particularidades cadastradas. Aplicativo no celular para detectar sobre perigos (queda, pressão, diabetes etc) e avisa órgão, vizinho ou familiar sobre emergência; | Aplicativo |
| Sensores | Cuidados de saúde | Ambiente de monitoramento que possa identificar perigo e disparar alerta; | Sensor de Perigo /Queda |
| Sensores | Cuidados de saúde | Câmeras e sensores de voz que acionem algum hospital; | Sensor de Perigo /Queda |
| Drone e Wearable | Cuidados de saúde | Drone acionado pelo próprio usuário em situações de perigo, através de uma pulseira com botão de pane, conectado à uma Central e alerta pessoas cadastradas; | Drone e Botão de Emergência |

# Apêndice VI –Registro das Atividades Participativas

Todas as atividades participativas tiveram registros. Algumas fotos podem ser visualizadas a seguir:

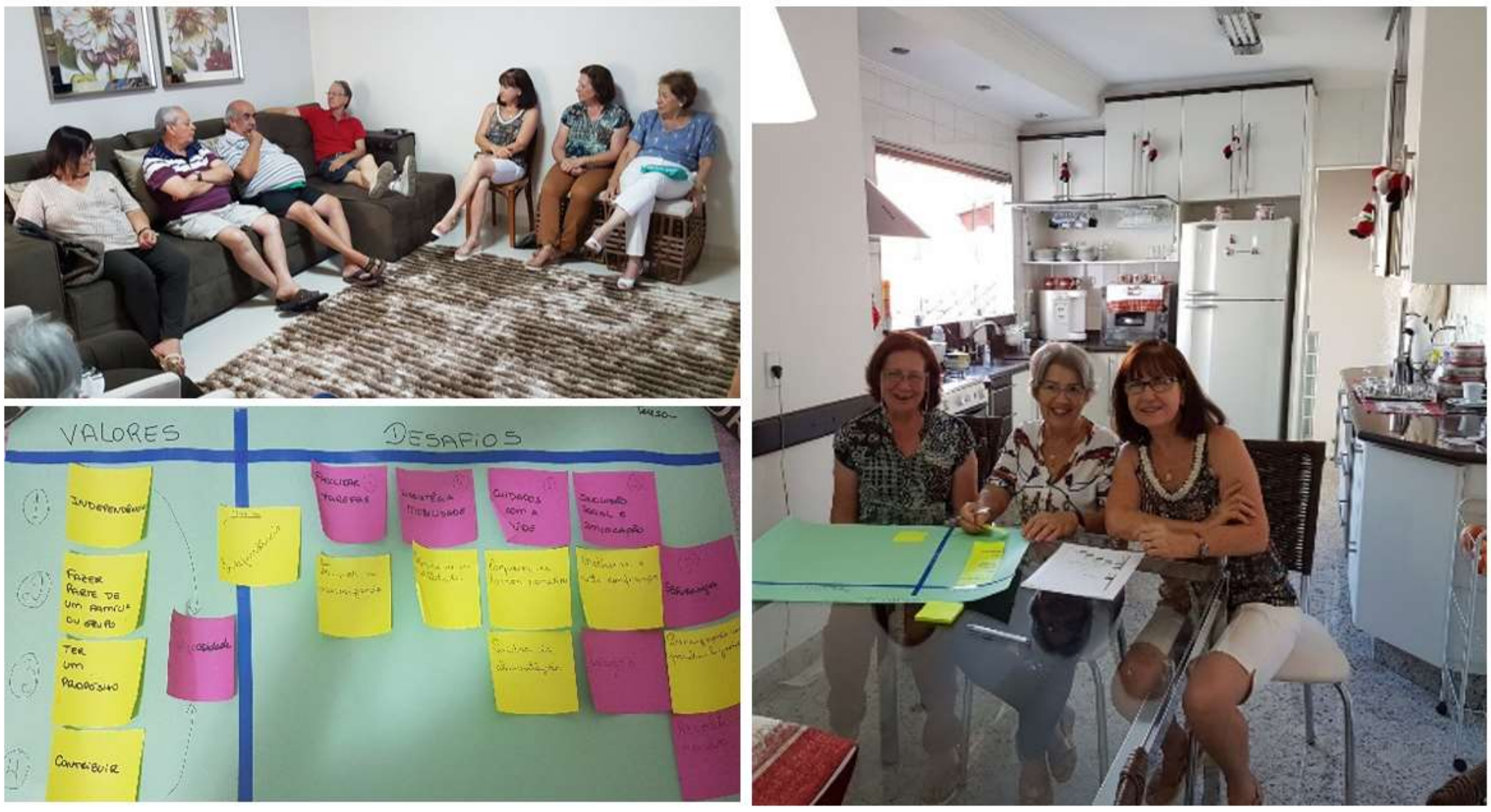

**Figura 55. Etapa 1: *Workshop* de Valores e Necessidades**

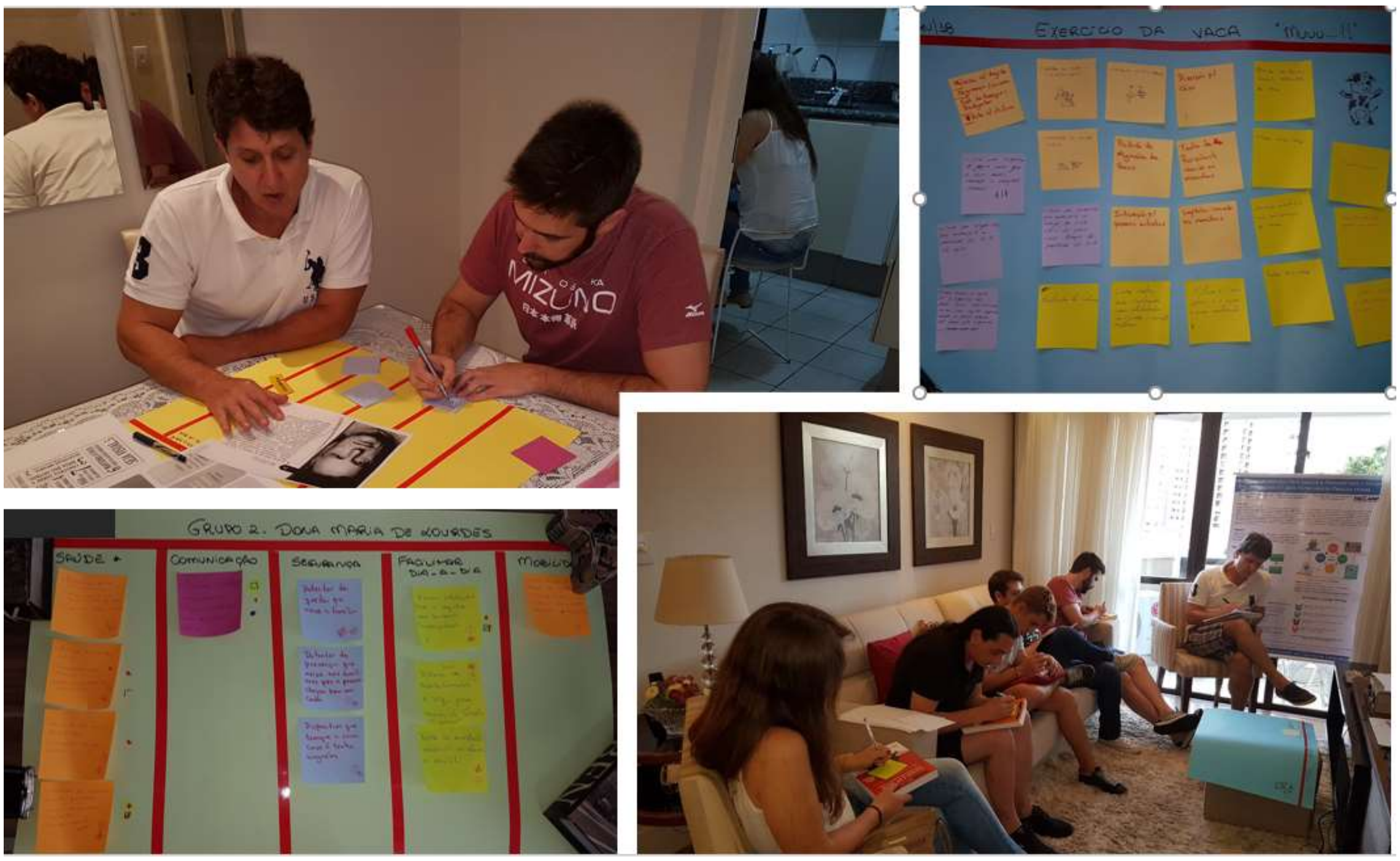

**Figura 56. Etapa 2: *Workshop* de Ideação**

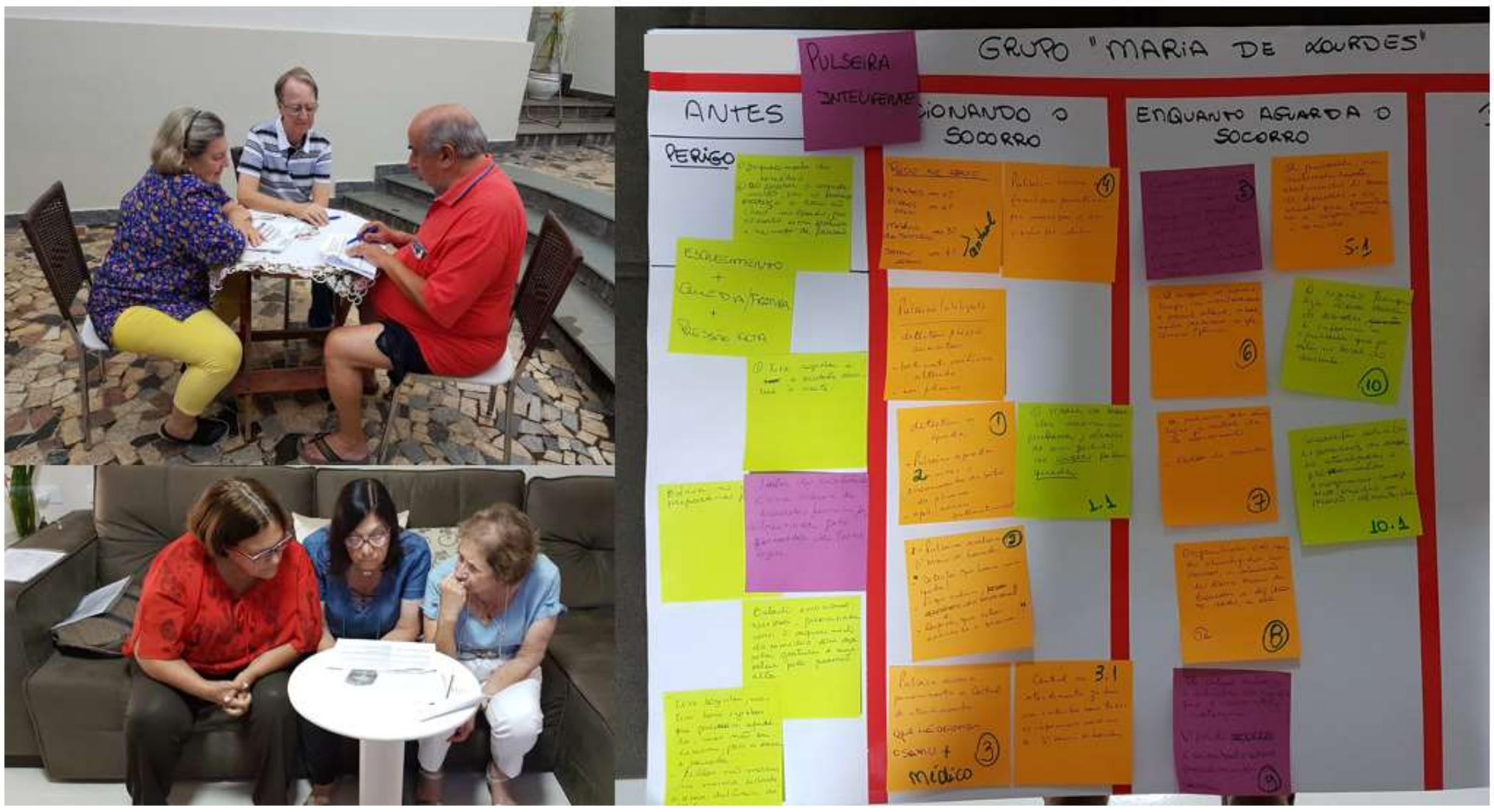

**Figura 57. Etapa 2: Workshop de Prototipação**